\documentclass{aa}
\usepackage{graphics}
\usepackage{tabularx}
\usepackage{hhline}
\usepackage{rotating}
\usepackage{fancybox}
\usepackage{epsfig}

\usepackage{natbib}
\bibpunct{(}{)}{;}{a}{}{,} 

\makeatletter

\begin{document}

\def\beq {\begin{equation}}
\def\eeq {\end{equation}}
\def\bea{\begin{eqnarray}}
\def\eea{\end{eqnarray}}
\def \bef {\begin{figure}}
\def \eef {\end{figure}}
\def\dslash#1{ \hbox{$#1 \hspace{-0.16cm} \slash$} }
\def\ghost#1{ }
\def\d {\partial}
\def\nn {\nonumber}
\def\etal {{\it{et al.}} }
\def\eg {{\it{e. g.}} }
\def\ie {{\it{i.e.}} }
\def\cf {{\it{c.f.}} }
\def\grad{{\bf \nabla}}
\def\div{{\mbox{div}}}
\def\g{{g'}}
\def\sv{{\langle {\sigma v} \rangle}}
\def\svb{{\langle \overline{\sigma v} \rangle}}
 \def\stat{{\epsilon }} 
 \def\hu{{h_{\rm \scriptscriptstyle 70}}} 
 \def\Om{{\frac{\Omega_m \hu^2}{0.3}}} 
 \def\Odm{{\frac{\Omega_{dm} \hu^2}{0.25}}} 
 \def\Ob{{\frac{\Omega_{b} \hu^2}{0.05}}} 
 \def\Th{{{\rm \scriptscriptstyle Th}}} 
 \def\SB{{{\rm \scriptscriptstyle SB}}} 
 \def\nb{{\widetilde {\dslash n} }} 
 \def\xs{{$\svb$ cross-section}} 
 \def\snu{{s_\nu}} 
 \def\srec{s_{rec}} 
 \def\seq{{s_{eq}}} 
 \def\sth{{s_{\Th}}} 
 \def\st{{\ \rm{ref}}} 
 \def\DM{{Dark Matter}}

\title{Constraints on   Dark Matter   interactions from 
structure formation: Damping lengths}
\author{C\'eline Boehm \inst{1},  Richard Schaeffer \inst{2}} 
\institute{Department of Physics, Theory division, CERN, CH-1211 Geneva 23, Switzerland;
LAPTH, 9 Chemin de Bellevue, B.P. 110 F-74941 Annecy-Le-Vieux Cedex, FRANCE 
\and  SPhT, CEA Saclay 91191 Gif-sur-Yvette, France }
\titlerunning{  Dark Matter   damping lengths}
\authorrunning{Boehm {\it{et al.}}}
\date{20/10/2004}
 

\abstract{
Weakly Interacting Massive Particles are often said to be the best Dark Matter candidates. 
Studies have shown that large Dark Matter-photon or  Dark Matter-baryon interactions 
could be allowed by cosmology. Here we address the question of the role of the 
Dark Matter interactions in more detail to determine at which extent Dark Matter has to be necessarily 
weakly interacting. To this 
purpose, we compute the collisional damping (and free-streaming) scales of generic interacting  Dark Matter   
candidates  and investigate the effects on structure formation. 
Our calculations are valid provided the Dark Matter particles have experienced a phase of 
statistical equilibrium at some stage during their evolution.   
By comparing these damping lengths to the scale of the smallest 
primordial structures known to exist in the Universe, we obtain  necessary conditions 
that any candidate must satisfy. These conditions are  expressed in terms of the   Dark Matter 
particles' mass and either the total  Dark Matter interaction rate or 
the interaction rate of Dark Matter with a specific species. The case of Dark Matter interacting with neutrinos or photons is
considered in full detail. Our results are valid even for energy dependent cross-sections and for any 
possible initial fluctuations spectrum. We point out the existence of new Dark Matter scenarios
and exhibit new damping regimes.
For example, an interacting candidate may bear a similar damping than that of collisionless Warm Dark Matter particles. 
The main difference is due to the Dark Matter coupling to interacting (or even freely-propagating) species. Our approach  
yields a general classification of  Dark Matter  candidates which extends the definitions
of the usual Cold, Warm and Hot   Dark Matter   scenarios when interactions,  
weak or strong, are considered.}

\maketitle


\section{ Introduction.}

The damping of primordial fluctuations has been considered 
more than 30 years ago (\cite{joe67,joe68, misner}) when baryonic matter was popular. 
In the early eighties, when the existence of   Dark Matter   became
a serious possibility, it was quickly understood  
that fluctuations made of Weakly Interacting Massive 
Particles will render galaxy formation much easier 
(\cite{gunn, peebles82, bond83}). If the   Dark Matter   interactions are 
weak enough and the   Dark Matter   mass heavier than a few keV, one indeed expects no collisional damping effects, and no free-streaming on scales relevant 
for galaxy formation. Such particles (\eg neutralinos) 
are now embedded in the well-known Cold Dark Matter (CDM) scenario.
As a limiting case, particles which marginally comply with the free-streaming constraints 
 are usually referred to as Warm   Dark Matter   
(WDM) candidates, the archetype  being (\cite{davis, blumen}) particles with a mass of $\sim 1$ keV 
whose damping length is about $100$ kpc.

While both the CDM and WDM initial conditions reproduce rather well 
the mass distribution of the 
observed large scale structures, the CDM scenario is in danger to 
yield an excess of small galaxies ( \cite{rijoe1,cdmcrisis,cdmcrisisb}). 
The WDM model, known to  
reproduce better the observed faint end of the galaxy multiplicity function (\cite{ rijoe2}), 
was revived (\textit{e.g.} \cite{wdm1}) as a possible solution to the 
discrepancy between observations and  CDM numerical simulations. However, 
the collisionless WDM scenario does not seem to be the solution to all discrepancies 
(\cite{julien}, \cite{ostriker}), especially if one considers the constraints from the late 
reionization signal claimed to be found by WMAP( \cite{WMAP,yoshidareio}). 

Some of the potential problems of the CDM 
scenario may possibly be alleviated by invoking astrophysical 
processes (\textit{e.g.} \cite{chiu,stoehr,deckel}) or perhaps by redoing data analysis 
of the dwarf and Low Surface Brightness rotation curves 
(\textit{e.g.} \cite{swaters,autres}). Also, the Dark halo profile of the Milky Way may (\cite{MWnew}) be 
more cuspy than previously found (\cite{binney}). Should the still remaining discrepancy 
between numerical simulations and observations be due to the nature of Dark Matter or not, a thorough 
investigation of the   Dark Matter   properties seems worthwhile.  

In particular, one may wonder whether taking into account the   Dark Matter   
interactions  -- generally considered to be negligible -- 
would help to solve the Dark Matter crisis at small scales.

This is discussed by \cite{sperstein} who consider 
strongly self-interacting   Dark Matter   to  
comply with the shape of the galaxy rotation curves 
(\cite{moore}). 
However, according to further investigations 
 (\cite{yoshida}, \cite{clusters}), this suggestion, at least in its original form, seems 
not to be the answer.   
 
One can then address a more general question to determine which kind of 
interaction  is allowed for   Dark Matter, with which particles, and with which strength. 

The original idea that led to the introduction of weakly interacting Dark Matter particles (that are now embedded in the Cold Dark Matter scenario) was to avoid  Silk damping as well as prohibitive free-streaming.  
In the present paper, we extend this approach to other Dark Matter interactions, say interactions of Dark Matter 
with any other species, including itself
and seek constraints on the allowed Dark Matter particles' parameter space.
First insights were given in \cite{bfsL,brhs,chen}. We here  undertake an exhaustive and extensive analysis by 
considering also new aspects, considering in particular the way the   Dark Matter particles achieve their relic density.

We start by relaxing the assumption of collisionless   Dark Matter   to estimate the   Dark Matter   collisional 
and free-streaming damping effects. Requiring that these two damping mechanisms do not prevent the formation of 
observed structures allows us to define the range of   Dark Matter   mass and interaction rates that are compatible 
with observations. This is independent of any particle physics. We  then classify all generic   Dark Matter  
candidates in a 2-dimensional parameter space (interaction rate versus mass) and determine what regions of this 
parameter space are allowed by the present observations.  

Our calculations are valid for massive neutrinos, provided their mass is within the known experimental and observational limits. 
For the sake of comparison with the general case we treat here, we use at various places a "reference", the cosmology in which i) the radiation has two components (namely photons and massless neutrinos, both having the standard temperature) and ii) the matter is made of two species, the baryons and the Dark Matter.

As pointed out in \cite{bfsL}, the collisional damping scale of the Dark Matter fluctuations 
can be expressed in terms of two separate contributions, namely the  
\emph{\,self-damping\,} and \emph{\,induced-damping\,} contributions. 

The self-damping is related to the collisional propagation 
of the   Dark Matter   particles. It may be directly compared with their 
free-streaming scale. 

The induced-damping 
corresponds to the (collisional 
or free-streaming) damping 
acquired by a species $i$ and transmitted to 
the   Dark Matter   fluctuations as long as the coupling between   Dark Matter   and this  
species is large enough. The induced-damping effect can  be seen 
as a generalization of the Silk damping (of baryons coupled to photons), here applied to the case
of   Dark Matter   particles coupled to any possible species $i$. We  especially focus our attention on the 
coupling to neutrinos and  photons, seen to provide a quite large damping and  yield rather stringent  constraints.
 
Our results are given in two separate papers. 
The present one (referred to as paper I) is dedicated to 
the calculation of the free-streaming and collisional damping scales 
associated with any interacting  Dark Matter   particles. We then derive 
the mass and interaction rates of the   Dark Matter   candidates 
allowed by structure formation. 
This is a rather technical paper. The second one (paper II), deals with the 
physical and astrophysical relevance of the candidates whose  
damping scales, calculated here, are not prohibitive.
These two papers have been written in a self-contained way and can be 
read independently.

The present paper is organized as follows:
in Section \ref{sec:transp}, 
we recall the expressions of the different transport coefficients for 
an imperfect fluid made of several species (relativistic 
or not), and take this opportunity to discuss issues for which there is still some confusion in the litterature. 
In Section \ref{sec:dampscales}, we derive the expressions of the 
self-damping and induced-damping lengths. We also point out the existence of a new damping effect, that we call 
 \textit{mixed-damping}. The latter is a special case of induced damping. It describes the 
damping acquired by Dark Matter particles that would be coupled to a species which is free-streaming. In Section 
\ref{sec:sdfsdamp}, we derive the constraints on the Dark Matter properties from the calculation of the 
free-streaming and self-damping scales.  For the  specific case of Dark Matter coupled to
neutrinos or photons, the calculation of the 
induced-damping lengths and the derivation of the associated constraints on interaction rates and cross-sections 
are done in Sections \ref{sec:idnudamp} and  \ref{sec:idphotdamp} respectively. 

The definitions of the quantities used in this paper are given
in Appendix \ref{app:def}.  We derive in Appendix \ref{app:denssym} expressions for the evolution
of the Dark Matter density, slightly simplified to be systematically used in our analytical calculations. The interaction rates appearing in the present work are given in the same appendix.  The regions of the Dark Matter parameter space (Dark Matter particles' mass and interaction rates) where the various 
damping lengths take different forms are explicitly given in Appendix \ref{app:bord} . Finally, in Appendix \ref{app:anndamp}, we display analytical expressions for the damping scales and for the associated limits on the Dark Matter interaction rates and cross-sections.

\section{ Transport coefficients. \label{sec:transp} }

To study the collisional damping of Dark Matter  fluctuations,  we first need to 
determine the transport coefficients of an imperfect fluid  in statistical 
\footnote{We use the word thermal equilibrium for particles whose temperature and velocities equilibrate 
at least locally, but which are not in chemical equilibrium. Thermodynamical equilibrium is meant for particles 
in both thermal as well as chemical equilibrium. Statistical equilibrium stands for particles either in thermal or 
in thermodynamical equilibrium.} equilibrium. The relevant coefficients are the shear viscosity, the heat conduction 
and the bulk viscosity. They appear in the expression of the energy-momentum 
tensor of an imperfect fluid, see \textit{e.g.} \cite{weinberg}.

Typically, we are lead to evaluate the transport coefficients of a mixture 
made of several species $i$ 
(including   Dark Matter). 
The key which will enable us to obtain constraints resulting from these transport coefficient on any kind of Dark Matter  will be to write these transport coefficients common to all species in the mixture as a sum over specific contributions. The latter may be viewed as the share of each of the species, contributing to the viscosity of the whole fluid.
\footnote{It is worth emphasizing that the statistical equilibrium of the fluid ensures that each of the species experiences the same transport coefficients, given by the sum of all contributions.}
The requirement, in a second stage, that {\it each} of the contribution in the sum, separately, must not be the source of prohibitive damping is then much simpler and will yield {\it necessary conditions}. 

We proceed now to get the transport coefficients in the form we seek.

\subsection{Shear viscosity.
\label{sec:shear}}

To derive its expression, we can start from the form given by 
\cite{chapman}: $ \eta = \,\sum_i \eta_i\,$  valid for 
a composite fluid made of \textit{non-relativistic} species $i$ 
in statistical equilibrium. The 
individual contribution \,$\eta_i $ of each  species $i$ 
is given by  $\eta_i= p_{i}/\Gamma _{i}$ where 
\beq 
{\Gamma _{i}} \ = \ \sum_j\ {\Gamma _{ij}} 
\eeq
represents the total interaction rate of species $i$. 

The specific interaction rate of 
species $i$  with species $j$ will be written as
\beq 
{\Gamma _{ij}} = \svb_{ij} \ n_j
\label{svb}
\,  , 
\eeq
where $n_j$ is  the number density of the species $j$ and $\,\svb_{ij}\,$ 
is the {\it suitably weighted} statistical average of the interaction 
cross-section between $i$ and $j$.
Under this form, it seems natural to expect that this expression can be extended to the case of 
\textit{relativistic}  
particles. Therefore one can write
\beq
\eta  \ =\  \sum _{i}\ \:\frac{p_{i}}{\Gamma _{i}} \,  , \label{viscp} \eeq where the sum over $i$ runs over all the species in statistical 
equilibrium with the fluid.
The pressure $p_i$ may be written as 
$ p_i = \frac{1}{3}\  v_i^2 \,\rho_i $, where  $v_i$ is the particle's 
 r.m.s. velocity in units of $c$.
This yields the shear viscosity coefficient 
of a fluid made of \textit{relativistic and non-relativistic} 
species in statistical equilibrium:
\begin{equation}
\eta  \ =\ \sum _{i}\ \frac{\rho _{i}\: v_{i}^{2}}{3\ \Gamma _{i}} \  . \label{visc} 
\end{equation} 

One can understand the origin of the expression (\ref{visc}), and convince oneself that it is valid also for 
relativistic species, by rederiving it directly 
from  kinetic theory. 
The transport of momentum is induced by 
a gradient in the collective fluid velocity $\bf{V}$ 
carried by all species of the fluid. 
The momentum flux induced by this  
transport of momentum may be written $\eta \grad \bf{V}$, and corresponds to a term  $\eta ( \partial V^a/\partial x_b+\partial V^b/\partial x_a-\frac{2}{3} \delta_{ab}\div{\bf V}) $  in the spatial part  $T^{ab}$ of the energy-momentum tensor.

Each species $i$ is expected to contribute its own share to the total transport of momentum, 
which we denote as  $\eta_i \grad \bf{V}$. We may derive the quantity $\eta_i$ by considering
the  momentum flux carried by each of the microscopic particles of species $i$. 
This momentum flux is  ${\bf v} \ e {\bf V}$ where ${\bf v} = {\bf q}/e$, ${\bf q}$ is the particle's velocity, 
${\bf q}$ the particle's momentum and $e$ its energy: each particle of species $i$ hence carries  
the collective momentum $e {\bf V}$.
It is transported over a distance of the order of the particle's 
mean-free-path v$ / \Gamma _{i}$, in the direction ${\bf {v}}$, inducing a net transport 
$ {\bf v} e {\bf V}({\bf x}) \  - \  {\bf v}  e {\bf V}({\bf x} -{\bf v} / \Gamma _{i}) $. 
The latter may be rewritten as $\frac{q^2}{3e} /\Gamma _{i}\ \grad \bf{V}$, taking advantage of the 
smallness of the mean-free-path and the isotropy of the momenta. After summing over all momenta of  
particles of a species $i$, the flux due to a gradient in $\bf{V}$ is then seen to be ${p_i}/{ \Gamma _{i}} 
\grad \bf{V}$, which translates into ${p_i}/{\Gamma _{i}} \   
( \partial V^a/\partial x_b+\partial V^b/\partial x_a-\frac{2}{3} \delta_{ab}\div{\bf V})  
$ in  $T^{ab}$. The coefficient 
$\eta_i$ associated with species $i$
is thus given by $ \eta_i \, = \  {p_i}/{ \Gamma _{i}} $.
This is similar to the usual result for the shear viscosity of a single 
species $i$. The interaction rate $\Gamma _{i}$, however, is the total 
interaction rate, \textit{i.e.} 
it includes all possible interactions with other species present in the 
composite fluid. 
Since the latter arguments do not depend on whether species $i$ 
is relativistic or not, the result holds for both 
\textit{relativistic and non-relativistic} species.

The total momentum transported by all species in the composite fluid is the sum 
of the amount of momentum 
transported by each species. Therefore, the shear viscosity 
coefficient of a composite fluid  is given by 
$ \eta = \,\sum_i \eta_i$, which justifies (\ref{viscp}) and (\ref{visc}).

\subsection{Heat conduction.
\label{sec:heat}}

One can also derive the expression of the heat conduction coefficient 
for a composite fluid (including relativistic and non-relativistic 
species) by  using kinetic theory. 
The expression of this coefficient for non-relativistic particles has 
not been given in the appropriate form by Chapman and Cowling. However, they 
mentioned that it should be similar to the one obtained for the shear viscosity coefficient.

A given species $i$ contributes to the transport of 
energy induced by a gradient of temperature through its own 
conduction coefficient $\lambda_i$. The latter is proportional to the 
particle's mean-free-path $ v_i / \Gamma _{i}$, times a flux factor $v_{i} \d \rho_i /  \d T$. 
The difference with the previous case 
(\ref{sec:shear}) is that, here, we consider the energy flux 
due to a change of temperature. Specifically, we get
\footnote{The derivatives with respect of the temperature are to 
be taken at fixed volume.}
\beq
 \lambda_i T =  \frac{\rho _{i} v_{i}^{2}}{3\ \Gamma _{i}} 
\frac{\d  \ln \rho_i}{\d  \ln T} \ , 
\eeq
which holds 
whether the fluid is relativistic or not. The total energy transported 
by all species is the sum of the individual contributions. Yet, the 
heat conduction coefficient of a composite fluid is given by 
$\lambda = \sum_i \lambda_i$.
This finally  yields
\beq
\label{heat}
\lambda  T \ = \
\sum_i \ \frac{\rho _{i}\: v_{i}^{2}}{3\ \Gamma _{i}} \ \ 
\frac{\d  \ln \rho_i}{\d  \ln T} \,.
\eeq

\subsection{Bulk viscosity.
\label{sec:bulk}}

The bulk viscosity  is related to departures from  equilibrium induced by a non-zero divergence of the 
velocity field. 
The latter induces a compression or expansion of the fluid elements which may  result 
in local conditions differing from the global ones. The former
have to be compensated to recover the general statistical equilibrium 
conditions. This required adjustment  provides for another source of viscosity.

Departures from chemical equilibrium also induce bulk viscosity contributions. 
The latter have already been applied to other, specific, problems (e.g. \cite{hasch}), and may be included by using similar
techniques. 
They allow to constrain the chemical reaction rates, in addition to the collision rates we
consider in the present paper. 

We consider in detail below the departures from thermal equilibrium, specifically. All the conditions that we obtain in this paper on the collision rates thus are necessary conditions.  
They must be satisfied separately, keeping in mind that there may exist  additional constraints 
on the chemical reaction rates.

It is worth to note, at this stage, that our calculation of the bulk viscosity due to departures from thermal equilibrium turns out to go much beyond what may be found in the existing literature.

\subsubsection{Single fluid. \label{sec:singbulk}}

A straightforward resolution of the transport equations, as outlined in \cite{bernstein}, yields for a single 
species $i$ \beq \label{singbulk} \zeta_i^{(single)} = \frac{{\d  \rho_i}/{\d  \ln T}}{\Gamma_i}  \ 
\left[ \omega_i -  \left(\frac{{\d  p_i}/{\d  \ln T}} {{\d  \rho_i}/{\d  \ln T}} \right)^2 \right] \eeq

The rate $\Gamma_i$ is the collision rate for particles of species $i$ with themselves. The coefficient $\omega_i$ is a statistical average over momentum-dependent functions, describing how the stretching of the momenta by the expansion (or compression) directly affects pressure and density while the ratio $\frac{{\d  p_i}/{\d  \ln T}} {{\d  \rho_i}/{\d  \ln T}} $ describes how the final equilibrium distribution (i.e. the temperature) is affected. For a single fluid, the source of bulk viscosity is therefore 
the offset between the expansion (or compression) which affects the momenta producing a non-equilibrium distribution and statistical equilibrium which tries to cope with this change by readjusting the temperature. The Boltzmann equilibrium distribution turns out to be preserved for an ultrarelativistic fluid ($p \to p/a$ inducing exactly $T \to T/a$) as well as for a non relativistic fluid of conserved particles ($p \to p/a$ inducing exactly $T \to T/a^2$), so the bulk viscosity vanishes in both cases. In intermediate situations, this is not so: the deviations of  $\omega_i$ from $\frac{{\d  p_i}/{\d  \ln T}} {{\d  \rho_i}/{\d  \ln T}}$ however  are quite small. A numerical estimate shows that  ${\zeta_i^{(single)} }/{\eta_i}$ in general barely exceeds  the {1} \% level. This ratio however is sizable for the case of zero chemical potential annihilating massive particles, for instance ${\zeta_i^{(single)} }/{\eta_i} \sim 0.5$ at $ T \sim m/20$, reaching asymptotically $2/3$ for the (unrealistic: it corresponds to a number density $n_i \propto e^{-m/T} \to 0$) case $m/T \to \infty $. 

The result (\ref{singbulk}) for a single fluid  is consistent with Bernstein's result \cite{bernstein}.
However, Bernstein considered only massive particles in the limit $m/T \ll 1$, that is ultra relativistic particles. 
The contribution (\ref{singbulk}) does not exist in the approach of \cite{weinberg} who  considered either 
strictly massless or massive non relativistic, stable ($m/T \gg 1$) particles. In the latter two cases, indeed,  
the contribution (\ref{singbulk}) vanishes. As we shall see in the next section, there is additional bulk viscosity 
generated in a composite fluid, which is the question addressed by Weinberg. 
Despite the comparison made in (\cite{bernstein}), these two authors  describe different physical phenomena, 
with no overlap of their results.

\subsubsection{Composite fluid. \label{sec:mixbulk}}
 
 For a composite fluid, the total bulk viscosity is not simply the sum 
of (\ref{singbulk}) over all species $i$ which are coupled together. 
There is an additional source of inelasticity (compared to the single fluid 
case) which originates from the constraint that each of the 
components $i$ should acquire the general temperature of the medium rather than 
the one it would acquire if it was isolated. This implies a rearrangement of the equilibrium distribution 
directly related to  exchanges among the various species of the fluid. This is in contrast with the transport of 
collective motion or temperature, for which each species contributes independently. 
 
 This calculation was first done 
by {\cite{weinberg}. The latter is enlightened by
 an interesting discussion due to  \cite{zimdahl}.
 
  With the assumption that the species $i$ are in statistical equilibrium, one gets, for a composite fluid~:
  \beq
 \label{sumbulk}
\zeta = \delta \rho  \ 
\left[ \omega -  \left(\frac{\delta  p} {\delta  \rho} \right)^2 \right] \eeq  with \[  \delta \rho = \sum_i  \frac{{\d  \rho_i}/{\d  \ln T}}{\Gamma_i}  \, \]
  \[  \delta p = \sum_i  \frac{{\d  p_i}/{\d  \ln T}}{\Gamma_i} \]  and
\[  \omega =   \sum_i \frac{{\d  \rho_i}/{\d  \ln T}}{\Gamma_i} \omega_i \ \ / \ \
\sum_i  \frac{{\d  \rho_i}/{\d  \ln T}}{\Gamma_i} \ . \]
 The weighting by the inverse of the interaction rates 
(\ie by the collision time of each species $i$ with all species within the 
fluid) is new. It arises from the actual resolution of the transport equations in the relaxation time approximation, 
the key being to use different collision times for each species (this is the very same approximation as in the 
previous cases for the shear viscosity and heat conduction) and to account for energy conservation.

The expression (\ref{sumbulk}) may be rearranged as
 \beq
 \label{sumzeta}
 \zeta =  \sum_i  \zeta_i \ ,
 \eeq
 with 
\begin{eqnarray}
\label{bulk}
\zeta_i  &=&  \frac{{\d  \rho_i}/{\d  \ln T}}{\Gamma_i}  \ 
\left[ \omega_i -  \left(\frac{{\d  p_i}/{\d  \ln T}} {{\d  \rho_i}/{\d  \ln T}} \right)^2 \right] \cr &+&  \frac{{\d  \rho_i}/{\d  \ln T}}{\Gamma_i}  \ 
\left(
 \frac{{\d  p_i}/{\d  \ln T}} {{\d  \rho_i}/{\d  \ln T}} - \frac{\delta  p} {\delta  \rho} 
 \right)^2 
 \  .
\end{eqnarray}

 The first term {corresponds to} the bulk viscosity {of a single fluid} due to the fact the temperature of species 
 $i$ has to adjust to any expansion/compression, which is a constraint when particles are neither fully 
 ultra-relativistic nor purely non relativistic. It differs however from   (\ref{singbulk}) by the interaction 
 rate $\Gamma_i$ which is now the collision rate of particles of species $i$ with all the species in the composite 
 fluid (and not only with themselves). Indeed, all  collisions contribute to the adjustment of the temperature.

 The second term of (\ref{bulk}) is a contribution specific to a composite fluid~:
  \beq 
\label{mixbulk}
\zeta^{(composite)}_i  \sim 
 \frac{{\d  \rho_i}/{\d  \ln T}}{\Gamma_i}   \ 
\left(
 \frac{{\d  p_i}/{\d  \ln T}} {{\d  \rho_i}/{\d  \ln T}} - \frac{\delta  p} {\delta  \rho} 
 \right)^2 
 \  .
\eeq
 It is due to the fact that the temperature of species $i$ also has 
to adjust to the change of the general temperature of the medium under the 
expansion/compression. Such a change represents another constraint when 
the equations of state of the medium (and of species $i$) are not the same;  
more specifically when 
$\frac{{\d  p_i}/{\d  \ln T}} {{\d  \rho_i}/{\d  \ln T}} \ne \frac{\delta  p} {\delta  \rho}$. 
Indeed, the contribution (\ref{mixbulk}) may be viewed as the pressure offset due to a departure from 
equilibrium, by  $\Delta f_i = \Delta T_i \ \d f_i/\d T$, of the Boltzmann distribution $f_i$ of species $i$, 
in-between two collisions, with  $\Delta T_i =  \frac{{1}}{\Gamma_i}  \ 
 \left( \frac{{\d  p_i}/{\d  \ln T}} {{\d  \rho_i}/{\d  \ln T}} - \frac{\delta  p} {\delta  \rho}  \right) \ 
 \div {\bf V}$.
  Note that this form induces no offset in the total energy-density, as required by energy conservation 
  \footnote{This can be explicitly checked: $\Delta \rho = \sum_i \int d^3p_i e_i \Delta f_i 
=  \sum_i {\Delta T_i}/{T} \ {\d  \rho_i}/{\d  \ln T}$ indeed vanishes due to the weighting, originally 
required to this purpose, by the factors ${{1}}/{\Gamma_i} $.}. 

The form (\ref{mixbulk}) is in agreement with the expression derived by \cite{zimdahl}. However, our result  
is obtained under the simple and unique assumption that all species are coupled and in statistical equilibrium. 
We also take into account, in accordance with the style of the present paper but to contrast with previous work, 
that  the coupling  of each species $i$  with the whole medium is, as a rule, different. On the other hand, 
\cite{zimdahl} makes much more restrictive assumptions, since they are sufficient to make his point. 
He considers species $i$ which are already in equilibrium with themselves at different internal temperatures 
$T_i$, under processes he does not show explicitly, and uses for his actual calculation a unique interaction rate 
for the various species just to equilibrate their temperature.

In  \cite{zimdahl}, the author noted his result was not the one obtained by  \cite{weinberg}, but did not 
provide the explanation for this. Our form (\ref{mixbulk}) readily shows that the contribution of the photons 
($i = \gamma$) to the sum (\ref{sumzeta}) specifically  is the (sole) term retained by \cite{weinberg}. It is 
implicit in the latter work that there is a second contribution due to the matter component  ($i = m$) which 
however was  omitted on grounds the matter interaction rate ($\Gamma_m$ in our notation) is assumed to be very 
large. But due to cancellations in the square of (\ref{mixbulk}), the pre-factor does not determine the size of 
the contribution. Indeed, this omitted second contribution is likely to be in most cases the dominant one. This 
is the explanation for the difference between these two authors, and will be discussed  in Section \ref{sec:mrbulk} 
using the same composite fluid these authors have considered.

 The contribution  (\ref{mixbulk}) to the bulk viscosity vanishes for a single fluid. 
 So, as a rule, in (\ref{mixbulk}), when the coefficient $ \frac{{\d  \rho_i}/{\d  \ln T}}{\Gamma_i} $ in front of 
 the bracket  is large, the two terms within the bracket nearly cancel each other since the composite fluid is 
 dominated by the single component $i$. Conversely, when the bracket is of order unity, it is necessarily for a 
 subdominant species for which there is no cancellation. So, $\zeta$ in general never gets very large as compared to 
 $\eta$.  {This is discussed  in Section \ref{sec:mrbulk} for the generic example of a composite fluid made of matter 
 and radiation. } In all cases, we find $\zeta \le  \frac{4}{3} \eta $, if not $\zeta \ll  \frac{4}{3} \eta $. 
 As we will see below, in Section \ref{sec:dampscales}, the viscosity terms enter the damping expression with a 
 weight $ \zeta + \frac{4}{3} \eta $.  For the sake of the present study,  we simply conclude that  the bulk 
 viscosity will not induce any new constraint as compared to the shear viscosity, a property which has been 
 already used in \cite{bfsL}. So, we do not need to consider the bulk viscosity for the limits on the interaction 
 rates we seek.

Our result are far more general than what is needed in this paper 
and more accurate than the ones which may be found in the literature. We give indeed 
the expression of  the bulk viscosity for a composite fluid as a sum of 
several well-identified contributions, due to different physical phenomena. This sets into perspective the various 
calculations done by the different authors. Also, we take into account the difference 
between the collision rates of various species, a move that is required by our aim to establish limits on these rates. 
We point out several inaccuracies in the literature.  Assuming an average interaction rate, 
as done in previous work, leaves the latter conclusions unchanged. Needless to say, our findings are in agreement with 
the statements made in \cite{bfsL}.

\subsubsection{Mixture of matter and radiation.\label{sec:mrbulk}}

 As an illustration of the above results, we discuss here explicitly the bulk viscosity of a composite fluid made  of radiation (massless, relativistic, particles with zero chemical potential and a unique
 collision rate $\Gamma_r$) and matter (non relativistic particles with  a unique rate $\Gamma_m$), assumed to form a unique fluid  in statistical equilibrium. We discuss the various relevant cases, and provide the explanation for the different, and contradictory, results found in the literature. We hope this will clear out the matters.

 The  bulk viscosity for the composite fluid is the {\it sum} 
\beq
\zeta = \zeta_r  +   \zeta_m 
\eeq
and has the same value for the matter as well as the radiation. 
Note that  -- although $\zeta_r$ is the contribution to the 
bulk viscosity from the relativistic species -- 
 the bulk viscosity coefficient of the relativistic species is not $\zeta_r$ but $\zeta$. Similarly, the contribution of the non-relativistic species to the bulk viscosity is 
$\zeta_m$, but its bulk viscosity  of course also is $\zeta$, the two species being part of the same fluid.

This bulk viscosity may be usefully compared to the shear viscosity of the composite fluid~: 
\beq 
\eta = \eta_r +\eta_m \ , 
\eeq 
related to the pressure $p_r$ of the radiation and $p_m$ of the matter by  $\eta_r = p_r/\Gamma_r$ and 
$\eta_m = p_m/\Gamma_m$.

\paragraph{The case of conserved matter particles.}
We first consider the case of stable matter particles, with a conserved particle number. There is then no 
contribution of the form (\ref{singbulk}). From (\ref{mixbulk}), we get
 \begin{eqnarray}
 \zeta_r &=& \frac{{\d  \rho_r}/{\d  \ln T}}{\Gamma_r}   \ 
\left( \frac{1} {3} - \frac{\delta  p} {\delta  \rho}  \right)^2 \\ &=& \nonumber
  \frac{4}{3}  { \eta_r}
   \left(\frac{\eta_m /8 } {\eta_r  + \eta_m / 8} \right)^2
   \ ,
\end{eqnarray}
and
\begin{eqnarray}
 \zeta_m &=&  \frac{{\d  \rho_m}/{\d  \ln T}}{\Gamma_m}   \ 
\left( \frac{2} {3} - \frac{\delta  p} {\delta  \rho}  \right)^2 \\ &=& \nonumber \frac{4}{3} { \eta_m}/{8 }
   \left(\frac{\eta_r} {\eta_r  + \eta_m / 8} \right)^2
   \  ,
\end{eqnarray}
since
\[  \delta \rho =  {{12\eta_r}} +  \frac{3}{2 } \eta_m \, ,\]
  \[  \delta p =  {{4\eta_r}} +  {{\eta_m}}  \,  . \]
This yields
\beq
\zeta = \frac{4}{3} \frac{\eta_r \ \eta_m/8 }{\eta_r + \eta_m/8} \ . \label{zetarm} \eeq

For  $ \eta_r  \gg  \eta_m$ we have $ \zeta_r \sim \eta_m^2/48\eta_r  \, \, (\ll \eta_m) \ll  \eta_r $  and 
$\zeta_m \sim  \eta_m/6 < \eta_m$. One can see in this example that, when the bound $\zeta_i \sim \eta_i$  is close 
to be reached, it is for a subdominant species. Indeed, we have $\zeta  \propto \eta_m \ll \eta  \sim \eta_r$. 
For the converse, somewhat less realistic assumption  $ \eta_r  \ll  \eta_m $, we would get 
$\zeta \sim \zeta_r \propto \eta_r \ll \eta \sim \eta_m$. This implies in particular $\zeta \ll \frac{4}{3} \eta$. 
The bulk viscosity is, as a rule, much smaller than the shear viscosity.

\paragraph{Annihilating matter.}

Here we consider the case of annihilating matter, when the chemical potential of the massive particles also vanishes.  
More specifically we consider the limit of large $ m/T$ (having in mind $m/T \sim 20$) where $m$ is the mass of the 
annihilating particles and $T $ their temperature.

There is in this case a sizeable contribution of the form (\ref{singbulk}) to $\zeta_m$ as discussed in the 
corresponding section~: 
\beq 
\zeta_m \le \frac{2}{3} \eta_m \label{zetasm} \ . 
\eeq Were the matter not coupled 
to radiation, this would be the only contribution to the bulk viscosity, with then  $\zeta < \frac{4}{3} \eta$ 
since $\zeta = \zeta_m$ and $\eta = \eta_m$ in this case.

For matter coupled to radiation,
there is also a contribution of the form (\ref{mixbulk}) specific to a composite fluid. We get, with 
$b  = \frac{m^2}{12T^2}$,  
\begin{eqnarray}
 \zeta_r &=& \frac{{\d  \rho_r}/{\d  \ln T}}{\Gamma_r}   \ 
\left( \frac{1} {3} - \frac{\delta  p} {\delta  \rho}  \right)^2 \\ &=& \nonumber
    \frac{4}{3} \eta_r
   \left(\frac{b  \eta_m } { \eta_r  + b  \eta_m} \right)^2
   \ ,
\end{eqnarray}
and
\begin{eqnarray}
 \zeta_m &=&  \frac{{\d  \rho_m}/{\d  \ln T}}{\Gamma_m}   \ 
\left( \frac{T} {m} - \frac{\delta  p} {\delta  \rho}  \right)^2 \\ &=& \nonumber
  \frac{4}{3} b  \eta_m
   \left(\frac{ \eta_r } { \eta_r  + b  \eta_m} \right)^2
   \  ,
\end{eqnarray}
since
\[  \delta \rho =  {{12\eta_r}} +  \frac{m^2}{T^2 } \eta_m \, , \]
  \[  \delta p =  {{4\eta_r}} + \frac{m}{T } {{\eta_m}}  \,  . \] 
This yields for the contribution (\ref{mixbulk}) to (\ref{bulk})~: 
\beq 
\zeta \sim  \frac{4}{3} \frac{\eta_r \ b \eta_m }{\eta_r + b  \eta_m} \ . \label{zetarma} 
\eeq

It is readily seen, for the contributions (\ref{mixbulk}) to $\zeta$, that $\zeta_r \le  \frac{4}{3} \eta_r$. 
One has also  $\zeta_m \le  \frac{4}{3} b  \eta_m$, but this is not a useful relation since $b  \gg 1$. 
A better bound can be obtained by noting that, whatever the value of $b  \eta_m$, one has 
$\zeta_m \le  \frac{1}{3} \eta_r$, so the damping constraints obtained from $\zeta_m$ do not lead to any new 
constraint once the damping due to $\eta_r$ is taken into account. This is natural since (\ref{zetarma}) implies 
$\zeta < \frac{4}{3} \eta$~: so, when the constraints obtained from $\eta$ are satisfies, 
there are no new ones which appear due to $\zeta$.

The bound $\zeta < \frac{4}{3} \eta$ still holds when the contribution (\ref{zetasm}) is added to $\zeta$:  
the bulk viscosity is not expected to bring in any stronger constraint than the shear viscosity. 

\subsection{Decoupling and interaction rates. \label{sec:symrates}}

In Section \ref{sec:transp}, we gave the transport coefficients as a sum 
over all species $i$ to which    Dark Matter   is coupled. 
The coupling condition  may be written as $\Gamma_{dm-i} > H$. 
More generally, a species $j$ is coupled to a species $i$ when $\Gamma_{ji} > H$. 
This condition ensures that the fluctuations of the particles $j$ 
follow those of particles $i$. More specifically this ensures that  
the fluctuations of particles $j$ will be erased at the same scale 
as those of species $i$.

The interaction rates
$\Gamma_{dm-i}$ and $\Gamma_{i-dm}$ are  in principle not equal. There are several reasons for this. First, the number of Dark Matter particles and of particles belonging to species $i$ may be different. Also, the relevant cross-section $\sv_{dm-i}$ is weighted by the momentum which is exchanged during the collision. This transfer may be quite different, depending on whether one considers the momentum transfer to a Dark Matter particle hitting a particle of species $i$ or whether it is the converse.

From the conservation of momentum, the relation between those two interaction rates can be written as
\begin{equation}
\dslash{\rho}_i \,{\Gamma }_{ij}   \ = \ 
\dslash\rho_j \,{\Gamma }_{ji} \, ,
\label{ratmom}
\end{equation}
with  $ \dslash \rho_i = \rho_i + p_i\,$.   
So, the interaction rates $\Gamma_{ij}$ and $\Gamma_{ji}$ are quite naturally seen to be different. The condition $\Gamma_{dm-i} > H$, for instance, does not necessarily imply $\Gamma_{i-dm} > H$.
This leaves also the possibility for  $\Gamma_{i-dm} < H$ and $\Gamma_{dm-i} > H$, or more specifically  
$\Gamma_{i-dm} < \Gamma_{i} < H < \Gamma_{dm-i}$. 
In this case, when the coupling 
between    Dark Matter   and species $i$ is efficient enough to modify the properties 
of the   Dark Matter   fluid but not that of species $i$, one  expects the 
damping of the freely-propagating species $i$ to be transmitted 
to the   Dark Matter   fluctuations. This, of course, occurs only in some range of 
parameters.

The relation (\ref{ratmom}) is due to the internal weightings within 
the statistical averages $\svb$ defined by eq. (\ref{svb}) since the interaction probability is 
weighted by the momentum transferred. This is
to contrast with what we denote as the $\sv$  cross-section which is the usual cross-section which counts 
the number of interactions. The averages $\svb_{ij}$ and $\svb_{ji}$ in general are not equal~:
\begin{equation}
 \dslash \rho_i \ n_j \,\svb_{ij}    \ =\ 
\dslash \rho_j \ n_i \,\svb_{ji} \ \ ,
\label{sigmom}
\end{equation}
so that 
$ E_i \ \svb_{ij} \, \sim \, E_j \ \svb_{ji}$\ where $E_i$ is the average energy per particle. 
Whether species $i$ is relativistic or not, one can write $E \sim  3T  c^2/v^2$, so for species $i$
and $j$ at the same temperature, there is the rule
$ v_j^2 \ \svb_{ij} \, \sim \, v_i^2 \ \svb_{ji}$.
In the rest frame of the universe (where the blackbody radiation is isotropic), light  particles hitting a much 
heavier particle will undergo 
substantial scattering, so each collision is associated to a transfer of momentum, and $\svb$ is of the order of  
$\sv$. On the other hand, heavy 
particles hitting the light ones are barely scattered and there is little momentum transfer, so $\svb$ is, as 
compared to $\sv$, down by a factor 
$v^2/c^2$, much lower than unity.
\footnote{The same  phase-space arguments obviously hold for the interaction rates governing the transport coefficients. This is illustrated by a 
well-known result for non relativistic particles (see \eg \cite{balian}). The rate $\svb_{ij} \sim \sv\vert_{ij} 
\equiv \sv\vert_{ji}$ appears in 
the Lorentz approximation for light particles  $i$ hitting heavy ones $j$. The rate to be used 
for brownian motion, when one considers heavy particles $j$ moving in the medium of the light ones $i$, 
is  $\svb_{ji} \sim m_i/m_j \ \sv\vert_{ij}$. Indeed, since $m_i v^2_i \sim m_j v^2_j$  
for species at the same temperature,  the latter cross-section is down by $v^2_j/v^2_i$, as above. } %


\section{Damping scales associated with   Dark Matter   fluctuations. 
\label{sec:dampscales}}

The transport coefficients defined previously  are associated with dissipative effects. 
The latter wash out all the   Dark Matter   primordial fluctuations having 
a size smaller than a well-specified scale, that we aim to calculate in this section.

\subsection{The collisional damping length. \label{sec:colldamp}}

Our basic (and sole assumption) is that there exists an epoch  where the   Dark Matter   particles are in statistical 
equilibrium; this implies that $\Gamma_{dm} =\sum_i \Gamma_{dm-i}> H$ is, at some time, satisfied. Due to the evolution 
of the Universe, one expects $\Gamma_{dm}$ to decrease and eventually to get smaller than the Hubble rate at a time 
$t_{dec(dm)}$.  

During all the period  where the condition $\Gamma_{dm} > H$ is 
satisfied,   Dark Matter   is collisional and all the associated primordial fluctuations 
with a size smaller than the collisional damping length are expected to be erased. 

This length may be obtained explicitly in terms of the transport 
coefficients (\cite{weinberg}). We write it as
\begin{equation}
\label{lweinb}
l_{cd}^2 = \pi^2 \int^{t_{dec(dm)}} 
\frac{\zeta + \frac{4}{3} \eta + \lambda T 
\frac {\rho _{m}^{2}} {4 \,\dslash{\rho} \, \rho _{r}} } {\dslash{\rho}  a^2} dt. \end{equation} %
Here $\dslash \rho = \sum_i \dslash \rho_i$. 
We have adopted the normalization of  \cite{efst} so that  
the mass-scale associated with the length $l_{cd}$ is given by 
$\,M_{cd} =(4\pi/3)\,\rho_m\, l^3_{cd}\,$.  
The integral runs over  all the period during which   Dark Matter   is collisional. 
For any reasonable energy-dependence of the cross-sections (see discussion in Paper II), the integral 
(\ref{lweinb}) is dominated by late times, so that the value of  $l_{cd}$ is determined by the   Dark Matter 
decoupling.  

Each transport coefficients $\zeta, \eta, \lambda T $ 
is given by a sum over $i$. Here $i$ denotes 
the species to which   Dark Matter   is coupled including   Dark Matter   itself. 
A given species drops out of the sum at a decoupling 
time $t_{dec(dm-i)} (\le t_{dec(dm)})$. This time corresponds 
to the epoch at which   Dark Matter   ceases to be coupled to $i$. 
The damping scale then reads~:
\begin{equation}
l_{cd}^2 = \pi^2 \sum_i 
\int^{t_{dec(dm-i)}} \frac{\zeta_i + \frac{4}{3} \eta_i + \lambda_i T 
\frac {\rho _{m}^{2}} {4 \,\dslash{\rho} \, \rho _{r}} 
}{\dslash{\rho}  a^2} dt,
\end{equation}
where $i$ runs over all the species to which   Dark Matter   is coupled
(note that for $i = dm$ one has to replace $t_{dec(dm-i)}$ by 
$t_{dec(dm)}$). 

As discussed in Section \ref{sec:bulk},
we can omit the bulk viscosity coefficient 
which is either negligible or at most of the same efficiency than the shear viscosity coefficient. 
Therefore, it will not provide any new constraints. 

Three different physical mechanisms can be at the origin of  damping due to interactions, as outlined in \cite{bfsL}:

\begin{enumerate}
\item  the damping due to the  interaction and displacement of the   Dark Matter   particles  within the fluid.
 This is the {\it self-damping}. 

\item  the damping due to a coupling between   Dark Matter   and a \textit{collisional} species $i$ which experiences its own 
collisional damping. The   Dark Matter   then, within conditions we will discuss, acquires the damping of this species.
This is the {\it induced-damping}.  

\item  the damping due to a coupling between   Dark Matter   and a \textit{freely-propagating} species $i$, as discussed in Section \ref{sec:symrates}. The origin of the damping of species $i$ in this case is the usual free-streaming damping. This damping is in the present case expected to be transmitted to   Dark Matter  . We will call such a damping  the {\it mixed-damping}. It can be interpreted as a part 
(or a particular case) of the induced-damping.
\end{enumerate}

\subsection{Self-damping and induced-damping.}

To get the corresponding damping length, it is  convenient to separate the sum over all species $i$ into a sum 
over the   Dark Matter   contribution $dm$ and over $i \ne dm$, that is over all species except Dark Matter. 
Eq.(\ref{lweinb}) can then be split into two parts. 
One which only depends on the   Dark Matter   properties 
(corresponding to the self-damping length) 
while the other one depends on the characteristics of species 
$i \neq dm$ (corresponding to the induced-damping
length).  This yields
\beq
 l_{cd}^{2}  =  l_{sd}^{2}  +  \sum_{i \ne dm}  l_{id}^{2}  \ ,
\eeq    
noting that the damping contributions combine quadratically. 
Using the explicit expressions of $\eta_i$ and $\lambda_i T$ given by 
eq.(\ref{visc}) and eq.(\ref{heat}), one  finds
\begin{equation} 
\label{intlsd} l_{sd}^{2}  =  \frac{{2\pi ^{2}}}{3} 
\int _{0}^{t_{dec(dm)}}\frac{{\rho _{dm}   v_{dm}^{2} t}}
{\dslash{\rho}  a^{2} \Gamma _{dm}} \left( 1+\Theta _{dm}\right) \ \frac{dt}{t} \  , 
\end{equation}  
and for each species $i \ne dm$~: 
\begin{equation} \label{intlid} l_{id}^{2}  =  \frac{2 \pi ^{2}}{3} 
\int ^{t_{dec(dm-i)}}_{0}\frac{\rho _{i}  v_{i}^{2} t } {\dslash{\rho}  a^{2}  \Gamma _{i}} 
\left( 1+\Theta _{i}\right) \ \frac{dt}{t}  \ .
\end{equation}The factor 
\begin{equation}
\label{theta} 
\Theta _{i} \,=\,\, \frac {\rho _{m}^{2}} {4 \,\dslash{\rho} \, \rho _{r}} 
\frac {\d  \ln \rho _{i}} {\d  \ln T_{i}} \,  ,
\end{equation}
obtained by using (\ref{heat}) and (\ref{lweinb})  
is associated with the heat conduction coefficient. 
Here, it has been calculated neglecting the pressure of non-relativistic particles. 
As already seen in (\cite{bfsL}), this yields 
$ \lambda_j \propto \frac {\d  \ln \rho _{j}} {\d  \ln T_{j}}
= \frac {\d  \ln m_j n_j } {\d  \ln T_{j}} = 0$ for a 
non-relativistic species $j$.
The heat conduction coefficient is then 
 non-zero only for
 \footnote{The complete expression including the 
pressure of the matter component can be derived along the lines of 
\cite{weinberg} but it is not needed in the present paper.} 
a relativistic species $i$ 
($\Theta_i \propto \frac {\rho _{m}^{2}} {\dslash{\rho} \, \rho
_{r}}$) and is important (\textit{i.e.} $\Theta_i$ larger than unity) 
only in the matter-dominated era where we have
 for any relativistic species $i$ the relation 
 $ \Theta_i \sim \frac {\rho _{m}} { \rho _{r}} \gg 1$ and 
$ \frac {\rho _{i}} { \rho _{m}}   \Theta_i 
\sim \frac {\rho _{i}} { \rho_{r}} \le  1 $.

It turns out  to be quite useful
 -- and sufficient
within the accuracy of our calculations -- 
to approximate the integrals (\ref{intlsd}) and (\ref{intlid}), written as $\int K \frac{dt}{t}$ by the maximum the factor $K$ reaches within the integration interval: \begin{equation}
\int K \frac{dt}{t} \simeq  {\rm Max}   [ K]
\label{appintcd}
\end{equation}
It is then straightforward to see that 
these integrals are dominated by the late times.
Indeed, the decoupling corresponds to an epoch where   
the system passes from a coupled to an uncoupled regime 
 while the Universe is expanding. This implies that $K$ is a growing function 
of time, as can be readily inferred from a short calculation counting the powers 
of $t$. 

The approximation (\ref{appintcd}), which aims to compare processes that differ by orders of magnitude, is good within factors of order unity. It considerably simplifies, at a very low cost, the overall classification of all kinds of Dark Matter we undertake in this paper.

As a result,  the self-damping scale may be written as \begin{equation} l_{sd}^{2} \sim \frac{{2\pi ^{2}}}{3} 
\frac{{\rho _{dm}  v_{dm}^{2} t}}
{\dslash{\rho}  a^{2} \Gamma _{dm}}\left( 1+\Theta _{dm}\right)  \vert_{{dec(dm)}} \  , \end{equation} while the induced-damping due to species $i$ reads 
\begin{equation}
l_{id}^{2} \sim \frac{{2\pi ^{2}}}{3} 
\frac{{\rho _{i}  v_{i}^{2} t}}
{\dslash{\rho}  a^{2} \Gamma _{i}} \left( 1+\Theta _{i}\right)  \vert_{{dec(dm-i)}} \ . \end{equation} It is convenient to rewrite these approximate expressions by introducing explicitly the Hubble rate (written as 
$\,H =\dot a/a= \alpha/t\,$,
\,with $\alpha$ = $1/2$ or $2/3$ depending on whether the Universe is in its radiation or matter-dominated era). 
This yields
\begin{equation}
\label{lsd}
l_{sd} \sim   \pi r_{dm} 
\left( \frac{H}{\Gamma_{dm}}\right)^\frac{1}{2} 
\frac{v_{dm} t}{a} \vert_{{dec(dm)}}\  ,
\end{equation}
and
\begin{equation}
\label{lid}
l_{id} \sim\ \pi \ r_{i} 
\left( \frac{H}{\Gamma_{i}}\right)^\frac{1}{2} 
\frac{v_i t}{a} \vert_{{dec(dm-i)}} 
\end{equation}
where we have kept the factor $H/\Gamma$ (although it is unity at the Dark 
Matter decoupling) 
in order to extend it to the specific situation discussed in 
section \ref{sec:III}. 
Here, we have introduced the dimensionless ratios 
\begin{equation} r_{dm} = \left[\frac{2\,\rho_{dm}}{3 \alpha\ \dslash \rho} \ 
\left(1+\Theta_{dm}\right) \right]^\frac{1}{2}  \ , 
\end{equation} and 
\begin{equation} r_{i}  = \left[\ \frac{2\,\rho_{i}}{3 \alpha  \dslash \rho} \ 
\left(1+\Theta_{i}\right) \right]^\frac{1}{2}  \ . 
\end{equation} 
{}From the relation $\rho_{i} \le \dslash \rho $ and the discussion 
in section \ref{sec:colldamp} of the 
possible values that the factors $\Theta_{dm}, \Theta_i$ may take, 
one can readily see that  $r_{dm}$ and $r_{i}$ are smaller or equal to unity.

Note  that our formulas 
are valid for  energy-dependent cross-sections, that is 
for temperature-dependent  averages $\svb$.


\subsection{Free-streaming.
\label{sec:fs}}

 After its decoupling at time $t_{dec(i)}$, species $i$ is freely-propagating. 
 Its primordial fluctuations experience the so-called free-streaming damping. 
 The damping acquired at a given time $t$ is proportional to the average distance traveled 
 by a particle of the species, that is in comoving units~: 
\begin{equation}
\label{lfsint}
l_{fs(i)} \propto  \int_{t_{dec(i)}}^t \frac{v_{i} \,dt }{a} \ . \label{intfs} \end{equation} 
It has been argued, and checked numerically, 
that an interesting approximation to this damping length is 
given by (\cite{bond83})~:  
\begin{equation}
\label{lfs}
l_{fs(i)} \ \simeq  \pi \, {\rm Max} \left[ \frac{v_{i} t }{a} \right] \ , \label{appintfs} \end{equation} 
where ${\rm Max}$ denotes the maximum value of the free-streaming scale 
within the integration interval $\,[t_{dec(i)},\,t]\,$
of eq.(\ref{lfsint}). This is a matter of convention about what one calls free-streaming length. 
 In the regime where we write our constraints, the two approximations are not dramatically different, and one might as well use the simpler one (\ref{appintfs}).  With the normalization (\ref{appintfs}), 
 the free-streaming length $\,l_{fs}\,$ is associated (\cite{bond83})
with the damping of all the fluctuations of mass scale
\beq
\,M\,=\,(4\pi/3)\ \rho_m \,l^3_{fs}  \   .
\label{mlrel}
\eeq
The quite simple  approximation (\ref{appintfs}) is well-adapted for the comparison with the collisional 
damping length which is also approximately evaluated by means of a similar simplification (\ref{appintcd}).

As a rule, close to the parameter values (mass and interaction rate) adapted to our constraints, (\ref{intfs}) 
yields length scales measured in the same dimensioned factors as the ones obtained from (\ref{appintfs}), but 
with numerical coefficients larger by a few units due to logarithmic contributions (but then it is not clear 
whether the mass-scale/length-scale relation is still (\ref{mlrel}) or the one where $l_{fs}$ is to be divided by $2$). For the sake of comparison, we give  in Appendix \ref{app:anndampsdfs} the expression of the free-streaming lengths calculated by means of the approximation (\ref{intfs}). The latter might be more suited than (\ref{appintfs}) far from the parameter values we handle in this work.


\subsection{Mixed-damping. \label{sec:mixed}}

A freely-propagating species may
transfer its own damping   to the   Dark Matter   fluctuations 
if the condition $ \Gamma_{dm-i} > H > \Gamma_i > \Gamma_{i-dm}$ 
is satisfied. 
This in particular 
requires that 
$\,\dslash \rho_i > \dslash \rho_{dm}$, due to the  symmetries of $\,\Gamma_{i-j}\,$ under the 
permutation of $i$ and $j$, as can be inferred from 
Eq.\,(\ref{ratmom}).
One can then decompose the induced-damping scale into two parts. %
\begin{eqnarray} 
l_{id}^{2}  &=& \frac{{2\,\pi ^{2}}}{3}
\int ^{t_{dec(i)}} \frac{\rho _{i}\: 
v_{i}^{2}}{\dslash{\rho} \: a^{2}\, \Gamma_i}
\left( 1+\Theta _{i}\right) \, dt   
\nonumber \\
  &+&  
\frac{{2\pi ^{2}}}{3}
\int _{t_{dec(i)}}^{t_{dec(dm-i)}}   \frac{\rho _{i}\: 
v_{i}^{2}}{\dslash{\rho} \: a^{2}\, H}
\left( 1+\Theta _{i}\right) \, dt \ . 
\label{intlre}
\end{eqnarray} 

The first term, valid as long as $ \Gamma_i  > H $, up to the time at which $i$ ceases to be collisional, 
is the (collisional) 
damping acquired by $i$ which is
communicated to the   Dark Matter  . 
The second one is an estimate 
of the damping acquired by $i$ during the period where   Dark Matter   remains 
coupled to this species, now freely-propagating since $ \Gamma_i < H $. It is obtained by simply replacing 
$ \Gamma_i $ by $H $ to get an expression where the distance covered by a particle of species $i$ while 
it was collisional is replaced by the distance it covers when being free-streaming.
The mixed-damping length may finally be written as: \begin{eqnarray} 
l_{md}^{2}  &=&
\frac{{2\pi ^{2}}}{3}
\int _{t_{dec(i)}}^{t_{dec(dm-i)}}   \frac{\rho _{i}\: 
v_{i}^{2}}{\dslash{\rho} \: a^{2}\, H}
\left( 1+\Theta _{i}\right) \, dt \ . 
\label{mixed}
\end{eqnarray} 
With the same approximation as in Eqs. (\ref{lid}) or (\ref{lfs}), 
we get:
\begin{equation}
\label{lidre}
l_{md} \sim \pi r_{i} 
\frac{v_i t}{a} \vert_{{dec(dm-i)}}  \ .
\end{equation}
Omitting the factor $r_{i}$, one easily recognizes the free-streaming  
scale of the species $i$ indicating that the damping acquired by $i$ 
is indeed transferred to the   Dark Matter. In fact, because we are 
considering a composite fluid, it is not the free-streaming of species $i$ 
which is communicated to the   Dark Matter   fluctuations but the free-streaming 
weighted by the factor $r_{i}$. This length finally contributes 
quadratically to the total damping scale $l_{cd}$.

One may however consider that (\ref{lidre}) is a too naive estimate. 
It is then worth to note that a lower bound to the damping length 
may be obtained by only considering the damping acquired by the 
  Dark Matter   fluctuations during the collisional regime of species $i$.
 This contribution corresponds to the first term in 
(\ref{intlre}) and represents a lower bound to the 
true damping length $l_{id}$ since
\begin{equation}
 l_{id}^{2}  >
 \frac{{2\,\pi ^{2}}}{3}
\int ^{t_{dec(i)}} \frac{\rho _{i} 
v_{i}^{2}}{\dslash{\rho} \: a^{2}\, \Gamma_i}
\left( 1+\Theta _{i}\right)\  dt   
 \  .
\end{equation}
This integral may be written, with  the same approximation as in 
\ref{sec:colldamp}~:
\begin{equation}
\label{lidsb}
 l_{id} >  \pi \ r_{i} 
\frac{v_i t}{a} \vert_{{dec(i)}} \  
\  .
\end{equation}
Such a result solely relies on  
the standard physics involved in the collisional regime. It differs from the mixed-damping estimate, eq.(\ref{lidre}),  
just by the time at which the factors are 
evaluated. The {collisional bound} is estimated at $t_{dec(i)}$ at which the species $i$ 
enters the free-streaming regime while the mixed-damping is estimated at $t_{dec(dm-i)}$ 
the true epoch at which the   Dark Matter   is no longer influenced by $i$. 
This collisional lower bound may be much 
smaller than the actual contribution, that we expect to be closer to the mixed-damping estimate.  
The latter will provide substantially 
stronger constraints. So, for our estimates we will use only this length, although it relies on 
less conventional physics.

\subsection{  Dark Matter   damping due to its own free-streaming.}

After its total decoupling,   Dark Matter   freely propagates 
erasing all fluctuations having a size smaller than its own free-streaming length. From (\ref{lfs}), 
this is at time $t > t_{dec(dm)}$ 
\begin{equation} \label{lfsdm} l_{fs(dm)} \ \simeq  \pi \, {\rm Max} \left[ \frac{v_{dm} t }{a} \right] \ , 
\end{equation}
where the maximum is to be taken within the time interval $[t_{dec(dm)}, t]$ where   Dark Matter 
is freely-propagating.

\subsection{Comparison of the free-streaming scales with the self-damping and induced-damping scales. 
\label{sec:fscd}}

There is a striking similarity between the expressions of the free-streaming scale (\ref{lfs}) and the 
self-damping scale (\ref{lsd}). This is simple to understand. 
Since the collisional damping scale is based on diffusion 
processes, the associated damping length is proportional to the distance 
traveled by collisional particles during their random walk. 
In a non expanding Universe, this distance is usually much smaller than the 
distance traveled during the free-propagation. 
However in the present case, due to the expansion of the Universe, 
it gets increasingly large up to 
the decoupling time. The collisional damping length is hence dominated by 
the late times, where the random walk marginally corresponds to the free 
propagation. 

The self-damping length (\ref{lsd}) is however still somewhat 
smaller than the   Dark Matter   free-streaming length (\ref{lfs}) 
since the transport of heat or momentum is done with an efficiency 
translated in terms of the factor $r_{dm}$ (smaller or equal to unity). 
We therefore have the relation 
\beq
l_{sd} \ \le\ l_{fs(dm)}  \ \,.
\eeq
The factor $r_i$ being also smaller or equal to unity,
we infer that the induced collisional damping due to species $i$ satisfies 
\beq l_{id} \ \le \ l_{fs}(i)  \, , 
\eeq 
where $ l_{fs}(i)$  is the free-streaming length (\ref{lfs}) of the species
$i$. Therefore, the induced-damping scale of   Dark Matter   fluctuations is not 
bounded by 
the distance traveled by the   Dark Matter   particles but rather by the distance 
covered by the most rapid particles to which   Dark Matter   is coupled. 
This is very important for massive   Dark Matter   particles in statistical 
equilibrium with relativistic species since, in this case,  
the   Dark Matter   induced-damping scale is bounded by $ct$ and
not by $v_{dm}t$. In this case, although the thermal 
velocity of such   Dark Matter   particles is relatively small, 
their coupling to 
relativistic species allows them to undergo substantial damping.


\section{Constraining the   Dark Matter   properties from free-streaming and self-damping.
\label{sec:sdfsdamp} }

The self-damping and free-streaming lengths  happen to depend only on two physical   Dark Matter 
parameters: the   Dark Matter   particles' mass 
$m_{dm}$ and the   Dark Matter   interaction rate with the medium $\Gamma_{dm}$. The dependence  
of the free-streaming damping length on the Dark 
Matter particles' mass is well-known, its dependence on the interaction rate is less obvious and 
is in some sense the new feature examined in this section.  

The constraints we get 
in this section are obtained  (there is one exception discussed later) at the  time  $t_{dec(dm)}$ where 
Dark  Matter decouples to 
any species $i$, including itself.
The rate ${\Gamma}_{dm}$ depends among others on the number density of the particles in the medium. Due to the 
expansion, this density, and whence 
the rate, decreases with increasing decoupling time, that is with increasing interaction strength. It turns out 
to be convenient to work with an 
interaction  rate which, taken at this epoch of reference, is an increasing function of the corresponding 
cross-section. Therefore we will use (see also Section \ref{app:redrates}) the  "reduced" interaction rate 
$\widetilde \Gamma_{dm} = \Gamma_{dm} a^3$, where the most obvious 
time-dependence due to the number density evolution is removed.

\subsection{Classification of the   Dark Matter   self-damping processes. \label{sec:regsdfs}}

We subdivide the plane ($m_{dm} , \widetilde \Gamma_{dm}$)
into regions within which the free-streaming and self-damping lengths have different expressions. 
The relevant time-scales which determine the damping lengths of the   Dark Matter   
primordial fluctuations 
are 
\begin{itemize}
\item the   Dark Matter   decoupling time $t_{dec(dm)}$ 
(at which   Dark Matter   has decoupled from all species)
\item the epoch at which   Dark Matter   becomes non-relativistic $t_{nr}$
\item the usual epoch of equality $t_{eq(\gamma +\nu )}$. 
\end{itemize}
These times may be translated in terms of 
scale-factors 
 $a_{dec(dm)}$,  $a_{nr}$ and 
$a_{eq(\gamma +\nu )}  = {\rho _{\gamma +\nu }(T_{0})}/{\rho _{m}(T_{0})}$. Their relative ordering 
defines six regions in the   Dark Matter   [mass/interaction rate]  parameter space:
\\
\begin{center} 
\doublebox{
\begin{tabular}{lc}
Region I  &$ a_{dec(dm)}< a_{nr}< a_{eq(\gamma +\nu )} $
\\
Region II  &$ a_{nr}< a_{dec(dm)}< a_{eq(\gamma +\nu )} $
\\
Region III  &$ a_{nr}< a_{eq(\gamma +\nu )}< a_{dec(dm)} $
\\
\\
\hline 
\\
Region IV  &$ a_{dec(dm)}< a_{eq(\gamma +\nu )}< a_{nr} $
\\
Region V  &$ a_{eq(\gamma +\nu )}< a_{dec(dm)}< a_{nr} $
\\
Region VI  &$ a_{eq(\gamma +\nu )}< a_{nr}< a_{dec(dm)} $ \end{tabular}} \end{center} 
This classification provides a natural way  to separate heavy from light, late from 
early decoupling   Dark Matter   particles. This also allows one to distinguish particles which are able to 
annihilate and those which are not.  
Region I includes in particular the standard Hot   Dark Matter   (HDM) scenario (like for instance 
neutrino   Dark Matter  ). Region II, includes, among others,  annihilating 
particles (\eg Weakly Interacting Massive Particles). 
Region III is devoted to strongly interacting particles. The 
last three regions, for which $a_{eq(\gamma +\nu )}< a_{nr}$, correspond to the less likely
very light   Dark Matter   particles, with masses below a few eV.

These Regions are shown in Fig. \ref{fig:sdgam}.  Explicit equations for their borderlines  are 
given in Appendix \ref{app:bordsdfs}. We  now give the expressions of the damping lengths in these regions, and 
the associated constraints, whose values also are displayed in Fig. \ref{fig:sdgam}. Analytical expressions of 
the damping lengths may be found in Appendix \ref{app:anndampsdfs}.

{\bef
{\par\centering \resizebox*{7.2cm}{9.6cm}{\includegraphics{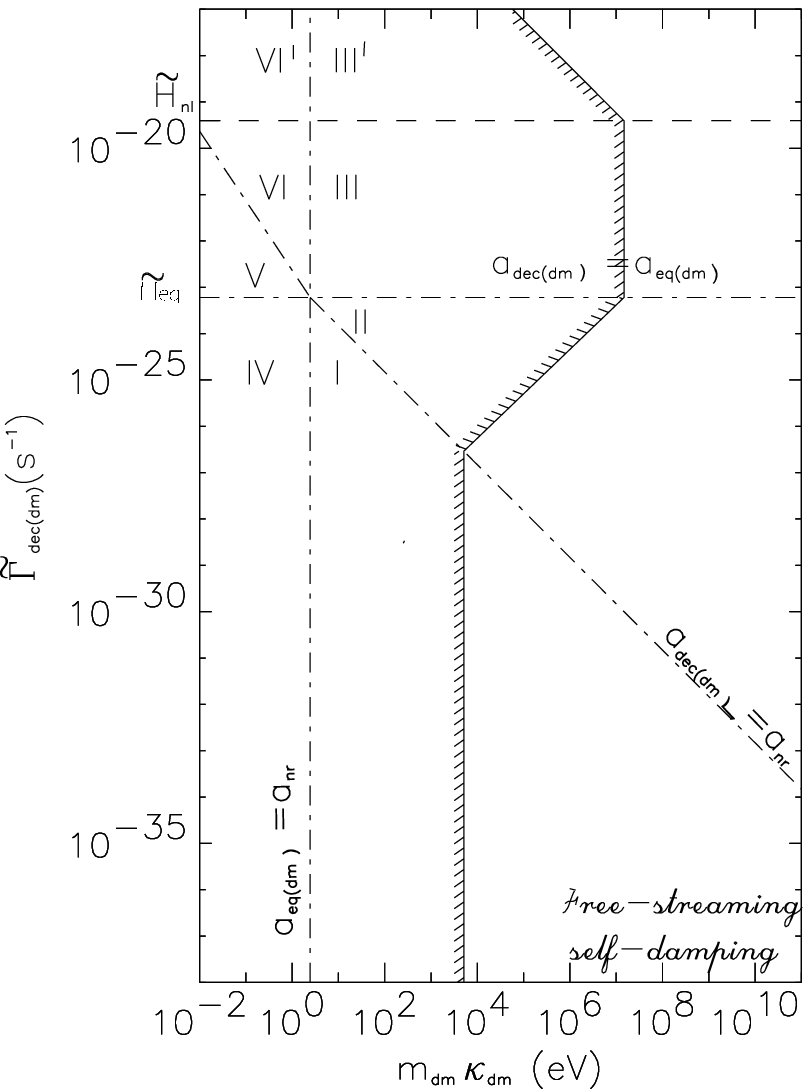}} \par} 
\caption{ 
Bounds in the [Dark Matter   particles' mass,    Dark Matter   interaction rate] parameter space obtained from self-damping and free-streaming. The interaction rate $\widetilde\Gamma_{dec(dm)}$ is $ \Gamma_{dec(dm)} a^3$ taken at the   Dark Matter decoupling. The labels I to VI correspond to different  scenarios. The hatches indicate which part of the parameter space is forbidden. It corresponds to the regions where the free-streaming and 
self-damping scales are above $100 kpc$ (\textit{i.e.} $\sim 10^8 M_\odot$) scale. This is seen to provide a constraint involving both the interaction rate and the mass. The indications in this figure are schematic. Some factors  in general of order unity are omitted for simplicity.  The reader interested by these constraints should use the expressions in the text, where all  factors are given explicitly.
}
 \label{fig:sdgam} 
\eef}
\vspace{0.3cm}

\subsection{Self-damping and free-streaming lengths. \label{sec:sdfslength}}

\subsubsection{Region I : $ a_{dec(dm)}< a_{nr}< a_{eq(\gamma +\nu )} $ \label{sec:I}}

Since $a_{dec(dm)}$ corresponds to the total decoupling of   Dark Matter  ,
annihilation processes by definition are frozen afterwards. The   Dark Matter  
comoving number density then remains constant~: 
$n _{dm}(T_{nr})\, a_{nr}^{3}=n _{dm}(T_{0})$. 
The former can then be translated in terms of energy
density $ \rho _{dm} =m_{dm} \, n_{dm} $, which, at the non-relativistic transition ($a = a_{nr}$), 
is required to be equal to the energy-density
\footnote{The factors entering the energy-densities are defined precisely in Appendix \ref{app:def} .}   
in the relativistic regime~: $ \rho _{dm} =\frac{{1}}{2}\stat g_{dm}a_{B}T_{dm}^{4} 
\equiv \frac{{1}}{2} g_{\star dm} a_{B}T_{\gamma}^{4} $. This 
turns out to be a very stringent requirement.
To be satisfied, it requires a quite small effective  $g_{*dm}$ to 
 ensure that the number density of   Dark Matter   particles
at the present epoch matches the observed relic abundance. 
Specifically  
\beq
g_{*dm} =
\frac{2 m_{dm} \, n_{dm}(T_{nr})}{\rho _{\gamma}(T_{nr})}
= 3.36  \frac{a_{nr}}{a_{eq(\gamma +\nu )}}  
\ .
\eeq
This condition may be rewritten in terms of $\kappa _{dm}$, the factor which determines the ratio of the photon to 
Dark Matter temperature, as 
\(\kappa^3 _{dm}(T_{nr}) = \stat \frac{ g_{dm}}{3.36} \frac{a_{eq(\gamma +\nu )}}{a_{nr}\kappa _{dm}(T_{nr})} \), 
that is 
\begin{equation} \label{URFOden} 
\kappa _{dm}(T_{nr}) \sim 62 \, \left( \stat \frac{ g_{dm}}{2}\right) ^\frac{1}{3} 
\left(\Om\right)^{-\frac{1}{3}}
 \left( \frac{m_{dm}}{1 MeV}\right) ^\frac{1}{3}     .
\end{equation}

In this scenario, the Dark Matter number density is to be set at a quite remote epoch, and all changes in the 
particle number are frozen out, much before the non-relativistic transition. We will thus call this way of matching 
the present relic density the Ultra-Relativistic Freeze-Out
 (URFO) scenario.
 
  The requirement (\ref{URFOden}) may be achieved through 
 various ways. One one hand,  $\kappa _{dm}$ may be large (quite larger than unity) if 
the decoupling between the   Dark Matter   (which may remain interacting)
and the thermal bath occurs at a remote enough time, generically much before its total decoupling. On the other hand, 
one can have a small $\stat$ if the chemical reactions among   Dark Matter   particles
which do not conserve the   Dark Matter   
number density  freeze out at early enough an epoch, the contact with the thermal bath being maintained. The particle 
number conservation from this epoch till nowadays then ensures that the   Dark Matter   density  
has the appropriate value at the non-relativistic transition 
and at the present epoch.

The expressions of the self-damping  and free-streaming scales in this 
region read respectively~:
\beq
l_{sd}^{(I)} \ = \ \pi \, r_{dm} \, \frac{c t}{a} \, \vert_{dec(dm)} \ 
 \label{lsd1}
\eeq
and 
\beq
 l_{fs}^{(I)} \ = \ \pi\, \frac{c t}{a} \, \vert_{nr}  \  .  \label{lfs1} 
\eeq
Numerically, one finds the results given in Table \ref{tab:tablereg1} where $\widetilde \Gamma_{{dec(dm)}}$ 
is the reduced   Dark Matter   interaction rate at $t_{dec(dm)}$. The free-streaming scale  corresponds to the 
result of \cite{bond80,davis}.

\begin{center}
\begin{table}[h]
\caption{\label{tab:tablereg1}}
\[
\begin{array}{l}
\hline 
\hline
\\
l_{sd}^{(I)} = 0.35 \, kpc\, r_{dm}(T_{dec(dm)}) \ \g_*^{-1}(T_{dec(dm)}) \  
\frac{\widetilde \Gamma_{{dec(dm)}}}{10^{-29} s^{-1} } 
\nonumber\\ 
\nonumber\\
l_{fs}^{(I)} =
0.51 \, kpc\,
\g_*^{-\frac{1}{2}}(T_{nr}) 
 \left(\frac{m_{dm} \kappa _{dm}(T_{nr})}{1 MeV}\right)^{-1} 
\\
\\
\hline
\hline
\end{array}
\]
\end{table}
\end{center}

The comparison of (\ref{lsd1}) with (\ref{lfs1}) shows that the 
self-damping contribution is down by a factor
\beq  
\frac{\frac{ct}{a} \vert_{dec(dm)}}{ \frac{ct}{a} \vert_{nr}} = 
 \frac{a_{dec(dm)}}{a_{nr}} < 1.
\eeq
The self-damping is therefore generically smaller than the free-streaming 
in Region I and will not be the most  constraining effect. 
The coefficient 
$r_{dm} \sim \left[\rho _{dm}(T_{dec(dm)}) / \dslash \rho (T_{dec(dm)}) \right]^{1/2}$,
is also below unity as long as   Dark Matter   is coupled to radiation.
Indeed (as may be inferred more readily using the ratio of the energy-densities given in Sect. 
\ref{app:Gamsym}) $\rho _{dm} / \dslash \rho \sim \rho _{dm} / \rho_r \le a_{nr} / a_{eq} $ 
implies 
$r_{dm} \le 1$. 
On the other hand, if the decoupling from radiation occurs 
before the   Dark Matter   total decoupling, the   Dark Matter   fluid does 
not contain any radiation component and one gets: 
$\dslash \rho \sim \rho_m$ and $r_{dm} \sim 1$.

{}From (\ref{lfs1}), we get the usual condition \begin{equation} m_{dm} \kappa _{dm}(T_{nr})  >  5.1 \ keV
 \g_*^{-\frac{1}{2}}(T_{nr}) 
\left( \frac{l_{struct}}{100 kpc} \right)^{-1}
\ .
\end{equation}
Implementing the estimated value (\ref{URFOden}) of $\kappa _{dm}$, yields the condition 
\bea
m_{dm}   >  0.9 \ keV
 \, \left( \stat \frac{ g_{dm}}{2}\right) ^{-\frac{1}{4} }  \g_*^{-\frac{3}{8}} 
 \cr \left(\Om\right)^{\frac{1}{4}} \left( \frac{l_{struct}}{100 kpc} \right)^{-\frac{3}{4}} \ , 
 \eea 
at the origin of the so-called "$1 keV$" mass limit (\cite{davis}).

\subsubsection{Region II : $ a_{nr}< a_{dec(dm)}< a_{eq(\gamma +\nu )} $. \label{sec:II}}

The    Dark Matter   (total) decoupling in Region II occurs after the non-relativistic transition.
There are in this case two ways of satisfying the relic density requirement~: 
\begin{itemize} 
\item The first one, as in Region I, is to consider the {URFO} scenario where 
the constraint (\ref{URFOden}) has to be imposed.

\item The second one is to consider particles which remain in chemical equilibrium up to their non-relativistic 
transition (actually slightly after, as we shall see below). The allowance 
for this second possibility opens a new window. Indeed, one may have a large energy 
density before the non-relativistic transition. Then the   Dark Matter   
particles,  still in chemical equilibrium, annihilate and the number density 
exponentially decreases. Once it reaches  the appropriate  value, one assumes that 
the reactions freeze-out. We will call this way of matching the present relic density the Non-Relativistic Freeze-Out 
(NRFO) scenario. 
\end{itemize} 

In the NRFO scenario, to match the observed relic density, a specific value of the freeze-out temperature, 
and in turn of the annihilation cross-section (see \eg \cite{leew,griest}) is required. We only shortly discuss its derivation, which is well-known. The conservation of the   Dark Matter 
number density in the NRFO scenario is relevant only after the chemical decoupling.
By equating the comoving   Dark Matter   number density during its exponential 
decrease (just before freeze-out)  
to the comoving density just after freeze-out, with 
\begin{equation}
  T_{dm}^{fo}= \frac{ m_{dm}}{x_{fo}} \ ,
\label{Tfo}
\end{equation}
one sees that  $x_{fo}$ obeys the equation 
\beq 
x_{fo}^\frac{3}{2} e^{-x_{fo}} = 
(2 \pi)^\frac{3}{2} \left(\frac{\hbar c }{T_0} \right)^3
\frac {\widetilde n_{dm}}
{ \, g_{*dm}/\kappa^3_{dm}(T_{fo})}
 \, .
\eeq    
 yielding the very simple rule
\begin{equation}
x_{fo}  \sim  14 + \ln  \frac{m_{dm}\kappa_{dm}(T_{fo}) }{1MeV} 
 \ , 
\label{xfo}
\end{equation}
where we have neglected  unimportant logarithmic factors. 
For $ \sim 100 \, GeV$ particles, $x_{fo}$ is of the order of $\sim 20$: 
the annihilation reactions freeze out somewhat after the non-relativistic transition, at 
$  T_{dm}^{fo} = \frac{ m_{dm}}{20} $. 
This implies a well-defined value for the cross-sections to maintain the 
chemical equilibrium up to $x=x_{fo}$ 
and not after.
It yields the standard constraint on the   Dark Matter   parameters, which in this case
must be satisfied instead of (\ref{URFOden}). Larger values of 
 $\sv_{ann}$  are prohibited\footnote{The values we mention here slightly 
 differ in case of co-annihilation.}, as well as lower values, unless they are small 
enough so as to comply with the {URFO} scenario and 
therefore satisfy condition (\ref{URFOden}).
In the special case where the chemical equilibrium is 
reached via direct annihilation, for instance, the 
requirement is found to be
\begin{eqnarray}
\label{NRFOann}
\sv_{ann} \, &=&
\frac{H_r x_{fo} T_0}{m_{dm}\kappa_{dm}(T_{fo}) \tilde n_{dm}} \cr &\sim& 7.6 \ 10^{-26} cm^3s^{-1} \frac {\g_*^\frac{1}{2}(T_{fo})}{\Odm\kappa_{dm}(T_{fo})}
\frac{x_{fo}}{20}
\ .
\end{eqnarray}
This value is nearly a universal constant owing our good knowledge of the density parameter $\Omega_{dm}$ and the Hubble constant $H_0$. In particular it is almost independent of the Dark Matter particles' mass. Any Dark Matter particle that is able to annihilate must satisfy this condition in order not to overclose the Universe. This is true, in particular, for any particle having a mass above the MeV range.

The condition 
(\ref{NRFOann}) however
may be alleviated, to allow larger (but not smaller) values of $\sv_{ann}$. This is the case provided the 
particles and anti particles (with respective densities $n_{dm+}$ and $n_{dm-}$) exhibits an asymmetry 
before the non-relativistic transition~: 
\beq
\beta = \frac{ n_{dm+} - n_{dm-}}{ n_{dm+} + n_{dm-}} \ . 
\eeq
Indeed, the density at the relativistic side then is $n_{dm}\vert_r = n_{dm+} + n_{dm-}$ and particle number 
conservation then shows the number density at 
the non-relativistic side to be
$n_{dm}\vert_{nr} = n_{dm+} - n_{dm-}$. To produce the observed relic 
number density, it may be shown  (see Appendix \ref{app:nrtran}) that a jump 
$\frac{n_{dm}\vert_{nr}}{ n_{dm}\vert_r} = \frac{a_{nr}}{a_{eq(dm)}}$ 
is required at the non-relativistic transition.
The relic density  condition (\ref{NRFOann}) then is replaced by the condition 
\begin{equation}
 \beta = \frac{a_{nr}}{a_{eq(dm)}} \ .
\label{NRFOasym}
\end{equation}

We now estimate the damping scale associated to   Dark Matter  
candidates belonging to region II.
 At total decoupling,   Dark Matter   
particles have a velocity
\begin{equation}
\label{vnr}
 v_{dm}(T_{dec(dm)}) = f  \left(\frac{{a_{nr}}}{a_{dec(dm)}}\right)^\frac{1}{2}
\end{equation}
where 
\beq
f=\left( \frac{\kappa _{dm}(T_{nr})}{\kappa_{dm} (T_{dec(dm)})}\right)^\frac{1}{2} \ . 
\eeq
We then get 
\beq
l_{fs}^{(II)} \ = \ l_{sd}^{(II)} /r_{dm} \ = \ \pi \, \frac{ v t }{a} \vert_{dec(dm)} 
\ .
\eeq
The ratio
$
 {\rho_{dm}(T_{dec(dm)})}/{\rho_r(T_{dec(dm)})} \sim {a_{dec(dm)}}/{a_{eq}} $ is still smaller than unity. 
 But $r_{dm}$ --which depends on the ratio $\rho_{dm}(T_{dec(dm)})/ \dslash \rho(T_{dec(dm)})$ of the species 
that 
are actually coupled-- 
may be unity if the   Dark Matter   is not coupled to radiation, as already noted for Region I.
The self-damping and free-streaming damping lengths in Region II finally 
 take the values given in Table \ref{tab:tablereg2}.

\begin{center}
\begin{table}[h]
\caption{\label{tab:tablereg2}}
\[
\begin{array}{ll}
\hline 
\hline \\
l_{sd}^{(II)}  \sim & 330 \, kpc \  
r_{dm}(T_{dec(dm)}) \ f \ \g_*^{-\frac{3}{4}}(T_{dec(dm)}) \, 
\\
&\hspace{2cm}
\left( \frac{m_{dm}  \kappa _{dm}(T_{nr})}{1 MeV}\right) ^{-\frac{1}{2}}\,  
\left(\frac{{\widetilde{\Gamma }_{{dec(dm)}}}}{ 6 \ 10^{-24}s^{-1}}\right)^\frac{1}{2} 
\label{lsd2}  \\
\\
l_{fs}^{(II)}  \sim  & 330 \, kpc \ f \ 
 \g_*^{-\frac{3}{4}}(T_{dec(dm)}) \, 
 \\
&\hspace{2cm}
\left( \frac{{m_{dm} \kappa _{dm}(T_{nr})}}{1 MeV}\right)^{-\frac{1}{2}}\, 
\left( \frac{\widetilde{\Gamma }_{{dec(dm)}}}{6 \ 10^{-24}s^{-1}}\right)^\frac{1}{2}
 \\
\\
\hline
\hline
\end{array}
\]
\end{table}
\end{center}

The condition $l_{fs}^{(II)} < l_{struct}$   implies
\begin{eqnarray}
\nonumber 
m_{dm}  \kappa_{dm}(T_{nr})  & >&   
11 \, \mbox{MeV} \ f^2 \ 
  \g_*^{-\frac{3}{2}}(T_{dec(dm)})
\nonumber \\ 
&& \frac{\widetilde{\Gamma }_{dec(dm)}}{ 6 \ 10^{-24}s^{-1}}
 \ 
\left( \frac{l_{struct}}{100 kpc} \right)^{-2}
\ .
\end{eqnarray}

\subsubsection{Region III : $ a_{nr}< a_{eq}< a_{dec(dm)} $. \label{sec:III}}

As in  the previous case, we have to  consider the {URFO} and  
 {NRFO} scenarios.
 To get the observed   Dark Matter   energy density  one has
to impose the same relations (\ref{URFOden}), and
(\ref{NRFOann}) or (\ref{NRFOasym}), 
respectively, as for Region II. 

The  velocity of the   Dark Matter   particles at  decoupling has the same expression 
as for
 region II. A straightforward calculation  (which takes into account that the  
 decoupling occurs while the universe is matter-dominated)  
provides the damping scales. The interesting point here is that, 
unlikely to region II, 
we expect the ratio 
$ r_{dm} \sim \left({\rho_{dm}}/{\dslash \rho}\right)^\frac{1}{2} $ 
to be of the order of unity 
(if not exactly equal to 1) since the density $ \dslash \rho $ refers 
only to the species coupled 
to the   Dark Matter   and the decoupling occurs in the matter dominated era.
Therefore self-damping and 
free-streaming are comparable in this region: 
\beq
l_{fs}^{(III)} \ = \ l_{sd}^{(III)} /r_{dm} \ = \ \pi \, \frac{ v t }{a} \vert_{dec(dm)}  \ , 
\eeq
which numerically yield the results given in Table \ref{tab:tablereg3}. Despite $l_{fs}$ and $l_{sd}$ are given by the same expression than in Region II, their evaluation  yields a different result since after equality, in Region III, the time of decoupling grows less rapidly with the interaction rate. As a result, the damping lengths turn out to be independent of this rate.

\begin{center} \begin{table}[h] 
\caption{\label{tab:tablereg3}} \[ \begin{array}{ll} \hline 
\hline \\
l_{sd}^{(III)} \sim  &  435 \, kpc   \left(\Om\right)^{-\frac{1}{2}}
r_{dm}(T_{dec(dm)})   f \left(\frac{{m_{dm} \kappa _{dm}(T_{nr})}}{1 MeV}\right)^{-\frac{1}{2}} 
\label{lsd3}\\
\\
l_{fs}^{(III)}  \sim  &  435 \, kpc \, \left(\Om\right)^{-\frac{1}{2}} \ f \, \left(\frac{{m_{dm} 
\kappa _{dm}(T_{nr})}}{1 MeV}\right)^{-\frac{1}{2}} \\ \\ \hline \hline \end{array} \] \end{table} \end{center}

This provides a range for the   Dark Matter   mass~:
\begin{eqnarray}
m_{dm} \kappa _{dm}(T_{nr})  & > & 
\nonumber
\\
 19 \,  \mbox{MeV}&  &\left(\Om\right)^{-1}  f^2   \,  
\left(\frac{l_{struct}}{100 kpc} \right)^{-2} .
\end{eqnarray}

\paragraph{A specific case: coupling up to the onset of structure formation.}

The previous expressions are only relevant provided 
decoupling occurs before structure formation, that is before the non-linear collapse of the primordial structures, 
assumed to take place at a scale-factor $  a_{nl}$. We leave the latter arbitrary, but it is expected to be roughly 
of the order of $  \sim 1/10 $ for objects with a size $ \le 10^8 M_{\odot}$. 

For 
  $ a_{dec(dm)} > a_{nl} $, which may be rewritten as  $\widetilde \Gamma_{dm}\vert_{a=a_{nl}} > \widetilde H_{nl} $ 
(the interaction rate
 $ \widetilde H_{nl}$ is given in Table \ref{tab:tyr}),
there is no free-streaming at all before structure formation. 
Only collisional damping is at work up to the non-linear collapse, with \beq l_{sd}/r_{dm}  = \pi 
\left({{\widetilde H_{nl}}/\widetilde \Gamma_{dm}}\right)^{\frac{1}{2}} \frac{ v t }{a} \vert_{nl} 
\eeq
that is
\begin{eqnarray}
\label{lsd3'}
l_{sd}^{(III')}  &\sim&    435 \, kpc \,  
\left(\Om\right)^{-\frac{1}{2}}
\, 
r_{dm}(T_{dec(dm)}) \, f
\nonumber \\ 
&&
\left(\frac{{m_{dm} \kappa _{dm}(T_{nr})}}{1 MeV}\right)^{-\frac{1}{2}} 
\left(\frac{\widetilde \Gamma_{dm}}{\widetilde H_{nl}}\right) ^{-\frac{1}{2}} \,  \ . 
\end{eqnarray} It is useful to note that 
$ l_{sd}^{III'}$ depends on the collision rate $ \widetilde \Gamma _{dm} (T_{nl}) $ taken at the time of the 
onset of the non-linear gravitational collapse and no longer at decoupling. 
 Also, $l_{sd}^{III'} = \left(
 \frac{\widetilde H_{nl}}{\widetilde \Gamma _{dm} } 
\right)^\frac{1}{2}  l_{sd}^{III} $ , eq.(\ref{lsd3'}),
is smaller than $ l_{sd}^{III} $, eq.(\ref{lsd3}) although the corresponding   Dark Matter   interaction rate, taken at a given epoch, is larger.

For the self-damping to be acceptable,
we must have
\begin{eqnarray}
\label{Gamsd3'}
m_{dm}  \kappa _{dm}(T_{nr})
 &> &
 19 \ MeV \  \left(\Om\right)^{-1}
r_{dm}^2(T_{dec(dm)}) 
 f^2   
\nonumber \\
&& \left(\frac{ \widetilde{\Gamma} _{dm}} { \widetilde H_{nl}} \right)^{-1}  \left(\frac{l_{struct}}{100 kpc}\right)^{-2} \ . 
\end{eqnarray}

Pending a more complete discussion of this issue in Paper II, we may  recall  that the present requirements are necessary conditions, as are all the bonds we establish in this paper, irrespectively of other astrophysical conditions which may be needed to achieve an acceptable scenario with the assumed Dark Matter parameters. 

\subsubsection{Region IV : $ a_{dec(dm)}< a_{eq}< a_{nr} $. \label{sec:IV}}

The   Dark Matter   particles are relativistic when they decouple. The 
analytic expression
of the damping scales is the same as in Region I. 
The free-streaming length $l_{fs}^{(IV)}$ is still
 given by (\ref{lfs1}). The numerical 
value of the collisional damping scale however is different. Since we have here $ \rho _{dm} \sim \dslash  \rho$, the collisional 
damping length is 
still smaller than the free-streaming length,
but by a factor
$  \frac{ct}{a} \vert_{dec(dm)} / \frac{ct}{a} \vert_{nr} =  a_{dec(dm)}/a_{nr} $ only.

Requiring the free-streaming damping scale to be smaller than 
$l_{struct} \sim 100 kpc$ 
 still calls for   Dark Matter   particle masses in the $keV$ range or above. 
This cannot be achieved in Region IV.
Region IV  is therefore excluded by the damping 
requirements.

\subsubsection{Region V : $ a_{eq}< a_{dec(dm)}< a_{nr} $. \label{sec:V}}

The calculation is strictly identical to the one done in the previous section 
for Region IV since $a_{dec(dm)} =  a_{eq}$, does no longer correspond to a change 
in the expansion regime. The transition from a radiation to a matter 
dominated universe at   Dark Matter   decoupling occurs indeed 
for  $a_{dec(dm)}  =  a_{nr}$.

Region V is therefore also excluded by the damping requirements.

\subsubsection{Region VI : $ a_{eq}< a_{nr}< a_{dec(dm)} $ \label{sec:VI}}

  Dark Matter   decouples in the matter dominated era. 
The collisional, as well as the free-streaming scales 
are dominated by the contribution near the decoupling time. 
Since we have still $ \rho _{dm} \sim \dslash \rho$, their 
contribution is nearly the same as long as decoupling occurs 
before the epoch of non-linear gravitational collapse where the 
structures actually start to build up. The same calculation as 
in Region III is relevant in the present case. The conditions 
for the $dm$ density to match the observed one still imposes  
(\ref{URFOden}) : anticipating the astrophysical discussion of Paper II, 
we  readily see that the {URFO} scenario is the only sensible 
case to consider. Indeed, in Region VI, the present-day number 
density of these light   Dark Matter   particles is much higher than 
the photon density.  Such an unusually large number density can only be 
reached under very special circumstances. It does not make much  
sense to assume that it results from an even much larger one, which is 
followed by a period of annihilation. However, strictly speaking, this is not forbidden.

The damping scales are given by their expressions in Region III, with 
comparable effects from collisions and free-streaming. 
Masses above the MeV scale are therefore required to avoid prohibitive 
damping. But such masses are not allowed in Region VI (fig. \ref{fig:sdgam}).

\paragraph{A specific case: coupling up to the onset of non-linear collapse.}

For $ \widetilde{\Gamma } (T_{nl}) > \widetilde H_{nl} $, which we call Region VI',
  there is, on the other hand, never decoupling until the onset of 
the non-linear gravitational collapse. Only collisional damping 
is at work so the free-streaming constraint drops out and 
expression (\ref{lsd3'}) of the self-damping 
length, as well as the constraint 
(\ref{Gamsd3'}) hold in this case.

This leaves an allowed window in the parameter space of Region VI', for extremely large an interaction rate.


\section{Constraining the   Dark Matter   candidates from the neutrino 
induced-damping.
\label{sec:idnudamp}}

The expression of the induced-damping scale (\ref{lid}) shows 
that relativistic particles yield a large damping effect. 
They are expected to provide the most stringent constraints 
on the   Dark Matter   
interaction rates. Since the obvious components of the radiation 
are photons and neutrinos, 
we specifically 
focus on these two species. 
These particles have the advantage that their dominant 
interaction (apart from the interaction with   Dark Matter   that we
aim to discuss in the present paper) is with electrons. It is  known and 
 provides the scales of reference for the discussion of 
the damping effect induced by these particles.

In this section, we
 compute the collisional damping of 
  Dark Matter   fluctuations due 
to a possible coupling of   Dark Matter   with neutrinos 
and evaluate the corresponding constraints.
The photon induced-damping of the   Dark Matter   fluctuations will be 
discussed in section \ref{sec:idphotdamp}.

\subsection{  Dark Matter   parameter space  for neutrino induced damping. \label{sec:regnu}}

The neutrino induced-damping scale depends on the total neutrino 
interaction rate with the medium  $\Gamma_\nu = \sum_j \Gamma_{\nu-j} = \Gamma_{\nu-e} + \Gamma_{\nu-dm} + ...$.  
In the standard scheme, this rate is dominated by the collisions 
with electrons, 
 $\Gamma_{\nu} \sim \Gamma_{\nu-e}$. The only possibility for 
$\Gamma_{\nu}$ not to be dominated by $\Gamma_{\nu-e}$ 
is when
$\Gamma_{\nu-dm} > \Gamma_{\nu-e}$.   

Thanks to our approximations (\ref{lid}), or (\ref{lidre}),
 the only other parameter which 
enters the calculation of the damping scale is the upper limit of the integral (\ref{intlre}),
namely the time $t_{dec(dm-\nu)}$ where   Dark Matter   decouples 
from neutrinos. This defines the {\it reference time} at which 
all our constraints on the interaction rate or on the cross-sections 
are obtained.

For the actual applications, it turns out to be convenient to note that both the interaction 
rate $\Gamma_{\nu-dm}$ and time  $t_{dec(dm-\nu)}$
 may be expressed as a function of 
the ``reduced'' (see Section \ref{app:redrates}) interaction rate  
$ \widetilde{\Gamma}_{dm-\nu} \equiv \Gamma_{dm-\nu} a^3$, 
or equivalently as a function of
 $\svb_{\nu-dm} \equiv  \widetilde \Gamma_{\nu-dm} /  \widetilde n_{dm} $. 
The other parameter entering the calculation is  the   Dark Matter   particles' mass
$ m_{dm}$. 
This defines a two-parameter space which allows  one
to classify all  kinds of   Dark Matter   particles 
which interact with neutrinos, say 
$  [m_{dm}, \widetilde{\Gamma}_{dm-\nu}  ] $ or $ [ m_{dm}, {\svb}_{\nu-dm} ] $.

In the following, we shall distinguish ``Regions" in this parameter-space which correspond to 
different analytical  expressions 
of the neutrino induced-damping scale. All together, they cover the whole parameter space.

\subsubsection{  Dark Matter   decouples from neutrinos while the latter are collisional.
\label{sec:regnucoll}} 

This is the case where 
\begin{center}
$\widetilde{\Gamma }_{dm-\nu} \le \widetilde{\Gamma }_{\nu }  \ \   \vert_{dec(dm-\nu)} \ . $
\end{center}
The   Dark Matter   then experiences  a collisional damping induced by its coupling to  neutrinos.
Since the neutrino interaction rate can be dominated by the collision rate with 
either the electrons or with Dark Matter, we are led to distinguish 
two sub-cases A and B.

\paragraph{Region A.}  

In this region, the neutrino interaction rate is dominated by  interactions with electrons. 
This corresponds to~: 
\begin{equation} \label{nuA}
\widetilde{\Gamma}_{\nu-dm }   \le
  \widetilde{\Gamma }_{\nu-e }  \vert_{dec(dm-\nu)} \ ,
\; \mbox{that is} \; 
\widetilde{\Gamma }_{\nu } \sim  \widetilde{\Gamma }_{\nu-e }   \   .
\end{equation}
In this case, the neutrino collisional damping is due to their interaction with electrons.

\paragraph{Region B.}  

Here the neutrino interaction rate is dominated by 
interactions with Dark Matter. This corresponds to~:
\begin{equation}
\label{nuB}
\widetilde{\Gamma}_{\nu-dm  }   >
   \widetilde{\Gamma }_{\nu-e }   \vert_{dec(dm-\nu)} \ , 
\; \mbox{that is} \;  
\widetilde{\Gamma }_{\nu } \sim  \widetilde{\Gamma }_{\nu-dm }  \  . \end{equation} %
The collisional damping is then due to the neutrino interactions with   Dark Matter.

\subsubsection{  Dark Matter   decouples from neutrinos 
while the latter are freely-propagating. \label{sec:regnumix}}

In this situation, there is only one region, namely: 

\paragraph{ Region C}
\begin{center}
\label{nufp}
$\widetilde{\Gamma }_{dm-\nu} > \widetilde{\Gamma }_{\nu } \ \ \vert_{dec(\nu)} \label{nufs} \ .$ \end{center}  
Neutrinos are  already  free-streaming when  
  Dark Matter   decouples from the latter. This is the mixed-damping regime.
The corresponding damping length has been estimated eq.(\ref{lidre}). 
Since the neutrino collision rate does not enter the mixed-damping expression 
there is no point in the present case to separate  
the case where $\widetilde{\Gamma}_{\nu-dm  } $ is smaller or larger than $ \widetilde{\Gamma }_{\nu-e }$.

\subsubsection{Borderlines.}

The borderlines between regions A, B and C are shown in 
Figs. \ref{fig:nugamurfo}, \ref{fig:nugamnrfo}, \ref{fig:nusigurfo}, and \ref{fig:nusignrfo}, 
depending on whether one considers the URFO or the NRFO scenario, and whether one displays the 
parameter-space in the $[  m_{dm}, \widetilde{\Gamma}_{dm-\nu} ]  $ or $ [ m_{dm}, {\svb}_{\nu-dm} ] $ 
coordinates. These borderlines are calculated explicitly in Appendix \ref{app:bordnu}. 
They are important since they define the conditions of validity of 
our expressions of the damping length.

%
%
{\bef
{\par\centering \resizebox*{7.2cm}{9.6cm}{\includegraphics{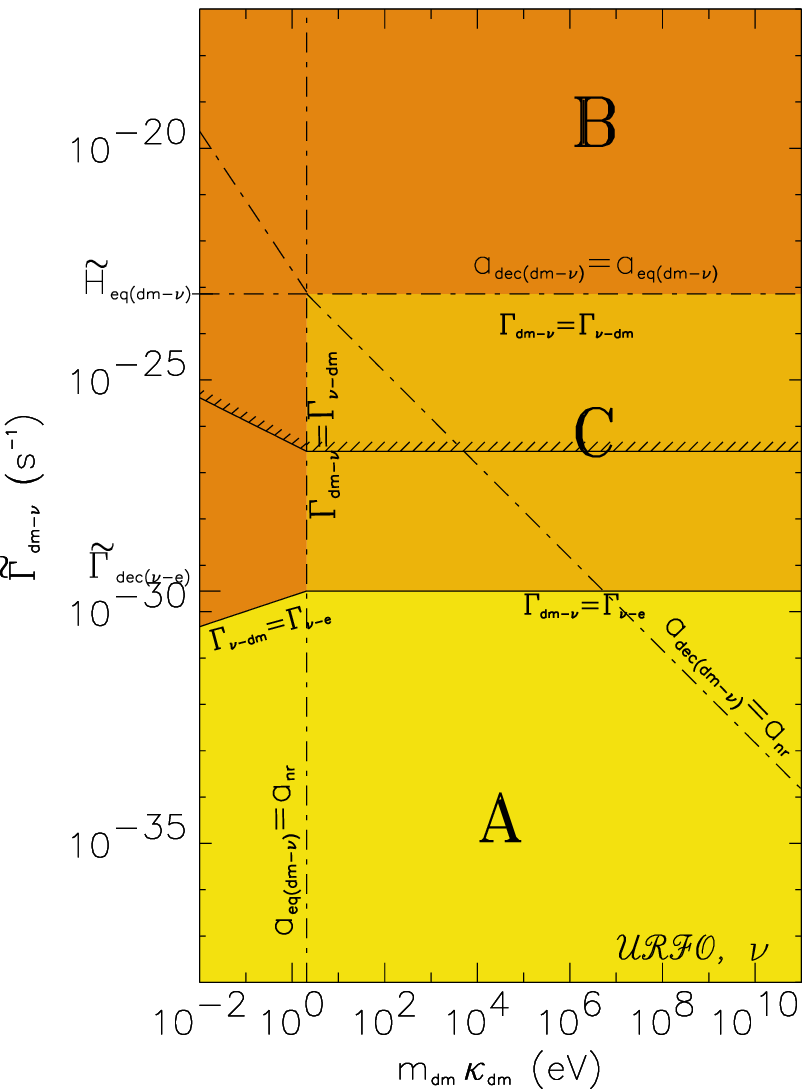}}
\par}
\caption{
Bounds in the [Dark Matter    particles' mass,  Dark Matter-neutrino interaction rate] parameter space obtained from neutrino induced-damping in the URFO scenario. The Regions A, B and C are distinguished by different colors. They correspond to different expressions, given in the text,  of the neutrino contribution to the Dark Matter   damping length. The dot-dashed lines separate the domains according to the ordering of the epoch of  the   Dark Matter - neutrino decoupling, the non-relativistic transition and  equality of the energy-densities. The hatches indicate in which part of the parameter space the neutrino induced-damping  scale is   greater than $100 kpc$ ($\sim 10^8 M_\odot$) scale. The indications in this figure are schematic. Some factors  in general of order unity are omitted for simplicity.  The reader interested by these constraints should use the expressions in the text, where all  factors are given explicitly.
}
 \label{fig:nugamurfo} 
\eef}
\vspace{0.3cm}

%
%
{\bef
{\par\centering \resizebox*{7.2cm}{9.6cm}{\includegraphics{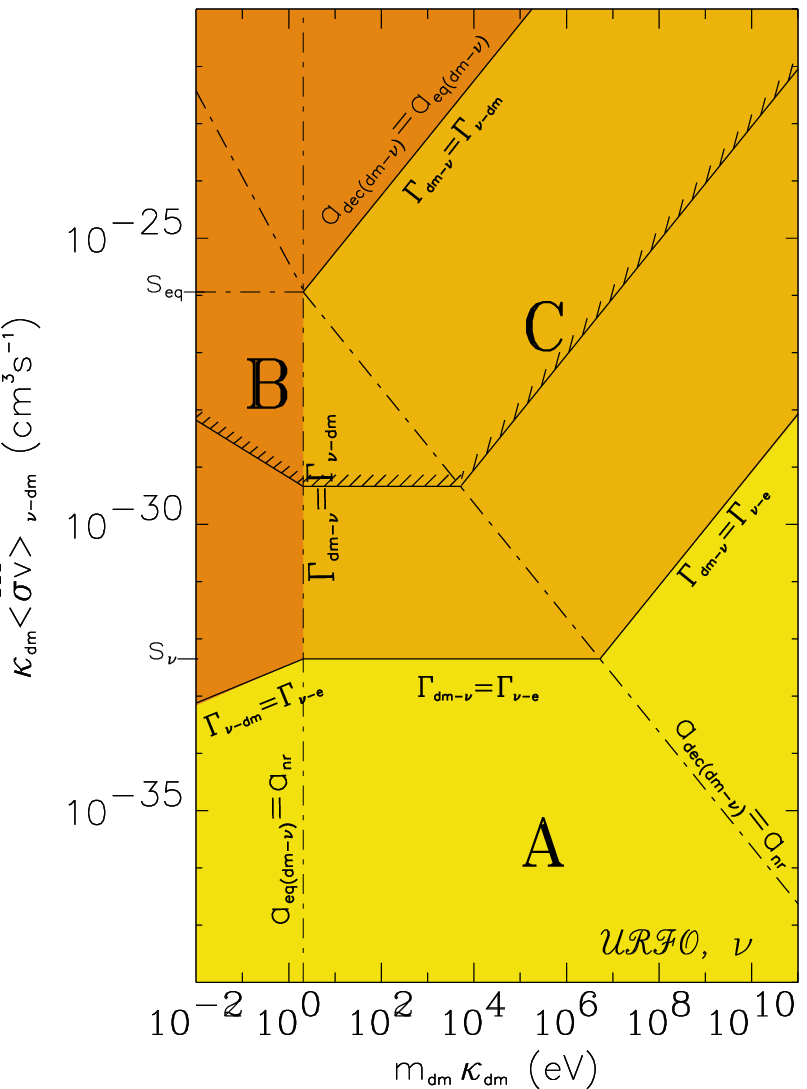}}
\par}
\caption{
Bounds in the  [Dark Matter   particles'  mass,   Dark Matter-neutrino cross-section] parameter space obtained from neutrino induced-damping in the URFO scenario. The Regions A, B and C are distinguished by different colorings. They correspond to different expressions, given in the text,  of the neutrino contribution to the   Dark Matter   damping length. The dot-dashed lines separate the domains in the parameter space where the ordering of the   Dark Matter   - neutrino decoupling, the non-relativistic transition, the epoch of equality of the energy-densities changes. The hatches indicate the region in parameter space which is forbidden because the neutrino induced-damping yields damping above $100 kpc$ ($\sim 10^8 M_\odot$) scale. The indications in this figure are schematic. Some factors  in general of order unity are omitted for simplicity.  The reader interested by these constraints should use the expressions in the text, where all  factors are given explicitly.
} 
\label{fig:nusigurfo} 
\eef}
\vspace{0.3cm}

%
%
{\bef
{\par\centering \resizebox*{7.2cm}{9.6cm}{\includegraphics{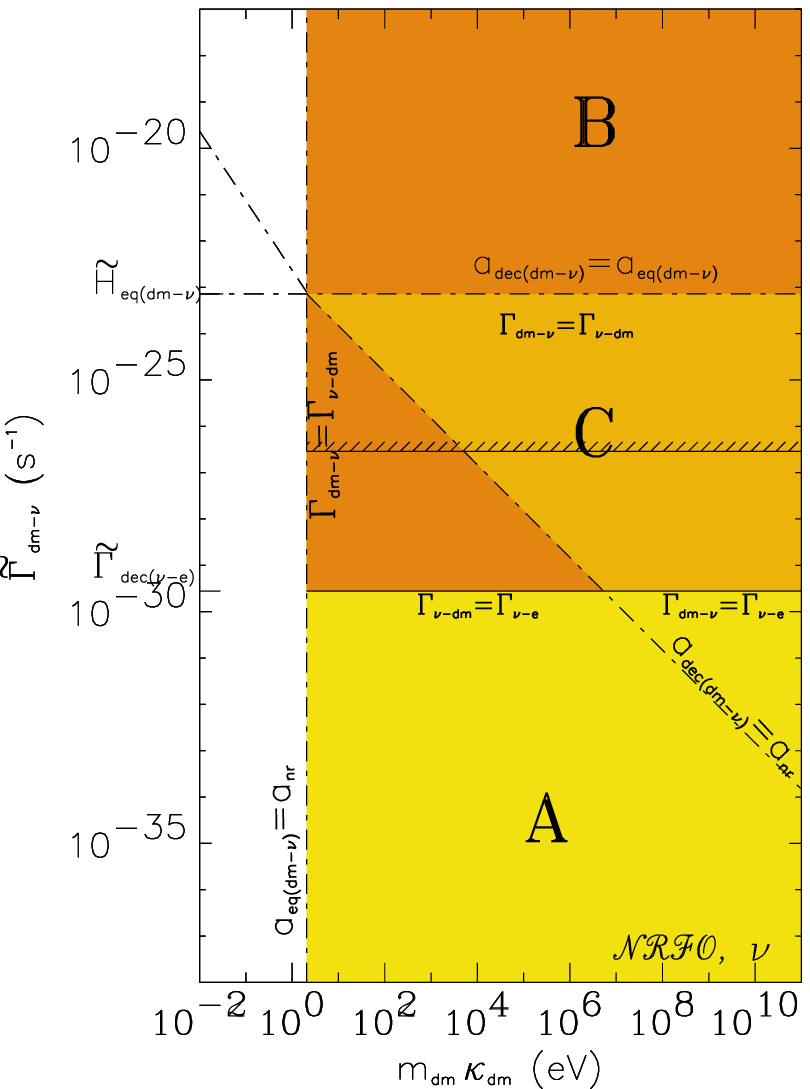}}
\par}
\caption{
Bounds in the  [Dark Matter    particles' mass,   Dark Matter-neutrino interaction rate] parameter space obtained from neutrino induced-damping in the NRFO scenario. The Regions A, B and C are distinguished by different colorings. They correspond to different expressions, given in the text,  of the neutrino contribution to the   Dark Matter   damping length. The dot-dashed lines separate the domains in the parameter space where the ordering of the   Dark Matter   - neutrino decoupling, the non-relativistic transition, the epoch of equality of the energy-densities changes.
The hatches indicate the region in parameter space which is forbidden because the neutrino induced-damping yields damping above the $100 kpc$ ($\sim 10^8 M_\odot$) scale. 
 The indications in this figure are schematic. Some factors  in general of order unity are omitted for simplicity.  The reader interested by these constraints should use the expressions in the text, where all  factors are given explicitly.
 }
   \label{fig:nugamnrfo} 
\eef}
\vspace{0.3cm}

%
%
{\bef
{\par\centering \resizebox*{7.2cm}{9.6cm}{\includegraphics{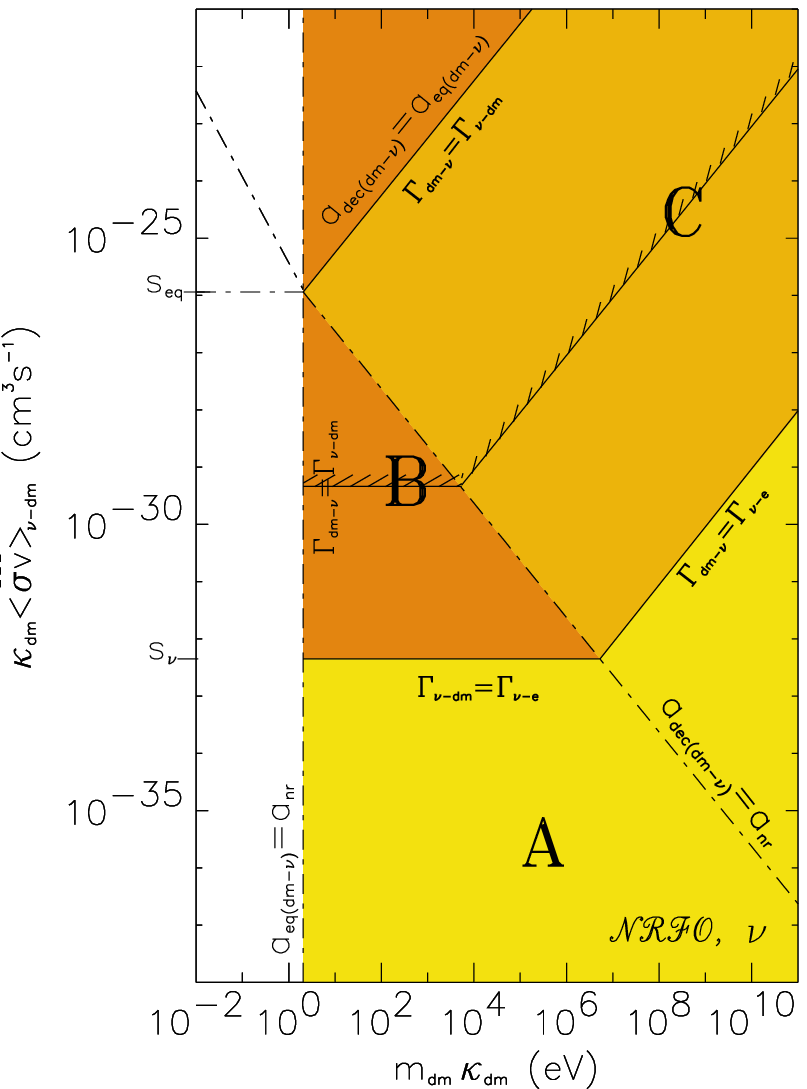}}
\par}
\caption{
Bounds in the [Dark Matter   particles' mass,   Dark Matter-neutrino cross-section] parameter space obtained from neutrino induced-damping in the NRFO scenario. The Regions A, B and C are distinguished by different colorings. They correspond to different expressions, given in the text,  of the neutrino contribution to the   Dark Matter   damping length. The dot-dashed lines separate the domains in the parameter space where the ordering of the   Dark Matter   - neutrino decoupling, the non-relativistic transition, the epoch of equality of the energy-densities changes.
The hatches indicate the region in parameter space which is forbidden because the neutrino induced-damping yields damping above the $100 kpc$ ($\sim 10^8 M_\odot$) scale.  The indications in this figure are schematic. Some factors  in general of order unity are omitted for simplicity.  The reader interested by these constraints should use the expressions in the text, where all  factors are given explicitly.
  } 
\label{fig:nusignrfo} 
\eef}
\vspace{0.3cm}

\subsection{Neutrino induced-damping scales and limits on the interaction rates. \label{sec:nulim}}

In the absence of significant   Dark Matter   interactions,
as it is commonly assumed,  
neutrinos are expected to decouple from the thermal bath in the radiation 
dominated era. The last interactions are with  electrons. 
If we adopt the standard neutrino-electron interaction (see Section \ref{app:typrates}), the neutrinos decouple at a 
temperature slightly above $1 MeV$, when their interaction rate is of the order of \(  \widetilde{\Gamma }_{dec(\nu-e) }  \sim  2.8 \  \ 10^{-30}  s^{-1} \ .
 \)
At neutrino decoupling, the damping scale is
\(
  l_{dec(\nu-e) } \  =  
\ \pi ct_r r_\nu \frac{\widetilde{\Gamma }_{dec(\nu-e) }}{H_r} \sim 97 \ pc    .
\)
The original calculation of this damping is due to  Misner (1967). It  corresponds to a mass scale somewhat below $ 1 \ M_\odot $. 
In the present case, the neutrino induced-damping scale 
takes this values in case the   Dark Matter    turns out to decouple from neutrinos just  at the epoch the
latter decouple from  electrons.

Let us now give the value of the damping lengths for all regions of the neutrino parameter-space. 
Analytical expressions of these lengths may be found in Appendix \ref{app:anndampnu}. 
Pending the astrophysical discussion in Paper II, we may already notice that the case $ \dslash \rho_{dm} \ge \dslash \rho_\nu$ makes little sense in case the relevant times to consider are close to the epoch of 
primordial nucleosynthesis, and is to be excluded. This is nevertheless possible somewhat before (provided 
annihilation occurs in time) or somewhat after (if   {  Dark Matter  }  particles are created late enough).

\subsubsection{ Expressions of the neutrino-induced damping scales in Region A and B. \label{sec:nucolldamp}}

\paragraph{Region A :}   
  
\beq
l_{\nu d} \  = \  r_\nu
\left(\frac{{\Gamma }_{dm-\nu}}{\Gamma_{\nu-e}}\right)^\frac{1}{2}
\pi \frac{ct}{a}  \vert_{dec(dm-\nu)}
\ ,
\eeq
or more explicitly
\beq
l_{\nu d}  \ =Ê\ 97  \ pc \  r_\nu
 { \kappa}  {\g_*^{-\frac{3}{2}}}
\left(\frac{ \widetilde{\Gamma }_{dm-\nu}}
{2.8 \ 10^{-30}  s^{-1}}\right)^\frac{5}{2}.
\label{lnuA}
\eeq
The largest value that this damping length 
can take is when $t_{dec(dm-\nu) } = t_{dec(\nu-e)}$. 
This case only 
provides  relevant limits  if we require primordial 
fluctuations to exist down to very small scales of less than $100 \, pc$ ($\sim 0.1\, M_\odot$). It can be obtained from eq.(\ref{lnuA}) by writing $l_{\nu d}  < l_{struct}$ but  will not be evaluated 
explicitly here as not of direct cosmological interest.

\paragraph{Region B :}

\beq
l_{\nu d}  =  r_\nu
\left(\frac{{\Gamma }_{dm-\nu}}{\Gamma_{\nu-dm}}\right)^\frac{1}{2}
\pi \frac{ct}{a}  \vert_{dec(dm-\nu)} \ .
\eeq
Only the case of a  radiation-dominated universe is relevant here 
since the converse leads to prohibitive damping.
Hence, as is readily seen in figs. (\ref{fig:nugamurfo}) to (\ref{fig:nusignrfo}), we need to consider only the case where   Dark Matter   is relativistic ($a_{dec(dm-\nu)} < a_{nr} $).
With the symmetry relations of the interaction rates 
(Sections  \ref{sec:symrates} and \ref{app:Gamsym}), we get the explicit relations displayed in Table \ref{tab:nuBURFO} and \ref{tab:nuBNRFO}. 
Note that we do not consider explicitly the {NRFO} scenario for $a_{eq(dm-\nu)} < a_{nr} $, a case which is possible but quite unlikely for the reasons discussed above, see Section \ref{sec:VI}.

Also, for the neutrino interaction rates relevant to our calculation, decoupling of the   Dark Matter   with neutrinos is to occur well before the standard epoch of matter-radiation equality and the fate of the neutrinos afterwards is irrelevant to our purpose: with the present limits on neutrino masses we remain well within the epoch where the neutrinos are fully relativistic.
%
\begin{center}
\begin{table}[h]
\caption{Region B (URFO scenario)} 
\[
\begin{array}{ll}
\hline
\hline 
\\
l_{\nu d} &= \, 68 \ kpc \  r_\nu   \left(\frac{4\g_{*\nu}}{3}\right)^\frac{1}{2}
\g_*^{-1} \left(\Odm\right)^{-\frac{1}{2}}
 \\
&
\hspace{.5cm} \left( \frac{m_{dm}  \kappa_{dm}}{1 MeV} \right)^\frac{1}{2} \frac{ \widetilde{\Gamma }_{dm-\nu}}{2.8 \ 10^{-30} \ s^{-1}} 
\\
\\
\widetilde{\Gamma }_{dm-\nu}  &<
4.1 \ 10^{-30} \ s^{-1} \ 
r_\nu^{-1}   \left(\frac{4\g_{*\nu}}{3}\right)^{-\frac{1}{2}}  \g_* 
\\
& \hspace{.5cm} \left(\Odm\right)^\frac{1}{2} 
 \left( \frac{m_{dm}  \kappa_{dm}}{1 MeV} \right)^{-\frac{1}{2}} 
\frac{l_{struct}}{100 kpc} 
\\
\\
\kappa_{dm} \svb_{\nu-dm}  &< 
6.6 \ 10^{-33} \ cm^3s^{-1} \ 
r_\nu^{-1}  \left(\frac{4\g_{*\nu}}{3}\right)^{-\frac{3}{2}}  \g_* 
\nonumber \\
&\hspace{0.5cm} \left(\Odm\right)^\frac{1}{2} 
\left( \frac{m_{dm}  \kappa_{dm}}{1 MeV} \right)^{-\frac{1}{2}} 
\frac{l_{struct}}{100 kpc}
 \\
\\
\hline
\hline
\end{array}
\]
\label{tab:nuBURFO}
\end{table}
\end{center}

\begin{center}
\begin{table}[h]
\caption{Region B (NRFO scenario)} 
\[
\begin{array}{ll}
\hline
\hline 
\\
l_{\nu d} &= 97 \ pc \ r_\nu 
\left( \frac{\g_{*\nu}}{ \g_{*dm}}\right)^{\frac{1}{2}}  
 \g_*^{-1} \
\frac{ \widetilde{\Gamma }_{dm-\nu}}{2.8 \ 10^{-30} \ s^{-1}} 
\\
\\
\widetilde{\Gamma }_{dm-\nu}  &<  
2.9 \ 10^{-27} \ s^{-1} \ 
r_\nu^{-1} 
\left( \frac{ \g_{*dm}}{\g_{*\nu}}\right)^{\frac{1}{2}}  
\g_*
\frac{l_{struct}}{100 kpc} 
\\
\\
 \kappa_{dm} \svb_{\nu-dm}  &<   4.6 \ 10^{-30} \  cm^3 s^{-1} \\
&r_\nu^{-1}   \left(\frac{4\g_{*\nu}}{3}\right)^{-1} 
\left( \frac{ \g_{*dm}}{\g_{*\nu}}\right)^{\frac{1}{2}}  
  \g_*
\frac{l_{struct}}{100 kpc} 
\\
\\
\hline
\hline
\end{array}
\]
\label{tab:nuBNRFO}
\end{table}
\end{center}

\subsubsection{Expressions of the neutrino-induced damping scales in Region C. \label{sec:numixdamp}}

The mixed-damping regime is at work in this region. We have  (Eq.\ref{lidre}) 
 \begin{equation} l_{\nu d} \sim 
 \pi  
r_\nu 
\frac{ c t}{ a} \vert_{dec(dm-\nu)}   \, .
\end{equation} 
Limits on $\widetilde \Gamma_{dm-\nu}$ are given in table \ref{tab:nuCURFO} and \ref{tab:nuCNRFO}, 
respectively, for the URFO and the NRFO scenario. These limits can again be transformed into limits 
for $\svb_{\nu-dm}$ by means of the symmetry relations between the interaction rates 
(Sections \ref{sec:symrates}  and \ref{app:Gamsym}).

This calculation does not require any knowledge of the number of electrons which are present, 
and holds whether the latter have annihilated or not. Also, all times of concern are well before 
the standard epoch of mater-radiation equality. It is worth to remember at this stage that our calculation are valid for massive neutrinos, provided their mass is within the known physical and astrophysical limits.


\begin{center}
\begin{table}[h]
\caption{Region C (URFO scenario)} 
\[
\begin{array}{ll}
\hline
\hline 
\\
l_{\nu d}  & \sim
97 \ pc \ r_\nu
 \g_*^{-1}
\frac{ \widetilde \Gamma_{dm-\nu}}{2.8 \ 10^{-30} \ s^{-1}}
\\
\\
\widetilde \Gamma_{dm-\nu} & <  
2.9 \ 10^{-27} \ s^{-1}  \ r_\nu^{-1}
 \g_*
\frac{l_{struct}}{100kpc}
\\
\\
a_{dec(dm-\nu)} > a_{nr}  \ \ \ &
\\
\\
\kappa_{dm} \svb _{\nu-dm}    &< 9.0 \ 10^{-28} \ cm^3s^{-1} 
\\ 
& r_\nu^{-2}  \left(\frac{4\g_{*\nu}}{3}\right)^{-1}  {\g_*^\frac{3}{2}}  
\frac{m_{dm}  \kappa_{dm} }{1 MeV}
\left( \frac{l_{struct}}{100kpc} \right)^2
\\
\\
a_{dec(dm-\nu)} < a_{nr} \ \ \  &
\\
\\
\kappa_{dm} \svb _{\nu-dm}    &< 4.6 \ 10^{-30} \ cm^3s^{-1} 
\\ 
& r_\nu^{-1 }  \left(\frac{4\g_{*\nu}}{3}\right)^{-1}    \g_*  
 \frac{l_{struct}}{100kpc} 
\\
\\
\hline
\hline
\end{array}
\]
\label{tab:nuCURFO}
\end{table}
\end{center}

\begin{center}
\begin{table}[h]
\caption{Region C (NRFO scenario)} 
\[
\begin{array}{ll}
\hline
\hline 
\\
l_{\nu d}  & \sim
97 \ pc \ r_\nu
 \g_*^{-1}
\frac{ \widetilde \Gamma_{dm-\nu}}{2.8 \ 10^{-30} \ s^{-1}}
\\
\\
\widetilde \Gamma_{dm-\nu}  & <  
2.9 \ 10^{-27} \ s^{-1}  \ r_\nu^{-1}
 \g_*
\frac{l_{struct}}{100kpc}
\\
\\
a_{dec(dm-\nu)} > a_{nr}  \ \ \ &
\\
\\
\kappa_{dm} \svb _{\nu-dm}    &< 9.0 \ 10^{-28} \ cm^3s^{-1} 
\\ 
& r_\nu^{-2}  \left(\frac{4\g_{*\nu}}{3}\right)^{-1}  {\g_*^\frac{3}{2}}  
\frac{m_{dm}  \kappa_{dm} }{1 MeV}
\left( \frac{l_{struct}}{100kpc} \right)^2
\\
\\
a_{dec(dm-\nu)} < a_{nr} \ \ \  &
\\
\\
\mbox{does not exist in this case}
\\
\\
\hline
\hline
\end{array}
\]
\label{tab:nuCNRFO}
\end{table}
\end{center}

%
%
%
%
{\bef
{\par\centering \resizebox*{7.2cm}{9.6cm}{\includegraphics{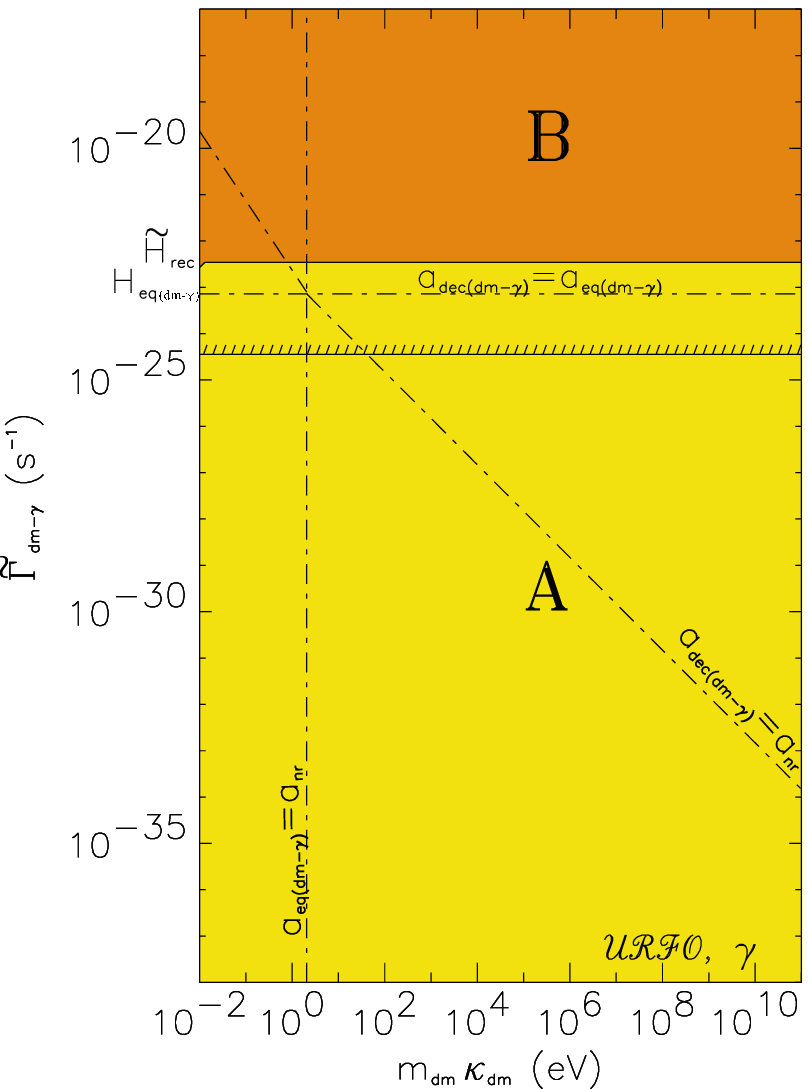}}
\par}
\caption{
Bounds in the  [Dark Matter   particles' mass,   Dark Matter  -photon interaction rate] parameter space obtained from photon induced-damping in the URFO scenario. The Regions A and B are distinguished by different colorings. They correspond to different expressions, given in the text,  of the photon contribution to the   Dark Matter   damping length. The dot-dashed lines separate the domains in the parameter space where the ordering of the   Dark Matter   - photon decoupling, the non-relativistic transition, the epoch of equality of the energy-densities changes.
The hatches indicate the region in parameter space which is forbidden because the photon induced-damping yields 
damping above the $100 kpc$ ($\sim 10^8 M_\odot$) scale.  The indications in this figure are schematic. Some factors  in general of order unity are omitted for simplicity.  The reader interested by these constraints should use the expressions in the text, where all  factors are given explicitly.
 }  
\label{fig:phgamurfo} 
 \eef}
\vspace{0.3cm}

%
%
{\bef
{\par\centering \resizebox*{7.2cm}{9.6cm}{\includegraphics{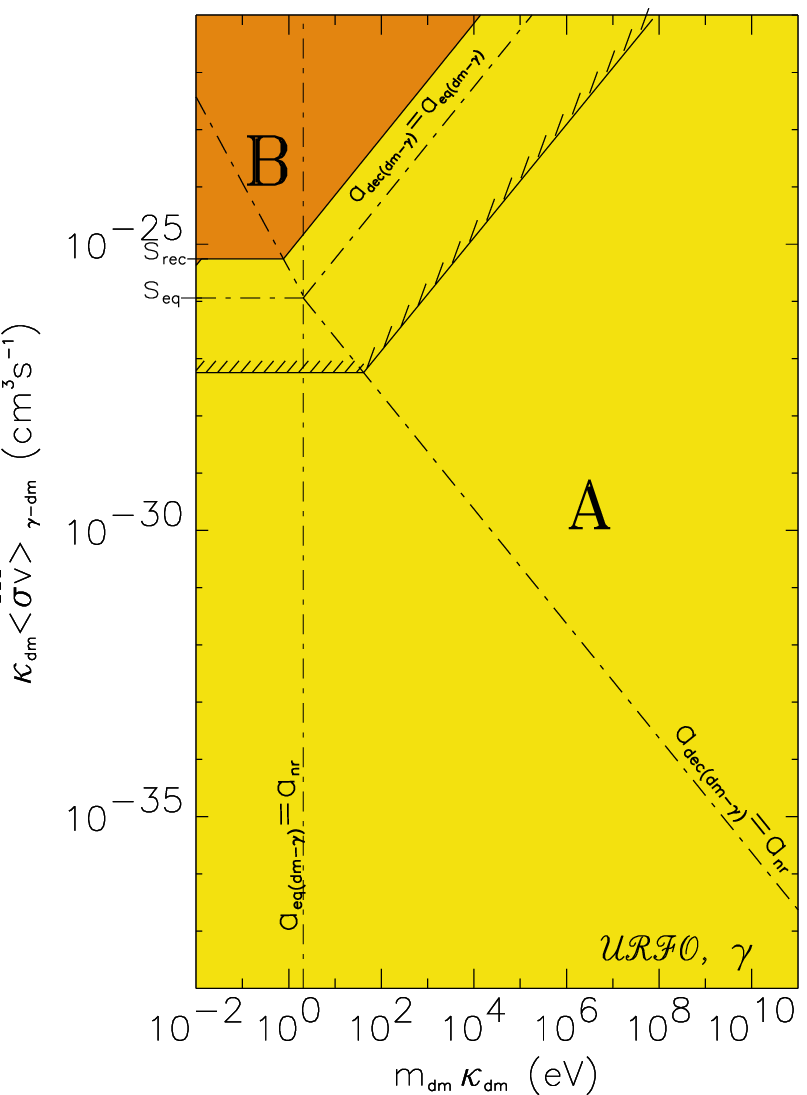}}
\par}
\caption{
Bounds in the [Dark particles' Matter mass,   Dark Matter  -photon cross-section] parameter space obtained from photon induced-damping in the URFO scenario. The Regions A and B are distinguished by different colorings. They correspond to different expressions, given in the text,  of the photon contribution to the   Dark Matter   damping length. The dot-dashed lines separate the domains in the parameter space where the ordering of the   Dark Matter   - photon decoupling, the non-relativistic transition, the epoch of equality of the energy-densities changes.
The hatches indicate the region in parameter space which is forbidden because the photon induced-damping yields damping above the $100 kpc$ ($\sim 10^8 M_\odot$) scale. The indications in this figure are schematic. Some factors  in general of order unity are omitted for simplicity.  The reader interested by these constraints should use the expressions in the text, where all  factors are given explicitly.
 } 
  \label{fig:phsigurfo} 
\eef}
\vspace{0.3cm}

%
%
{\bef
{\par\centering \resizebox*{7.2cm}{9.6cm}{\includegraphics{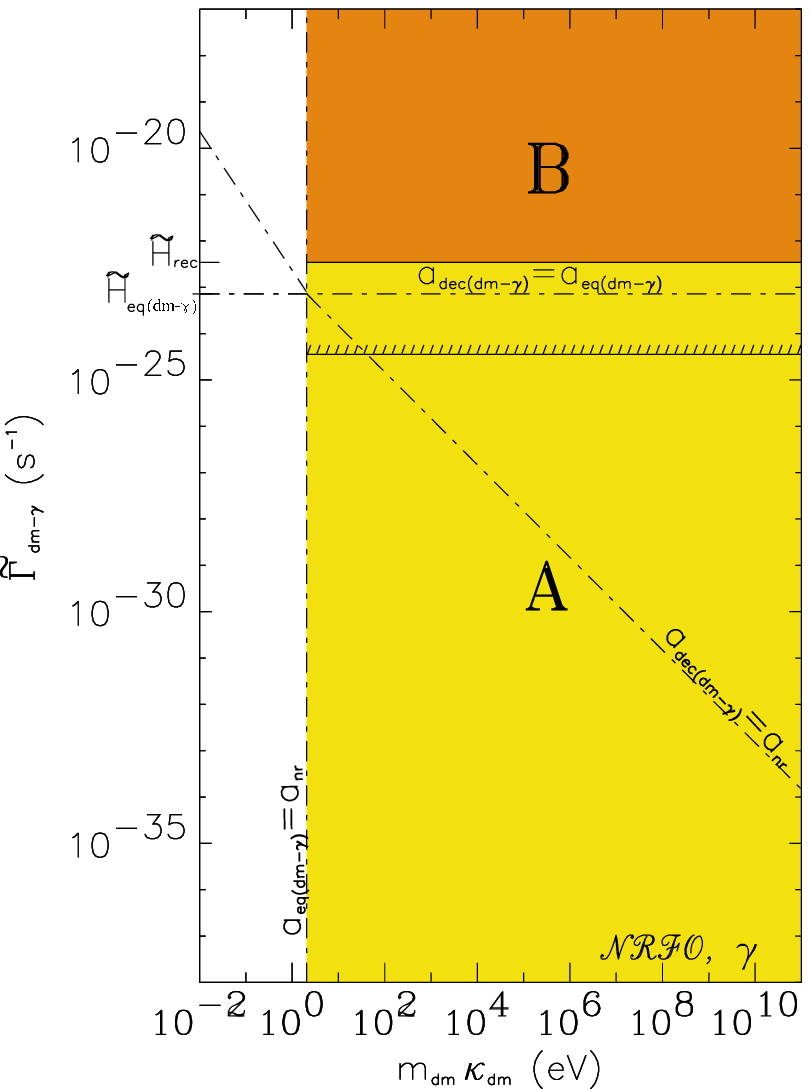}}
\par}
\caption{
Bounds in the  [Dark Matter   paticle's mass,   Dark Matter  -photon interaction rate] parameter space obtained from photon induced-damping in the NRFO scenario. The Regions A and B are distinguished by different colorings. They correspond to different expressions, given in the text,  of the photon contribution to the   Dark Matter   damping length. The dot-dashed lines separate the domains in the parameter space where the ordering of the   Dark Matter   - photon decoupling, the non-relativistic transition, the epoch of equality of the energy-densities changes.
The hatches indicate the region in parameter space which is forbidden because the photon induced-damping yields damping above the $100 kpc$ ($\sim 10^8 M_\odot$) scale. The indications in this figure are schematic. Some factors  in general of order unity are omitted for simplicity.  The reader interested by these constraints should use the expressions in the text, where all  factors are given explicitly.
}
 \label{fig:phgamnrfo} 
\eef}
\vspace{0.3cm}

%
%
{\bef
{\par\centering \resizebox*{7.2cm}{9.6cm}{\includegraphics{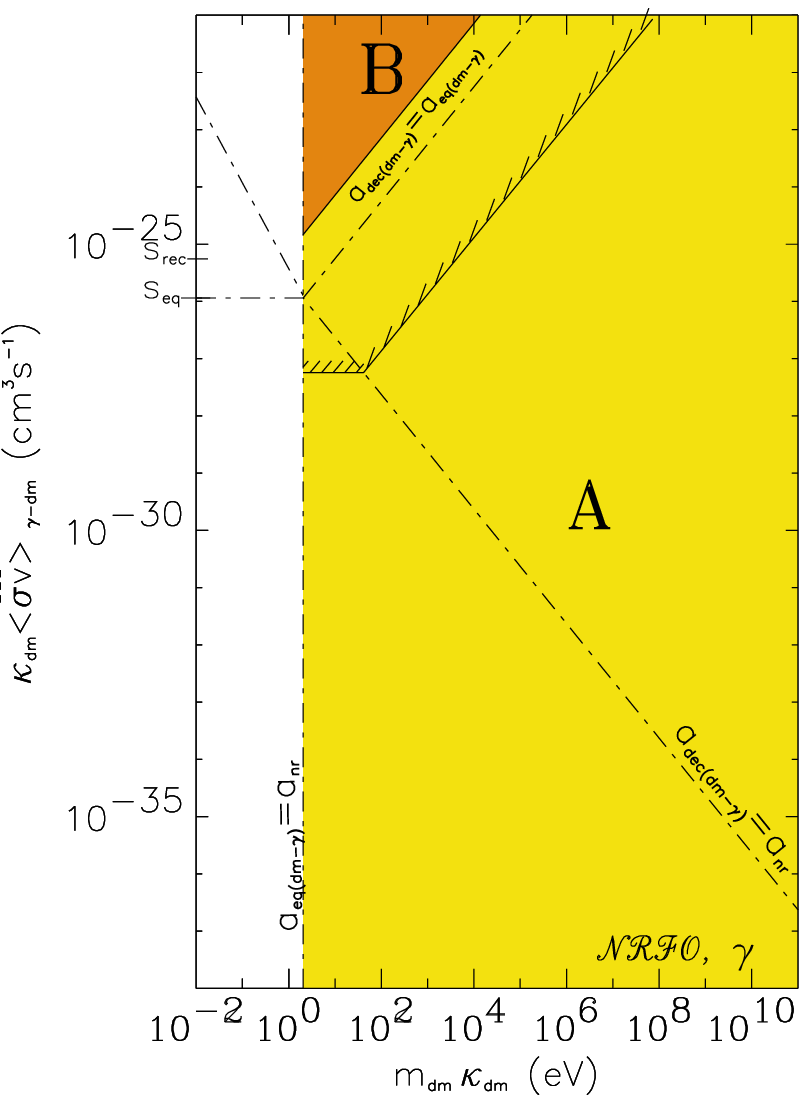}}
\par}
\caption{
Bounds in the [Dark Matter   particles' mass,   Dark Matter  -photon cross-section] parameter space obtained from photon induced-damping in the NRFO scenario. The Regions A and B are distinguished by different colorings. They correspond to different expressions, given in the text,  of the photon contribution to the   Dark Matter   damping length. The dot-dashed lines separate the domains in the parameter space where the ordering of the   Dark Matter   - photon decoupling, the non-relativistic transition, the epoch of equality of the energy-densities changes.
The hatches indicate the region in parameter space which is forbidden because the photon induced-damping 
yields damping above the $100 kpc$ ($\sim 10^8 M_\odot$) scale.  The indications in this figure are schematic. Some factors  in general of order unity are omitted for simplicity.  The reader interested by these constraints should use the expressions in the text, where all  factors are given explicitly.  
 }  
 \label{fig:phsignrfo} 
\eef}
\vspace{0.3cm}

\section{Constraining   Dark Matter   properties from the photon induced-damping scale.\label{sec:idphotdamp}}

As already mentioned, the possible coupling between   Dark Matter   and photons 
also induces a source of collisional damping. 
The associated damping scale $l_{\gamma d}$ can be inferred from the 
formula (\ref{lid}). The calculation, however, is somewhat different from the 
one 
 for $l_{\nu d}$ because the epoch at which the Universe becomes 
matter dominated and the epoch of the recombination now get into play. 
This introduces two new scales $\widetilde H_{eq(dm-\gamma)}$ and 
$\widetilde H_{rec}$ which are the "reduced" ($\widetilde H = H a^3$) Hubble rates at some specific time related to the epoch of equality and recombination, respectively (precise definition are given in Sect. \ref{app:typHrates}). 
Similarly to the neutrino case, from
$\Gamma _{\gamma } =
\Gamma _{\gamma-e}  + 
\Gamma _{\gamma-dm}  + .... $
we see that the relevant interaction rates are $\widetilde\Gamma _{\gamma }$, 
$\widetilde\Gamma _{\gamma-e} $ and $\widetilde\Gamma _{\gamma-dm} $  (see Section \ref{app:redrates}). So, the parameter space for the photon induced-damping may be taken as $  [m_{dm}, \widetilde{\Gamma}_{dm-\gamma} ]  $ or $[  m_{dm}, {\svb}_{\gamma-dm} ] $.

\subsection{  Dark Matter   parameter space \label{sec:regphot}}

"Regions"  are expected to appear, as in the neutrino case, depending on whether photons are collisional or collisionless.  The new discussion here is about the role of recombination and the relevance of the epoch of matter dominance. The borderlines of these regions are given explicitely in Appendix \ref{app:bordphot}.

\subsubsection{  Dark Matter   decouples from photons while the latter are collisional.
\label{sec:regphotcoll}}

This implies

\begin{center}
$\widetilde{\Gamma }_{dm-\gamma} \le \widetilde{\Gamma }_{\gamma }  \ \   \vert_{dec(dm-\gamma)} \ . $
\end{center}

\paragraph{Region A.} Photons collisions dominated by photon-electron scattering.

 $ \Gamma_{\gamma-dm} \le \Gamma_{\gamma-e}     \ \ \    \vert_{dec(dm-\gamma)} $

\paragraph{Region B.} Photons collisions dominated by photon-  Dark Matter   scattering.

 $ \Gamma_{\gamma-dm} > \Gamma_{\gamma-e}     \ \ \    \vert_{dec(dm-\gamma)} $

The separation between the two regions A and B may, among others, occur for $dm-\gamma$ decoupling before recombination, that is for a sizeable $\Gamma_{\gamma-e}$ rate. This requires simultaneously sufficiently small $\Gamma_{dm-\gamma}$ rate for the decoupling be early enough, and a sufficiently large $\Gamma_{\gamma-dm} $ rate to be larger than the $\gamma-e$ interaction.
These two, somewhat antagonistic, requirements can actually be satisfyed (see the evaluation in Appendix \ref{app:bordphot}), but for unrealistic values of the   Dark Matter   particles' mass $m_{dm}$ (of the order of $10^{-2} eV$). This separation is barely visible in figs. (\ref{fig:phgamurfo}) and (\ref{fig:phsigurfo}).

More realistically, the separation between the regions A and B simply occurs at recombination: before the latter the $\gamma-e$ interactions dominate, and after, only $\gamma-dm$ interactions are left. So,  Regions A and B in practical cases correspond just to the separation between $dm-\gamma$ decoupling before and after recombination (Table \ref{tab:gamreg}).

\subsubsection{  Dark Matter   decouples from photons
while the latter are freely-propagating. \label{sec:regphotmix}}

This is the analogue of region C in the neutrino case.  It requires $\Gamma_\gamma < H$ at the    Dark Matter  -photon decoupling  ($\Gamma_{dm-\gamma} = H$). The earliest this may happen, whatever the strength of the   Dark Matter   interactions, is at recombination: so,   Dark Matter  -photon decoupling necessarily occurs at an epoch where the universe is matter-dominated. One then  has necessarily $\Gamma_{dm-\gamma} \le \Gamma_{\gamma-dm}$ according to the relation (\ref{ratmom}). But the first relation implies $\Gamma_{\gamma-dm} < H$
and in turn the second $\Gamma_{dm-\gamma} < H$. 
Therefore   Dark Matter   cannot remain coupled to  photons while the latter are freely propagating.   So there is no Region C, \ie no mixed-damping regime.

\begin{center}
\begin{table}[h]
\caption{Definition of the two   Dark Matter   Regions useful for the {photon} induced-damping calculation 
\label{tab:gamreg}}
\[
\begin{array}{lll} \hline 
\hline
&&\\
\mbox{Region A} &\hspace{.3cm}    \widetilde{\Gamma }_{dm-\gamma} \le
\widetilde{\Gamma }_{\gamma-e} \ \ \vert_{dec(dm-\gamma)} 
& \hspace{.3cm}  \ a_{dec(dm-\gamma)} \le a_{rec} \\
\\
\hline
&&\\
\mbox{Region B} &\hspace{.3cm} \widetilde{\Gamma }_{dm-\gamma} \le \widetilde{\Gamma }_{\gamma-dm} \ \ \vert_{dec(dm-\gamma)} 
& \hspace{.3cm} \ a_{dec(dm-\gamma)} > a_{rec} \\
\\
\hline
\hline
\end{array}
\]
\end{table}
\end{center}

\subsection{Photon induced-damping scales and limits on the interaction rates. \label{sec:photlim}} %

We now compute the damping scales associated with the regions A and B 
for the two scenarios NRFO and URFO.  
Region A is slightly more complicated 
than region B because it can potentially exist in both a radiation 
and matter dominated Universe  while 
region B necessarily takes place in a matter dominated Universe. We give here the most relevant numerical results. A more systematic presentation of our analytical results is given in Appendix \ref{app:anndampphot}.

\subsubsection{Expression of the photon-induced damping scale for Region A.}

The condition 
$\widetilde{\Gamma}_{\gamma-e } > \widetilde{\Gamma}_{\gamma-dm }$ is in some sense the standard situation since, in this case,  
the photon decoupling is
set by the recombination epoch. The damping of the   Dark Matter   fluctuations is then induced by the 
interaction of photons with the electrons.
Below a temperature around 100 keV (a case which is relevant for 
$\widetilde{\Gamma }_{dm-\gamma} > 10^{-28} s^{-1}$ 
and is the only regime of practical interest), the photon-electron elastic 
scattering cross-section (\textit{i.e.} the 
Thomson cross-section) yields the reduced interaction rate 
$\widetilde\Gamma _{\Th } = \sigma_{\Th} \ c \ \widetilde n_e
\ \sim 5 \ 10^{-21} \ s^{-1}$, while the recombination corresponds to a (reduced) Hubble rate $\widetilde H _{rec} \sim 3 \ 10^{-23} s^{-1}$ . 

We get in this case
\beq
l_{\gamma d} = \pi \, r_\gamma \,
\left(\frac{\widetilde{\Gamma }_{dm-\gamma}}{\widetilde{\Gamma }_{\Th}}\right)^{\frac{1}{2}} \, \frac{ct}{a}\vert_{dec(dm-\gamma)} \, . \eeq 

We are led to consider the two following cases~: 
\begin{enumerate}
\item 
  Dark Matter    decouples from photons in a radiation dominated 
Universe, that is $ \widetilde{\Gamma }_{dm-\gamma} < \widetilde H_{eq(dm-\gamma)}$. In this case the shear viscosity dominates. This  yields 
the damping length
as well as the constraints given in Table \ref{grd1}. 

\begin{center}
\begin{table}[h]
\caption{Region A. Decoupling in a radiation dominated Universe 
($a_{dec(dm-\gamma)}< a_{eq}$) \label{grd1}}
\[
\begin{array}{l} 
\hline 
\hline  \\
 l_{\gamma d}  = 8.2 \ Mpc  \   r_\gamma   
{\g_*^{-1}}
  \left(\Ob \right)^{-\frac{1}{2}}
 \left(\frac{ \widetilde{\Gamma }_{dm-\gamma}}
{  6 \,10^{-24} \, s^{-1}}\right)^\frac{3}{2} \label{lgamA}  \\ \\ \widetilde{\Gamma }_{dm-\gamma}  <  3.2  \ 10^{-25} \  s^{-1}  
 \  r_\gamma^{-\frac{2}{3}} \ 
{\g_*^\frac{2}{3}}   \left(\Ob\right)^{\frac{1}{3}}
  \left( \frac{l _{struct}}{100 kpc} \right)^{\frac{2}{3}} \label{GgamA}  \\
   \\
a_{dec(dm-\gamma)} > a_{nr}   \\
\\
\kappa_{dm} \svb_{\gamma-dm}   < 
1.1 \ 10^{-23} \  cm^3s^{-1} 
   r_\gamma^{-\frac{4}{3}} 
{\g_*^{ \frac{5}{6}}(T)}  \ 
\nonumber \\
\hspace{2.3cm} 
\left(\Ob\right)^{\frac{2}{3}} 
 \left(\frac{m_{dm}\kappa_{dm}}{1MeV}\right) \
  \left( \frac{l _{struct}}{100 kpc} \right)^{\frac{4}{3}}   \\
\\
a_{dec(dm-\gamma)} < a_{nr}   \\
\\
\kappa_{dm} \svb_{\gamma-dm}  <  
 5.1  \ 10^{-28} \ cm^3 s^{-1}
 \  r_\gamma^{-\frac{2}{3}}  
{\g_*^\frac{2}{3}}{\left(\frac{4\g_{*\gamma}}{3}\right)}^{-1} 
\nonumber \\
\hspace{2.3cm} 
\left(\Ob \right)^{\frac{1}{3}}
  \left( \frac{l _{struct}}{100 kpc} \right)^{\frac{2}{3}}   \\
\hline
\hline
 \label{sgamdmA}
\end{array}
\]
\end{table}
\end{center}

\item 
  Dark Matter    decouples from photons in a matter dominated 
Universe, that is 
$\widetilde{\Gamma }_{dm-\gamma} > \widetilde H_{eq(dm-\gamma)}$. This is the case  
where the heat conduction dominates, so that 
\begin{eqnarray}
l_{\gamma d}  = 8.2 \, \mbox{Mpc} \   r_\gamma
\, & \left(\Om\right)^{\frac{1}{3}} \left(\Odm\right)^{-1}  \left(\Ob \right)^{-\frac{1}{2}}  \cr & \left(\frac{ \widetilde{\Gamma }_{dm-\gamma}} {6 \  10^{-24} \, s^{-1}}\right)^\frac{5}{6}. \end{eqnarray} This scale is typically above $1 Mpc$. So, when the {  Dark Matter  }-photon 
decoupling is in the matter dominated era,
the photon induced-damping is always prohibitive according to our criteria. \\ \\ 
It may be of interest to note that for the largest interaction rate allowed in Region A, that is for 
$\widetilde{\Gamma }_{dm-\gamma}
= \widetilde H_{rec }$, \textit{i.e.}  the   Dark Matter   decouples from the photons at the recombination epoch. 
The scale $l_{\gamma d}$ thus reaches $l_S = \pi c t_m r_\gamma \left( {\widetilde H_{rec } a_{rec }}/
{ \widetilde{\Gamma }_{\Th}} \right)^\frac{1}{2}$, the Silk damping length  which may be estimated as
 \beq 
l_S =4.9 \ Mpc \ ( \Omega_m \hu^2 )^{-\frac{1}{4}}  
( \Omega_b \hu^2)^{-\frac{1}{2}} \  .
\eeq
This length corresponds to a total mass (including the   Dark Matter  )
\beq
 M_S = 6.6 \ 10^{13} \ M_\odot \ ( \Omega_m \hu^2 )^{\frac{1}{4}}  
(\Omega_b \hu^2 )^{-\frac{3}{2}} \ .
\eeq
 It may be noted that, with the now well established values of $\Omega_m$, $H_0$ and especially $\Omega_b$, these are very large scales, of the order of $l_S = 30 \, Mpc$ and $M_S = 10^{15} M_\odot$. \end{enumerate}

\subsubsection{Expression of the photon-induced damping scale for Region B. \label{sec:photlimB} } %
The case where   Dark Matter   decouples from the photons after the recombination 
obviously leads to a damping which is larger than the Silk damping. 
We shortly treat this case for completeness. 
For cross-sections just above this threshold, that is for 
$\widetilde{\Gamma }_{dm-\gamma} \sim \widetilde H_{rec}$, the 
reduced photon-  Dark Matter   interaction rate is required to be
$\widetilde{\Gamma }_{\gamma-dm} = \widetilde{H }_{eq(dm-\gamma)} (\widetilde H_{rec}/\widetilde{H }_{eq(dm-\gamma)})^{5/3}$ and is smaller than $ \widetilde{\Gamma}_{\Th}$ . 
Since, after recombination, the photon-electron interactions 
become negligible, only $\widetilde{\Gamma }_{\gamma-dm}$ contributes  to $\widetilde{\Gamma }_{\gamma}$. 
Such a photon-  Dark Matter   interaction therefore induces a sudden increase 
of the transparency (and of the viscosity coefficients), inducing in turn an increase of the damping length~: 
\beq l_{\gamma d} = \ \pi \ r_\gamma \ \left(\frac{\widetilde{\Gamma }_{dm-\gamma}}{\widetilde{\Gamma }_{\gamma-dm}}
\right)^{1/2} \ \frac{ct}{a} \ \vert_{dec(dm-\gamma)} . 
\eeq 
More specifically, we obtain  (for non-relativistic Dark Matter, as is the case in nearly all of Region B):
\beq\nonumber
l_{\gamma d}  \sim  350 \ Mpc \ r_\gamma \ \left(\Om\right)^{-\frac{1}{2}}  \left(\Odm\right)^{-\frac{1}{2}} \ .
\eeq
The damping length stays constant   (and equal to the above
value) whatever the interaction rate. For  larger ${dm-\gamma}$ interaction rates, the viscosity coefficient gets smaller but the interaction occurs a longer time. 
The two effects happen to exactly compensate.

The damping in this case, obviously, is prohibitively large. This is also expected to be the case for the vanishingly small region in the parameter-space where the Dark Matter is relativistic.

\subsection{Limits on indirect   Dark Matter  -photon interactions. }
Even neutral   Dark Matter   may experience photon induced-damping. Indeed, if the   Dark Matter   happens to be directly coupled to any of the species $j$ which are themselves coupled to the photon fluid (baryons, electrons etc...), the above damping is relevant. Photon induced-damping then is at work as long as the   Dark Matter   is coupled to the fluid $j$, and the previous estimates are relevant, with the replacement
\beq
\widetilde{\Gamma }_{dm-\gamma} \ \to \  \widetilde{\Gamma }_{dm-j}  \ = \ 
 \svb_{dm-j } \ \widetilde{n}_j \ .
\eeq
This can be
used to get constraints on the relevant $dm-j$ cross-sections. We for instance get 
\begin{eqnarray} 
\nonumber 
\svb_{dm-b }  &<&  1.6  \ 10^{-18} \  cm^3s^{-1} \ r_\gamma^{-\frac{2}{3}}  
{\g_*^\frac{2}{3}(T)}  \\
&& {\left(\Ob\right)^{-\frac{2}{3}}}  
 \left( \frac{l _{struct}}{100 kpc} \right)^{\frac{2}{3}} \ . \end{eqnarray} 
 Exactly the same constraint holds for the $dm-e$ cross-section.

 \section{Conclusions.}

In this general evaluation of damping properties, 
we have developed an approach which allows to discuss systematically all possible kinds of   Dark Matter. 

We have written the transport coefficients for a composite fluid of species in terms of the interaction rate of 
each of these species with the medium. This is rather straightforward for the shear viscosity and can be readily 
extended to heat conduction. We have taken some care to discuss bulk viscosity along the same lines, although 
the latter turns out for the sake of the present paper  not to generate stronger constraints on the collision 
rates and cross-sections than shear viscosity. This provides a new  insight into the physical processes at work, 
and has been used as an opportunity to correct some inaccuracies and wrong statements in earlier papers. We finally 
wrote all the transport coefficients of a composite fluid as a sum of  contributions from the individual components, taking the coupling to the whole medium into account. This is the key step  which allows us to {\it discuss 
the damping of all possible kinds of   Dark Matter}.

We are led to separate, following \cite{bfsL}, the contribution due to \emph{self-damping}, that is the transport 
by the   Dark Matter   itself, and  the one due to the \emph{induced-damping} which is the transport by other species effective for damping the   Dark Matter   fluctuations. The classical examples of which being neutrinos and photons.  This work aims to put our previous results on a general and firm basis.

 Writing that each of these contributions alone must not produce prohibitive damping, 
we were able to get \emph{necessary} conditions separately for  the   Dark Matter   interaction rate with all coupled species, as well as for its specific interaction with each of the species it is coupled to. More generally, we identify the parameter space of relevant, namely interaction rate and   Dark Matter   particles' mass. Another parameter, which is seen to appear at the proper places, comes into play. The latter ($\kappa_{dm}$ in the text) measures the possible offset of the   Dark Matter   and photon temperatures. It is in practice of order unity or somewhat larger, but it must be stressed that in special circumstances (e.g. axion-like   Dark Matter) this parameter may get overwhelmingly large and provide the dominant physics.

As expected, the bounds due to self-damping are closely related to the damping due to   Dark Matter   free-streaming. 
Both bear similar properties: the velocity of the Dark Matter particles, that is their mass, is the most important parameter. However, there are regions of parameter space where the Dark Matter interaction plays a role. As a result, in these regions, the expression of the free-streaming length (expressed in term of the Dark Matter mass), 
which is generally used for WDM candidates, does not hold anymore. Instead the free-streaming length depends on both 
the mass and interaction rate of the Dark Matter candidate. 
Said differently, the well-known requirement of having   Dark Matter   particles with a mass above $\sim 1 keV$ 
is valid only in a restricted region of parameter space, which covers only a  fraction of the Dark Matter   models, 
where the Dark Matter interactions are really negligible. In regions where the Dark Matter interaction rate is not 
too large, the smallest acceptable value of the Dark Matter mass is above $1 MeV$. In other parts, where the 
interaction rate turns out to be very large,  extremely small masses (notably in the $ eV$ range) 
are allowed.

The bound due to the damping induced by a given species merely depends on the interaction rate of the   Dark Matter   
with this species, as well as on the species' particle velocity. Relativistic species are  the more efficient, 
damping the   Dark Matter   fluctuations nearly over the horizon size.

The neutrino induced-damping yields quite a rich  parameter-space diagram (Dark-Matter-neutrino interaction rate/
Dark Matter   particles' mass). The bounds due to the possible coupling of the   Dark Matter   to the neutrinos turn 
out to be quite novel. Indeed, one in particular refers to Dark Matter that is coupled to neutrinos which are already 
free-streaming! We have evaluated analytically this effect in what we have called the \emph{mixed-damping}. The 
difference with the \emph{collisional bound}, which is only a lower bound to the damping length but takes only into 
account the effect of standard collisions, is large. Our mixed-damping estimate is still rather rough, although quite 
natural a result. It definitely would deserve a check by numerical simulations. The difference with the {collisional 
bound} is so large that there should be a substantial effect even in a realistic calculation, and  the damping of a 
collisional fluid coupled to the free-streaming neutrinos should be the dominant contribution as we have stressed here. 
This is important, especially in the context where Warm   Dark Matter  scenarios. Indeed, we saw that, 
when our constraints are marginally satisfied, the damping induced by the neutrinos could yield a collisional 
Dark Matter candidate.

The bound due to a possible coupling of   Dark Matter   with photons (this may be the case through higher order 
reactions involving charged particles even if the   Dark Matter   is neutral), in the (Dark-Matter-photon interaction rate/
Dark Matter   particles' mass) parameter space, are somewhat less constraining. They are 
nevertheless relevant for a large fraction of the parameter space. Again, the possibility for Warm   Dark Matter  
scenarios, where the damping is not due to free-streaming, but to 
interactions, exists.

The major (and nearly the only) hypothesis is that the   Dark Matter   went through a phase of statistical equilibrium, at some stage of its evolution. 
There is also the (milder) assumption  that a sudden decrease by many orders of magnitude in the   Dark Matter particle density is not possible elsewhere than at its non-relativistic transition (for instance a phenomenon like the recombination for the electrons -where the latter are substracted from the medium at a temperature bearing no relation with the electron mass- is assumed not to occur for   Dark Matter  ).

Our bounds on the   Dark Matter   interactions with the most relevant species have been calculated  without any 
assumption on the form of the primordial spectrum. 
The calculations are valid for any Dark Matter interaction rate and any Dark Matter particle mass, the values of practical interest covering 20 orders of magnitude for the interaction rates or cross-sections and much over 13 orders of magnitude for the Dark Matter particles' mass.  Our estimates are valid within factors of order unity, a procedure which provides important simplifications in the classification. A thorough discussion of all possibilities allows to place any scenario within the relevant parameter space, with a precise statement on the important parameters and the actual physical process at work. The reader interested in a specific scenario can easily work out the relevant bound with more accuracy. He/she will find here a useful estimate of the result and a guide on how to proceed.
The present results provide, we believe, a new starting basis for a systematic 
discussion on the possible nature of   Dark Matter. The astrophysical relevance of these findings, including other 
constraints than just the damping condition, or more restrictive assumptions on the initial conditions, in particular 
those of (\cite{brhs})  which specifically restrict themselves to CDM initial conditions in order to yield more stringent limits, will be thoroughly discussed in Paper II.

\section*{Acknowledgement}

The authors would like to thank P. Fayet for stimulating discussions.

\bibliographystyle{aab}
\bibliography{references}




\appendix



\section{ Definitions  and notations.
\label{app:def} }

We give here the expressions actually used in the text. They may be obtained  from elementary textbooks (e.g. \cite{KT}).

\subsection{Present-day Universe.} 

We write the present-day Hubble constant as
\[
 H_0 = 70 \  \hu \ km \ s^{-1} Mpc^{-1} \ .
 \]

\subsubsection{Temperature. } 

We use for the present CMB temperature the value
\[ 
T_0= 2.73\  K = 2.35 \ 10^{-10} \ \mbox{MeV} \ , 
\]
that is
\[ 
 \frac{ \hbar c }{T_0} = 8.39 \ 10^{-2} \ cm \  .
\]

\subsubsection{Energy densities. }

The present critical, photon and neutrino (the latter for our "reference" cosmology) energy densities are \begin{eqnarray*} \nonumber 
\rho_{0c} &=& \  9.21\ 10^{-30} \ \hu^2  g \ cm^{-3} \cr
\ &=& \ 5.16 \ 10^{-3} \  \hu^2 MeV/c^2 \ cm^{-3}\cr
\ &=& \ 1.36 \ 10^{11} \  \hu^2 M_{\odot} \ Mpc^{-3}
\cr 
\cr
\rho_{0\gamma} &=& 4.67  \ 10^{-34} \  g \ cm^{-3} \ \cr \rho_{0\nu \st } &=& 3.19  \ 10^{-34} \  g \ cm^{-3} \cr \rho_{0(\gamma+\nu \st )} &=& 7.86 \ 10^{-34} \  g \ cm^{-3} \ . \end{eqnarray*}

The ratio of the total matter, 
  Dark Matter   and baryon energy densities to the critical density are denoted respectively
$\Omega_{m}$,  $\Omega_{dm}$ and $\Omega_{b}$~:
 \begin{eqnarray*} 
\nonumber 
\rho_{0m} &=&  \Omega_{m} \rho_{0c}
\cr
\rho_{0dm} &=&  \Omega_{dm} \rho_{0c} 
\cr
\rho_{0b} &=&  \Omega_{b} \rho_{0c} 
\ .
\end{eqnarray*}

\subsubsection{Number densities.}

The   Dark Matter  , photon, neutrino (the latter in our "reference" cosmology), and baryon number densities are respectively equal to~:
\begin{eqnarray*}
\nonumber 
n_{0\gamma} &=& 413 \  cm^{-3} \cr
n_{0\nu \st } &=& 338 \  cm^{-3}  \cr
n_{0dm}
&=& 1.29 \ 10^{-3} cm^{-3} \Odm \left(\frac{m_{dm}}{1MeV}\right)^{-1} \cr n_{0b} &=& 2.75 \ 10^{-7} cm^{-3} \Ob \ .
 \end{eqnarray*} 
The latter value  has been obtained assuming an average mass per baryon of $938 MeV$, that is the proton mass. It will also be used as the comoving number of baryons since we neglect  in this paper the slight average mass offset, upwards before nucleosynthesis, and downwards after.  

\subsection{Evolution.}

\subsubsection{Times. \label{app:times}}

We write the scale factor as 
\[
a \ = \ \left( \frac{t}{t_{\alpha }}\right) ^{\alpha } \ ,
\]
This expression holds even if $\Omega$ is not unity, and/or $\Lambda$ non-zero provided the scale factor $a$ is not too close to unity, say for $a < \frac{1}{5}$, if $\Omega$ is not much smaller than $0.2$, and $\Lambda$ not much larger than $1$. The parameter $\alpha$ is equal to $1/2$ in the radiation dominated era 
and equal to $2/3$ in the matter dominated era. 
The normalization ${t_{\alpha }}$ may be obtained 
by matching the above expression to the Friedman equations. 
In the radiation dominated era, $t_{\alpha}$ is equal to 
\begin{eqnarray}
\nonumber 
t_{r}\equiv t_{1/2} & = &
 \left[ \frac{16\, \pi \, G}{3}\, \frac{g_{* }(T)}{\kappa ^{4}(T)}\, \rho _{\gamma }(T_{0}) \right]^{-\frac{1}{2}} \nonumber \\  & = & 2.39 \, 10^{19} \, s \,  \, \g_* ^{-\frac{1}{2}}(T) \nonumber \\  & = & 2.32 \, 10^{5}\, Mpc/c \,  \, \g_* ^{-\frac{1}{2}}(T) \nonumber \, 
\end{eqnarray}
where (see Sect.\ref{sec:eden} for more detail) \(\g_* = g_*/3.36\). In the matter dominated era, $t_{\alpha}$  is equal to 
 \begin{eqnarray} t_{m}\equiv t_{2/3} & = & \left[ 6\, \pi \, G\, \rho _{m}(T_{0}) \right] ^{-\frac{1}{2}} \nonumber \\  & = & 5.37 \, 10^{17}s\, \left( \Om \right)^{-\frac{1}{2}} \nonumber \\  & = & 5.21 \ 10^3 \, Mpc/c \,  \left( \Om \right)^{-\frac{1}{2}}\nonumber \ . \end{eqnarray}

At the epoch of equality of matter and radiation, the expansion factor is not a power-law and the above approximations break down. They have been established so as to be exact far from this point. For this reason, it is slightly inaccurate to define the time of equality by equating the two expressions of the expansion parameter one obtains for $\alpha = 1/2$ and $\alpha = 2/3$. A better way of defining the epoch of equality in terms of the expansion factor $ a_{eq}$ will be given in the following section, equating the associated Hubble rates which are more directly connected with the energy densities of the various species. 
With the definition of $ a_{eq}$ given in Section \ref{app:Hrate}, we have  \[ t_m \, \propto \, t_r a_{eq}^{\frac{1}{2}}  \ . \]

\subsubsection{Hubble rates. \label{app:Hrate} }

We introduce the constant
\[
H_{\alpha }=\frac{\alpha }{t_{\alpha }} \ .
 \]
The latter is in the radiation dominated era equal to 
\[
H_{r} \equiv H_{1/2} = 2.10 \ 10^{-20} s^{-1} {\g_*^{\frac{1}{2}}(T)} 
 \]
and in the matter dominated era equal to 
\[ 
H_{m} \equiv H_{2/3 } = 1.24 \ 10^{-18} s^{-1}  \left( \Om \right)^{\frac{1}{2}}\ 
\]
This allows us to write the relation between the expansion factor and the Hubble parameter at a given epoch as 
\[ 
a = \left(\frac{H}{H_\alpha }\right)^{-\alpha} \ . 
\]
It will also be convenient, for further use, to define a "reduced" Hubble rate \[ 
\widetilde{H} = Ha^3 \ .
\]
Hence we have also
\[ 
a = \left(\frac{\widetilde H}{H_\alpha }\right)^{\frac{\alpha}{3\alpha-1}} \ . \] The epoch of matter-radiation equality will be defined by equating the two expressions of $a$ obtained for $\alpha ={1}/{2}$ and $\alpha = {2}/{3}$, so that the Hubble rate, which is proportional via the Friedman equations to the energy-densities, is continuous at the transition, reflecting the continuity of the energy densities. We then get at the epoch of equality \[ H = H_{eq} \equiv \frac{H_m^4}{H_r^3} = 2.59 \,10^{-13} \,Ês^{-1} \, {\g_*^{-\frac{3}{2}}(T_{eq})}  \left( \Om \right)^2 \] and \[ \widetilde{H} = \widetilde{H}_{eq} \equiv \frac{H_r^3}{H_m^2} = 5.96 \, 10^{-24} \,s^{-1}
\, {\g_*^{\frac{3}{2}}(T_{eq})}   \left( \Om \right)^{-1}  .
\]

This implies
\[
a_{eq} = \left(\frac{H_r}{H_m}\right)^2
\]
or, equivalently
\[
H_r = H_m a_{eq}^\frac{1}{2} \ .
\]

\subsubsection{Temperatures. \label{app:temp}}

The temperature of a given species $i$ with a thermal distribution   
\footnote{Even if the distribution is non-thermal, one can define by 
averaging over the actual distribution an effective temperature 
$T_i = <\frac{p^2}{3e}>$, and accordingly a parameter $\kappa_i(T)$, which 
for most if not all applications in the present paper, play the same role.}   
will be denoted $T_i$. 
We omit the index $i$ when dealing with photons. 
For each species $i$, we may define a parameter 
$\kappa_i(T) $ which relates
the temperature $T_i$ of the species $i$ to the scale-factor 
$a$~: 
\[ 
T_i = \frac{T_0}{\kappa_i(T) \ a} \ .
\]
This parameter $\kappa_i$ is constant in most cases.  Its value at a 
given time depends on the history of the system. It can  
be determined by using the comoving 
entropy conservation when the latter is indeed conserved.
As is well-known, the term $\kappa_i$ changes if, for instance,  the fluid to which  $i$ belongs contains another species which is under way of annihilating. 
 We use
$\kappa_{\gamma}(T) \equiv \kappa(T)$. 
The latter is equal to unity nowadays~: 
$\kappa(T_0) =1$, while $\kappa_{\nu}(T_0)$, 
which is associated with the present neutrino 
temperature, is, in case our "reference" cosmology is adopted, equal to $(11/4)^\frac{1}{3}$.

\subsubsection{Energy densities. \label{sec:eden}}

The energy density of a relativistic species is given by~: 
\[ \nn \rho_i = \stat \frac{g_i}{2}  a_{\SB} T_i^4 \ ,\]
with $a_{\SB}$ is the Stephan-Boltzmann constant, $g_i$ the 
spin-degeneracy factor and $\stat$ a factor 
equal to $7/8$ or $ 1$ for, respectively, Fermi or Bose-Einstein statistics  
of relativistic particles with negligible chemical potential.
\footnote {The factor $ \stat $ may differ from $7/8$ or unity 
if we consider, for instance, relativistic particles with 
non-zero chemical potential (as may be the case if they are chemically 
decoupled but still in thermal equilibrium).}

It turns out to be convenient to write the previous relation as 
\[ 
 \rho_i = \frac{g_{*i}}{2} a_{\SB} T^4 \ , 
\]
where 
\[ 
 g_{*i} = \stat g_i \frac{\kappa^4}{\kappa_i^4} \ . 
\]

We also define the quantity 
\[ 
 \g_{*i}(T) =  \frac{g_{*i}(T)}{3.36 \ \kappa^4(T)} \ , 
\]
which is more useful than $g_{*i}$ for the calculations of the 
damping scales.  This yields
\[
 \rho_i =  \g_{*i}(T)  \frac{\rho_{0(\gamma+\nu \st )}}{a^4}
\ .
\]
The parameter $ \g_{*i}$ is, under normal circumstances, of the order of unity. 
We have, at the present epoch for our "reference" cosmology, $\g_{*\nu \st } = 0.41$  and $ \g_{*\gamma \st } = 0.59$. 

We also have, summing over all relativistic species in the universe \[
 \g_*(T) = \sum_i \g_{*i}(T) = \frac{g_{* }(T)}{3.36 \ \kappa ^{4}(T)} \ , \] so as to get \[  \rho_r=  \g_{*}(T)  \frac{\rho_{0(\gamma+\nu \st )}}{a^4} \ . \] In our "reference" cosmology, this factor is unity after the electron annihilation. At early times, $  \g_*(T)  $ behaves  as $1.9 \ g_{* }^{-1/3}(T)$.

Finally, we define
\[
\dslash \rho_i = \rho_i + p_i
\]
where $p_i$ is the pressure of species $i$. For relativistic matter, $\dslash \rho_i \sim \frac{4}{3} \rho_i $ while for non-relativistic matter $\dslash \rho_i \sim \rho_i $.

\subsubsection{Number density \label{sec:numberd}} 

We use throughout this paper comoving number densities, namely~:   
\[ 
\widetilde n \ = \ n \ a^3 \, .
\]
The latter appear more convenient because they are constant as a function of 
time when the particle number is conserved, 
so we keep only the dependence due to the microphysics,
the cosmological dependence being removed.
In particular, the comoving number density 
of protons or electrons before or after their non relativistic 
transition, are respectively
\begin{eqnarray*}
\widetilde n_p(T_b > m_p/3) &=& \widetilde n_e(T_e > m_e/3) \ ,    \\
\cr
\widetilde n_b(T_b < m_p/3) \sim \widetilde n_p(T_b < m_p/3) &=& \widetilde n_e(T_e < m_e/3) 
 \ .  
\end{eqnarray*}
For the electron density, we neglect the slight offset due to the presence of Helium 
(that is neutrons) after the epoch of nucleosynthesis. Clearly also, the electron density is relative to free electrons only before recombination.

For a relativistic species $i$, including possibly the   Dark Matter   itself, 
it will turn out to be more convenient to use (e.g. in Appendix \ref{app:Gamsym}, to relate various interaction rates) the quantity \[ 
 \nb_i(T) \ = \ 
\frac{ \dslash \rho_i a^4}{3T_0} \ = \ 
 625 \ cm^{-3}  \ \frac{4\g_{*i}(T)}{3}
\ ,
\]
rather than the comoving number density $\widetilde n_i$ of 
species $i$, to which however $\nb_i$  is close.

\subsubsection{Reduced interaction rates. \label{app:redrates}}

As for the comoving number density, to remove the most obvious cosmological 
dependence of the interaction rates $\Gamma$ and leave only the 
temperature dependence related to the microphysics, we introduce  ``reduced'' interaction rates~: 
\[ 
  \widetilde{\Gamma }=\Gamma \, a^{3} \ . 
\]

We define the time $t_{dec(dm-i)}$ and the corresponding scale-factor 
$a_{dec(dm-i)}$, related to the epoch where  
  Dark Matter   particles decouple from the species $i$. It is, by convention,  taken to 
be the epoch where 
the interaction rate of the   Dark Matter   
with the species $i$, namely $\Gamma_{dm-i}$ is equal to the Hubble rate~: 
\[ 
 {\Gamma }_{dm-i} = H \  .
\]
The value 
of the reduced interaction rate $\widetilde{\Gamma }_{dm-i}$ at  
the   Dark Matter   decoupling from the species $i$ will be denoted 
it $\widetilde \Gamma _{dec(dm-i)}$. It corresponds to
\[ 
\widetilde \Gamma _{dec(dm-i)} \ = \ \Gamma _{dm-i} \ a^3 \, \vert_{a=a_{dec(dm-i)}}  \ .. \]

We also need 
the total interaction rate of a species $i$, which includes
the interactions with all species $j$ (including $i$) \[\widetilde \Gamma _{i} \ = \ \sum_j \widetilde \Gamma _{i-j} \ .\] 
  The quantity $\widetilde \Gamma _{dec(i)}$ will 
then denote the value of $\widetilde \Gamma _{i}$ at the 
time $t_{dec(i)}$ and  scale-factor $a_{dec(i)}$ where species $i$ 
totally decouples (even with itself). 

For generality, we denote by $\widetilde \Gamma _{dec(x)}$ and $\Gamma _{dec(x)}$ the reduced and normal interaction rate associated with $x$, where $x$ stands either for 
a particular coupling $i-j$ between the species $i$ and $j$ or simply for the 
full coupling of a species $i$ to the medium. The relation between $\widetilde \Gamma _{dec(x)}$ and 
$\Gamma _{dec(x)}$ may be written~: 
\[
\Gamma _{dec(x)} = H_{\alpha }
 \left( \frac{\widetilde{\Gamma }_{dec(x)}}{H_{\alpha }}\right) 
 ^{-\frac{1 }{3\alpha-1 }} \ .
\]
This relation takes two different forms~: 
\begin{center}
\[
\begin{array}{ll} 
\mbox{{Radiation dominated era}~}:\\
\Gamma _{dec(x)} = H_{eq}
\left( \frac{\widetilde{\Gamma }_{dec(x)}}{\widetilde H_{eq}}\right) ^{-2} \ 
 \ , \nonumber\\
\nonumber \\
\mbox{{Matter dominated era}~}:\\  
\Gamma _{dec(x)} = H_{eq} \, \left( \frac{\widetilde{\Gamma }_{dec(x)}}{\widetilde H_{eq}}\right) ^{-1}\,  \ .\\ \end{array} \] \end{center}

\subsection{Typical scales.}

\subsubsection{Interaction rates. \label{app:typrates}}

The neutrino-electron and the 
photon-electron interaction rate play a central role in the determination of the damping induced by respectively the neutrinos and the photons.  

Taking for the neutrino-electron cross-section (averaged over particles and antiparticles as well as over the three flavors) 
\[
\sigma _{\nu -e} c = 4.02 \ 10^{-34} \mbox{cm}^3 \mbox{s}^{-1} \ 
\frac{E_\nu E_e}{(1 MeV)^2} 
\]
we get, after statistical average
\[
\widetilde{\Gamma }_{\nu-e } \ = \ \langle\sigma_{\nu-e} \, v\rangle  \, \widetilde n_e \ , \] where $\widetilde n_e$ is the total (relativistic) electron density including electrons and antielectrons. This rate varies as $\propto T^2$ because of the energy-dependence of the cross-section. At $\nu-e$ decoupling (given by $\Gamma _{\nu-e } = H$), 
this rate takes the value $\widetilde{\Gamma }_{dec(\nu-e) } $. 

For non-relativistic electrons, up to recombination, the $\gamma-e$ 
interaction rate is a constant as a function of time~:
\[
\widetilde\Gamma _{\gamma-e } = \widetilde\Gamma _{\Th }
= \sigma_{\Th} \, c \, \widetilde n_e \ ,
\]
where  
$\sigma_{\Th} = 
\frac{8\pi \alpha_{\rm \scriptscriptstyle EM}^2}{3 m_e^2}$ is the Thomson cross-section.

 Numerical values of these rates are given in table \ref{tab:tyr}.

\subsubsection{Scale-factors (Table \ref{tab:a}). \label{app:typscafac} }

Several typical scale-factors may be defined. They  are 
associated with the  
time-scales used for the calculation of the damping effects. 

There first is $a_{nr}$, the epoch at which the 
  Dark Matter   becomes non-relativistic, taken by convention to correspond to
$ T_{dm} = m_{dm}/3$.

We are led to define the scale-factor 
$a_{eq(\gamma+\nu \st )}$ which characterizes the epoch of equality between  
matter and photon+neutrino energy-densities
in our "reference" cosmology and $a_{eq}$ the true epoch 
of matter-radiation equality in the general case, related to $H_r$ and $H_m$ already defined in Section \ref{app:Hrate}.

In order to get simpler expressions for the borderlines in the induced-damping case, we define also \[a_{eq(dm-i)} = {\dslash \rho_i a^4}/{\rho_{0dm}} \ , \] the ratio of the  energy-density 
of a relativistic species $i$ extrapolated to the present-day
to the present-day energy-density of the   Dark Matter  . 
If the   Dark Matter   is non-relativistic at $a_{eq(dm-i)}$, we have
$\dslash \rho_i/\dslash \rho_{dm} = a_{eq(dm-i)}/a$ . So, $a_{eq(dm-i)}$ can be loosely interpreted -the true definition being in all cases the relation above- as the epoch of equality
 between the energy-densities of the   Dark Matter   and species $i$. Its true interpretation is that it provides a constant value which measures the ratio of the energy density of a relativistic species to the present-day nonrelativistic   Dark Matter   energy-density.
At times where the   Dark Matter   is relativistic, we can 
define the ratio
\[a_{eq(dm)} = {\dslash \rho_{dm} a^4}/{\rho_{0dm}} \  , \]  which we denote just by analogy with $a_{eq(dm-i)} $ as
$a_{eq(dm)}$. It is the ratio of the relativistic   Dark Matter   energy-density 
extrapolated to the  present epoch to the actual (non-relativistic) energy-density.  While, again, the true definition is the relation above, when evaluated at $a=a_{nr}$, we see this quantity to be closely related to the offset of the energy-density extrapolated from the relativistic side and the energy-density extrapolation fron the non-relativistic side, that is to the annihilation factor at this epoch. It is introduced to provide a measure of this annihilation factor, and is of interest only in the NRFO scenario.

The epoch of decoupling of    Dark Matter   with species $i$ is denoted
$a_{dec(dm-i)}$. We have especially
$a_{dec(dm-\nu)}$ and $a_{dec(dm-\gamma)}$, the epoch at which 
   Dark Matter   decouples from neutrinos and photons respectively.
The epoch at which    Dark Matter   decouples from all species including itself
corresponds to  $a_{dec(dm)}$.
We introduce in addition the scale-factor which defines the 
epoch of decoupling of neutrinos with electrons, namely  
$a_{dec(\nu-e)}$. 

For completeness, we recall that
$a_{rec}$
is the scale-factor corresponding to recombination. Needless to say, its value is taken from the recent WMAP results. The epoch of the non-linear collapse, corresponding to $a_{nl}$,  
will also be needed for strongly interacting   Dark Matter   particles.

The analytical and numerical expressions of all of these scale-factors are 
summarized in Table \ref{tab:a}. 
The   Dark Matter   properties typically enter these expression by means
 of the   Dark Matter    mass 
$m_{dm}$ and interaction rate $\widetilde \Gamma_{dec(dm-i)}$ or 
$\widetilde \Gamma_{dec(dm)}$.

\begin{center}
\begin{table}[h]
\caption{Specific scale-factors \label{tab:a}.}
\[
\begin{array}{ll} \hline 
\hline\\
a_{nr}  & = \frac{3T_{0}}{ m_{dm}\kappa _{dm}(T_{nr})} \\ 
        & = 7.06 \, 10^{-10}\, \left(\frac{m_{dm}\, \kappa _{dm}(T_{nr})}{1 MeV}\right)^{-1}\\ & \\ \hline &\\
a_{eq(\gamma +\nu )}  & = \frac{\rho _{\gamma +\nu }(T_{0})}{\rho _{m}(T_{0})} \\
                      & = 2.85 \, 10^{-4}\, \left(\Om \right)^{-1}\  \\ &\\ \hline &\\
a_{eq}  &= H_r^2/H_m^2 \\
       &= \  \g_*(T_{eq})  a_{eq(\gamma +\nu)} 
\\ 
& \\ \hline &\\
& \mbox{(species {\it i} relativistic)} 
\\  \\
a_{eq(dm-i)}(T) &= \
\frac{\dslash \rho_i a^4}{\rho_{0dm}}  
 \\ 
                   &= \  \frac{4\g_{*i}(T)}{3} \ \frac{\Omega_m}{\Omega_{dm}} a_{eq(\gamma +\nu \st )} 
\\
                   & = \ 3.41 \ 10^{-4}  \  \frac{4\g_{*i}(T)}{3} \  \left(\Odm \right)^{-1} \\ & \\ \hline &\\
{\rm NRFO \   scenario:} \\
(\dslash \rho_{dm}\ \ \  \mbox{evaluated}&\mbox{at an epoch where it is relativistic})\\ 
a_{eq(dm)}(T) &= \ \frac{\dslash \rho_{dm} a^4}{\rho_{0dm}}  
 \\ 
                   &= \  \frac{4\g_{*dm}(T)}{3} \ \frac{\Omega_m}{\Omega_{dm}} a_{eq(\gamma +\nu \st )} 
\\
                   & = \ 3.41 \ 10^{-4}  \  \frac{4\g_{*dm}(T)}{3} \  \left(\Odm \right)^{-1} \\ & \\ \hline &\\
a_{dec(dm-i)} & = \left( \frac{\widetilde{\Gamma }_{{dec(dm-i})}}{H_{\alpha }}\right)^{\frac{\alpha }{3\alpha-1 }}\\
 & =  2.85 \, 10^{-4}\, 
\left( \frac{\widetilde{\Gamma }_{{dec(dm-i)}}}{5.96  \ 10^{-24}s^{-1}}\right)  \, \g_*^{-\frac{1}{2}}(T_{dec(dm)}) \\ 
&\hspace{1cm} \mbox{in the radiation era}\\
 & = 2.85 \, 10^{-4}\, \left( \frac{\widetilde{\Gamma }_{{dec(dm-i)}}}{5.96  \ 10^{-24}s^{-1}}\right) ^{2/3}\, 
\left(\Om \right)^{-\frac{1}{3}} \\ 
&\hspace{1cm} \mbox{in the matter era} \\
&\\ \hline &\\
a_{dec(\nu-e)}  &= \ \frac{\widetilde{\Gamma }_{dec(\nu-e)}}{H_r} \ \\ &= \, 1.33 \ 10^{-10}  \ {\g_*^{-\frac{1}{6}}} { \kappa^{-\frac{2}{3}}} \ \\ &\\ \hline &\\
a_{rec} &=1/1090 \\ & \\
&\\ \hline &\\
a_{nl} &\sim 1/10 
\, \, \, \, \, \, \, \, \, \, \, \, \, \, \, \mbox{(but left as a free parameter)} \\ & \\ \hline 
\hline
\end{array}
\]
\end{table}
\end{center}

\subsubsection{Typical Hubble rates (Table \ref{tab:tyr}). \label{app:typHrates}}

One can translate the above scale-factors into reduced Hubble rates $\widetilde{H}$ in order to define typical rates. 
The relation between $H$ and the scale-factor $a$ may be found 
in Sect.\ref{app:Hrate}. By convention, we define $\widetilde H_{eq(dm-i)} $ as $\widetilde H_{eq(dm-i)}  = a_{eq(dm-i)} H_r $. Together with the rates given in Section \ref{app:typrates}, 
this sets the typical scales, displayed in Table \ref{tab:tyr}, to which the Hubble rate $\widetilde{H}$ and the reduced interaction rates 
$\widetilde{\Gamma}$ of Sect. \ref{app:redrates} are to be compared.

\begin{center}
\begin{table}[h]
\caption{Specific reduced rates \label{tab:tyr}.}
\[
\begin{array}{ll} \hline 
\hline\\
\widetilde H_{nr} & = \, 1.48 \ 10^{-29} s^{-1} 
\g_* ^\frac{1}{2} \left(\frac{m_{dm} \kappa_{dm}}{1MeV}\right)^{-1}  \\ &\hspace{1cm} \mbox{in the radiation era} \\ \widetilde H_{nr} & = \, 2.33 \ 10^{-32} s^{-1} 
\left(\Om\right)^\frac{1}{2} \left(\frac{m_{dm} \kappa_{dm}}{1MeV}\right)^{-\frac{3}{2}}
\\
&\hspace{1cm} \mbox{in the matter era}  \\
\\
\widetilde H_{eq} &=\  5.96 \, 10^{-24} s^{-1}\, \g_* ^{\frac{3}{2}}(T_{eq}) \left(\Om \right)^{-1}  \ . \\ &\\ 
\widetilde H_{eq(dm-i)} 
\ &= \  7.16 \, 10^{-24} s^{-1}\, \frac{4\g_{*i}(T)}{3}\g_* ^{\frac{1}{2}} 
\  \left(\Odm \right)^{-1} 
 \\
&\\
{\widetilde{\Gamma } }_{dec(\nu-e)}
\  &= \, 2.80 \  \ 10^{-30}  s^{-1} \ 
{\g_*^{\frac{1}{3}}} { \kappa^{-\frac{2}{3}}}
 \ , \\
&\\
\widetilde H_{rec } & = 3.45 \ 10^{-23} s^{-1}
\ \left(\Om \right)^\frac{1}{2}  \ , 
\\
&\\
\widetilde{\Gamma }_{\Th}
\  &= \, \sigma_{\Th} c \widetilde n_e =
5.49 \ 10^{-21} \ s^{-1} \  \Ob \ , \\
&\\
\widetilde H_{nl} \ & = \ 
3.93 \ 10^{-20} s^{-1} \left(\frac{a_{nl}}{0.1}\right)^{\frac{3}{2}}  
\left(\Om \right)^{\frac{1}{2}} \,  \\
&\\
\hline 
\hline
\end{array}
\]
\end{table}
\end{center}

\subsubsection{Cross-section scales
(Table \ref{tab:xs}).
 \label{app:typxs}}

Our results may also be displayed in terms of the   Dark Matter     
$\svb$ cross-sections, more directly connected to 
particle physics.  
The latter are defined in terms of the interaction rate by~: 
\[
\widetilde{\Gamma }_{ij} = \svb_{ij} \ \widetilde n_{j} \ .
\]
 The values which set the scales of the $\svb$ cross-sections relevant 
to the limits we seek in this work (see \textit{e.g.} the relations given in Table \ref{tab:irxs} which are used in Appendix \ref{app:Gamsym}) turn out to be for neutrinos \[ \snu = \frac{ \widetilde{\Gamma }_{dec(\nu-e)}}{ \nb_\nu} \] while for photons we are led to use the values of the $\svb$ cross-sections related to the epoch of recombination 
\[ \srec= \frac{\widetilde H_{rec }}{\nb_\gamma} \,  ,  \]
or values of $\svb$ related to the Thomson cross-section
\[\sth= \sigma_\Th c \  . \]
Finally, we will also be led to use, for any relativistic species $i$, 
\[ \seq = \frac{\widetilde H_{eq(dm-i)}}{ \nb_i} \ . \]
While the definition of $\seq$ is always the above relation, it may be (somewhat loosely) interpreted as the $\svb$ cross-section needed for the   Dark Matter   to decouple from species $i$ just at the epoch of equality of the energy-densities $\dslash \rho_i $ and $\rho_{0dm}/a^3$.
Note that $\seq$, which may be readily evaluated as $\seq = \frac   {H_r }   {{\rho_{0dm}}/{3T_0 }}$ by means 
of the expressions of $\widetilde H_{eq(dm-i)}$ (Table \ref{tab:tyr}) and $ \nb_i $ (Section \ref{sec:numberd}), turns out to be independent of the relativistic species $i$ considered.

These typical cross-sections are given in Table \ref{tab:xs}.

\begin{center}
\begin{table}[h]
\caption{ Specific cross-sections \label{tab:xs}.}
\[
\begin{array}{ll} \hline 
\hline\\
\snu &= 4.48 \  \ 10^{-33}  cm^3 \ s^{-1} \ 
\g_*^{\frac{1}{3}} \left(\frac{4\g_{*\nu}(T)}{3}\right)^{-1}  { \kappa^{-\frac{2}{3}}} \ ,\\ \srec &= 5.53  \ 10^{-26} cm^3 s^{-1} \left(\frac{4\g_{*\gamma}(T)}{3}\right)^{-1} 
\ \left( \Om\right)^\frac{1}{2} \ , \\
\sth&= 1.99 \ 10^{-14} \ cm^3s^{-1}  \\
\seq&= 1.15 \ 10^{-26} \ cm^3s^{-1} \left(\Odm\right)^{-1} \\ &\\ \hline 
\hline
\end{array}
\]
\end{table}
\end{center}

\section{Density evolution and relations among interaction rates. \label{app:denssym} }

\subsection{Density evolution. \label{app:densevol}}

In case the   Dark Matter   is non relativistic at the epoch of reference, 
and if the number of particles is conserved till nowadays, one has \[ 
\widetilde n_{dm} = \frac{ \Omega_{dm} \ \rho_{0c}}{m_{dm}} \ . \] In the {NRFO} scenario, stritly speaking, this relation is valid only fr $ a > a_{fo} \sim 7 a_{nr}$.
In the {URFO} scenario, it is valid even for relativistic   Dark Matter  .

In the {NRFO} scenario, for relativistic   Dark Matter  , the number density may be written
\footnote{The exact relation differs by a numerical coefficient very close to unity.  We use the above approximation 
because it joins continuously to the non-relativistic expression at $a=a_{nr}$.
This turns out to be accurate enough for our purpose and yields substantially simpler practical calculations (it avoids in particular to distinguish fermionic and bosonic   Dark Matter  ).}
\[
\widetilde n_{dm} 
\  = \   \nb_{dm} \ \kappa_{dm}(T)
\ ,
\]
where $\nb$ is the density introduced in  Section \ref{sec:numberd}.
As expected, $\widetilde n_{dm}$ does not depend on the   Dark Matter   particles' mass.
It is independent of time except 
for possible variations 
of the factor  $\kappa_{dm}(T)$ or those due to the coefficient $g_{*dm}$ 
contained in $ \nb_{dm}$. 
The densities on the relativistic and the non-relativistic side in the {NRFO} scenario are related by \[ \widetilde n_{dm}(T<T_{nr}) \ = \ 
\nb_{dm} (T_{nr}) \frac {\kappa_{dm}(T_{nr}) a_{nr}}{a_{eq(dm)}} 
\ .
\]

Although the   Dark Matter   density does not depend on which species $i$ is present in the system, a quite useful 
relation relating the non-relativistic   Dark Matter   density to the  density $\nb_i$ (see Sect. \ref{sec:numberd}) of a relativistic species $i$ is 
\[
\widetilde n_{dm}(T<T_{nr}) \ = 
\nb_i \frac {\kappa_{dm}(T_{nr}) a_{nr}}{a_{eq(dm-i)}}
\ .
\]

\subsection{ Expression of the $\Gamma_{i-dm}$ rate and the $\svb_{i-dm}$ cross-section in terms of $\Gamma_{dm-i}$. \label{app:Gamsym}}

The relation (\ref{ratmom}) between the rates $\widetilde{\Gamma }_{i-dm}$ 
and $\widetilde{\Gamma }_{dm-i}$ is governed by the ratio of the energy densities 
$\dslash{\rho}_{dm}/\dslash{\rho}_i$ relative to these two species~: \[ \widetilde{\Gamma }_{i-dm} = \frac{\dslash{\rho}_{dm}}{\dslash{\rho}_i} \widetilde{\Gamma }_{dm-i} \ .  \]
The expressions of  $\widetilde{\Gamma }_{i-dm}$  
may be transformed into expressions of $\svb_{i-dm}$
by means of 
\[ \svb_{i-dm} \ = \ \frac{\widetilde{\Gamma }_{i-dm}}{\widetilde n_{dm} } \ .\] 
To study the correspondence between the interaction rates and cross-sections, 
we need to establish the evolution of the number density of   Dark Matter   
particles, that may or may not be the same before and after the 
non-relativistic transition.

We seek a relation between $ \widetilde{\Gamma }_{ i-dm} $ and $ \widetilde{\Gamma }_{ dm-i} $. In the special case it is  evaluated at an epoch $a$  such as the   Dark Matter   particle
number is conserved  afterwards, the latter can
be written in a more explicit form than the general relation (\ref{ratmom}). To this purpose, we  evaluate the ratio 
${\dslash \rho_{i}}/{\dslash \rho_{dm}}$, specifically 
under this assumption, for relativistic and non-relativistic 
  Dark Matter   particles.

\subsubsection{Relativistic species $i$.}

This is the case we explicitly need to consider in the present paper.

In case the   Dark Matter   particles are already non-relativistic, 
that is for $a > a_{nr} $, for both the {URFO} and the {NRFO} scenario, 
we have \( \frac{\dslash \rho_{dm}} {\dslash \rho_{i}}\ = \  \frac  {a}{a_{eq(dm-i)}} \).

If the   Dark Matter   particles are still relativistic, on the other hand, for
$a <  a_{nr} $, we get in the {URFO} scenario
\footnote{The exact relation differs by a numerical coefficient. The density $\widetilde n_{dm}$ in section \ref{app:densevol} (see the footnote there) is related to this expression of the energy-density by taking $3T_{dm}$ as the average energy of a particle independently of the statistics. This insures consistency with our choice to set the non-relativistic transition at 
$m_{dm}=3T_{dm}$.}
$\dslash \rho_{dm} \sim  \ \rho_{0dm}   a_{nr}/a^4$,
and thus
\( \frac{\dslash \rho_{dm}}{\dslash \rho_{i}} \ =  \  \ \frac{a_{nr}}{a_{eq(dm-i)}}  \).
 
 In the {NRFO} scenario, on the other hand, when $a$ correspond to an epoch where the   Dark Matter   still is relativistic, 
since the annihilation is still to come, the   Dark Matter   radiation density may be expected to be of the order of the density 
of any other relativistic species. More precisely, for $a <  a_{nr} $, one has  
\( \frac{\dslash \rho_{dm}}{\dslash \rho_{i}} = \frac{\g_{*dm}} { \g_{*i}} \). The ratio $ {\g_{*i}}/{\g_{*dm}}$, except in very exotic models,  
is expected (taking into account that we are in the {NRFO} scenario) to be of order unity.

These relation, which will be needed throughout the whole paper, and the inferred relation among interaction rates and cross-sections are recalled in Table \ref{tab:irxs}.

\begin{center}
\begin{table}[h]
\caption{Relation between $i-dm$ and $dm-i$ interaction rates and cross-sections for a relativistic species $i$. \label{tab:irxs}} \[ \begin{array}{lll} \hline 
\hline\\
 a > a_{nr} & &\mbox{URFO $\&$ NRFO} \\
 & &  \frac{\dslash \rho_{dm}} {\dslash \rho_{i}}\ = \  \frac {a} {a_{eq(dm-i)}} \\ \\  & &\widetilde{\Gamma }_{i-dm} = \ 
 \frac{a} {a_{eq(dm-i)}}  \ \widetilde{\Gamma }_{dm-i} \ . \\ 
& &\kappa_{dm} \svb_{i-dm  }  =  \frac{a} {a_{nr}}  
\frac{\widetilde{\Gamma }_{dm-i } }{ \nb_i} \\ 
& & \\ \hline \\
a < a_{nr}  & &\mbox{URFO} \\
& & \frac{\dslash \rho_{dm}}{\dslash \rho_{i}} \ =  \  \ \frac{a_{nr}} {a_{eq(dm-i)}} \\ \\ & &\widetilde{\Gamma }_{i-dm} \ = \  \frac{a_{nr}} {a_{eq(dm-i)}}  \ \widetilde{\Gamma }_{dm-i} \ .\\ & & \kappa_{dm} \svb_{i-dm  }  =  
\frac{\widetilde{\Gamma }_{dm-i } }{ \nb_i}\\
& & \\ \hline \\
a < a_{nr} & &\mbox{NRFO} \\
& & \frac{\dslash \rho_{dm}}{\dslash \rho_{i}} = \frac{\g_{*dm}} { \g_{*i}} \\ \\ & &\widetilde{\Gamma }_{i-dm} = 
\frac {\g_{*dm}}{\g_{*i}}  \ \widetilde{\Gamma }_{dm-i}
 \ .\\
& &\kappa_{dm} \svb_{i-dm  } 
 =  
 \frac{\widetilde \Gamma_{i-dm  }}{ \nb_{dm}}
 =  
 \frac{\widetilde \Gamma_{dm-i }}{ \nb_i}\\
& & \\ 
\hline 
\hline
\end{array}
\]
\end{table}
\end{center}

\subsubsection{Nonrelativistic species $i$.}

Similar expression may be established in this case. They will turn out not to be needed in the practical applications. We give them here (Table \ref{tab:inrxs}) for the sake of completeness: in the remainder of the paper, only the expressions for  relativistic  species $i$ will be given. To this purpose, it turns out to be convenient, rather than $a_{eq(dm-i)}$, which is a quantity adapted to  relativistic species $i$, to use another ratio suited to nonrelativistic species $i$~: \[f_{dm/i} = \frac{\rho_{0dm}}{ \rho_{i} a^3} \ . \] In case the number of particles $i$ is conserved, one has obviously $ f_{dm/i} = \frac{\Omega_{0dm}}{\Omega_{0i}}$.

\begin{center}
\begin{table}[h]
\caption{Relation between $i-dm$ and $dm-i$ interaction rates and cross-sections for a nonrelativistic species $i$. \label{tab:inrxs}} \[ \begin{array}{lll} \hline 
\hline\\
 & & \svb_{dm -i }  =    
 \frac{\widetilde{\Gamma }_{dm-i } }{ \widetilde n_{i} } \\
& & \\ \hline \\
 a > a_{nr} & &\mbox{URFO $\&$ NRFO} \\
 & &  \frac{\dslash \rho_{dm}} {\dslash \rho_{i}}\ = \  f_{dm/i} \\ \\  & &\widetilde{\Gamma }_{i-dm} = \ 
f_{dm/i} \ \widetilde{\Gamma }_{dm-i} \  \\ 
 & & \svb_{i-dm  }  =  
\frac{m_{dm} }{m_i} \ \frac{\widetilde{\Gamma }_{dm-i } }{ \widetilde n_{i}} =  
\frac{m_{dm} }{m_i} \  \svb_{dm -i }  
\\ 
& & \\ \hline \\
a < a_{nr}  & &\mbox{URFO} \\
& & \frac{\dslash \rho_{dm}}{\dslash \rho_{i}} \ =  \ f_{dm/i} \ \frac{a_{nr}} {a} \\ \\ & &\widetilde{\Gamma }_{i-dm} \ = \
 \ f_{dm/i} \ \frac{a_{nr}} {a}  \ \widetilde{\Gamma }_{dm-i} \   \\
 & & \svb_{i-dm  }  =  \frac{a_{nr}} {a}  \   \frac{m_{dm} }{m_i} \
\frac{\widetilde{\Gamma }_{dm-i } }{ \widetilde n_{i} }  = 
 \frac{a_{nr}} {a}  \  \frac{m_{dm} }{m_i} \  \svb_{dm -i }  
\\ 
& & \\ \hline \\
a < a_{nr} & &\mbox{NRFO} \\
& & \frac{\dslash \rho_{dm}}{\dslash \rho_{i}} =  \  f_{dm/i} \ \frac{a_{eq(dm)}}{a} \\ \\ & &\widetilde{\Gamma }_{i-dm} = 
\  f_{dm/i} \ \frac{a_{eq(dm)}}{a}  \ \widetilde{\Gamma }_{dm-i} \  \\
 & & \svb_{i-dm  }  =  \frac{a_{nr}} {a}  \   \frac{m_{dm} }{m_i} \
\frac{\widetilde{\Gamma }_{dm-i } }{ \widetilde n_{i} }    =  \frac{a_{nr}} {a}  \ 
\frac{m_{dm} }{m_i} \   \svb_{dm -i }  
\\ 
& & \\ 
\hline 
\hline
\end{array}
\]
\end{table}
\end{center}


\subsection{The non-relativistic transition 
in the NRFO scenario. \label{app:nrtran}}

In the {NRFO} scenario, there is a sudden decrease of the   Dark Matter   particle number when the latter becomes non-relativistic.
We approximate the transition around $a = a_{nr}$ 
(and more precisely between $a = a_{nr}$ and $a=a_{fo} \sim 7 a_{nr}$) by an abrupt transition at $a = a_{nr}$ where some physical quantities 
(such as for instance the density) switch abruptly from a ``relativistic'' 
value to a ``non-relativistic'' value.
This results in a great simplification in displaying our results, 
and is of no harm, 
provided the associated discontinuities 
at $a = a_{nr}$ are properly evaluated.

It is  readily seen from Section \ref{app:densevol} that in the {NRFO} scenario, 
 there is a drop of 
the density 
by a factor
\[
\frac{\widetilde n_{dm}( T_{nr})\vert_{nr}}{\widetilde n_{dm}(T_{nr})\vert_r}  \ = \ \frac{a_{nr}}{a_{eq(dm)}} \  . \] This ratio  represents the annihilation factor, due to the decay
of the   Dark Matter   particles between
$a_{nr}$ and $a_{fo}$, needed to insure cosmological consistency. To represent really annihilation it is obviously required to be smaller than unity. This makes sense only in Regions I, II, III ($ a_{eq(\gamma + \nu \st )} > a_{nr}$) since we expect $a_{eq(dm)} \sim a_{eq(\gamma + \nu \st )}$.

The ratio of
the non-relativistic value   $\dslash \rho_{dm}(a_{nr})\vert_{nr}$
to the relativistic value $\dslash \rho_{dm}(a_{nr})\vert_r$ at $a_{nr}$ is also, from our previous estimates (Section \ref{app:Gamsym}) of 
the ratio  $\dslash \rho_{i}/\dslash \rho_{dm}$
\[
\frac{\dslash \rho_{dm}(T_{nr})\vert_{nr}}{\dslash \rho_{dm}(T_{nr})\vert_r} \ 
 = \ 
\frac{ a_{nr}}{a_{eq(dm)}} \ .
\]
The question which has been treated in Section \ref{app:Gamsym} is, given the interaction rate $\widetilde \Gamma_{dm-i}$, taken at a fixed epoch, what is the 
expression of
the rate $\widetilde \Gamma_{i-dm}$ and the cross-section $\svb_{i-dm}$, taken at the same epoch. 
We give  explicitly
here the discontinuities of the latter around $a = a_{nr}$.

The rate $\widetilde \Gamma_{dm-i}$ is obviously continuous around $a_{nr}$. {}From the basic relation (\ref{ratmom}), we then readily see that \[ \frac {\widetilde \Gamma_{i-dm}( T_{nr})\vert_{nr}} {\widetilde \Gamma_{i-dm}( T_{nr})\vert_r} \ = \ \frac{ a_{nr}}{a_{eq(dm)}} \ . \] It bears the same discontinuity than the densities. {}From its definition, $\svb_{i-dm}=\widetilde \Gamma_{i-dm}/\widetilde n_{dm}$ is the ratio of two discontinuous quantities, but remains 
exactly continuous, as it should be, despite the approximations we have made. This is possible only thanks to the care 
we have 
taken in evaluating the approximate expressions of the energy-densities, and 
was by no means granted.

\section{Classification of the   Dark Matter   candidates. \label{app:bord}   }

As explained in Section \ref{sec:dampscales}, the calculation of the damping scales is separated into several contributions,  the self-damping $\&$ free-streaming, 
neutrino and photon induced-damping. We treat each of these effects 
separately in order to display the different constraints in a more appropriate way. 
The damping lengths turn out to depend on the   Dark Matter   particles' mass and on an appropriate interaction rate. Their values thus may be displayed in a two-parameter space, respectively
[$m_{dm},\widetilde\Gamma_{dec(dm)}$],  [$m_{dm},\widetilde\Gamma_{dec(dm-\nu)}$], 
[$m_{dm},\widetilde\Gamma_{dec(dm-\gamma)}$] for the self-damping $\&$ free-streaming, 
the neutrino induced-damping and photon induced-damping. 
The analytical expression of the damping takes different forms 
in the various "regions" of the relevant parameter space,
as is seen in Sections \ref{sec:regsdfs}, \ref{sec:regnu} and \ref{sec:regphot}, respectively. Here we give the borderlines of these regions.

\subsection{  Dark Matter   classification from 
self-damping $\&$ free-streaming. 
\label{app:bordsdfs} }

The borderlines of the regions defined in Section \ref{sec:regsdfs} 
within which the damping lengths take different forms
are given by the conditions
\begin{eqnarray}
\nonumber 
a_{eq(\gamma +\nu)}  \ &=& \ a_{nr} \ ,
\nonumber 
\\
 a_{dec(dm)} \ &=& \ a_{eq(\gamma +\nu )} \ ,
\nonumber 
\\ 
a_{dec(dm)} \ &=& \, a_{nr} \ \ . 
\nonumber 
\end{eqnarray}
They cross at one single point, 
and
define six regions, labelled from I to VI.

The discussion in terms of $a_{eq(\gamma +\nu )}$ is useful  because this scale-factor  
remains defined (and relevant) even in regions 
IV, V and VI (although, in this case, it bears no relation with the epoch 
of equality which occurs at $a = a_{nr}$ in these regions). 

The numerical values of the above scale-factors are given in 
Appendix \ref{app:typscafac}.
They imply the bordelines given explicitely in Table \ref{tab:sdbor}.

\begin{center}
\begin{table}[h]
\caption{Explicit borderlines for self-damping and free-streaming \label{tab:sdbor}} \[ \begin{array}{ll} \hline 
\hline\\
a_{eq(\gamma +\nu)}  \ &= \ a_{nr} \\
 \Leftrightarrow \ m_{dm}\, \kappa _{dm} \ &= \ 2.48 \, 10^{-6}MeV\, \Om \\ 
&\\ \hline
 a_{dec(dm)} \ &= \ a_{eq(\gamma +\nu )} \\ 
\Leftrightarrow  \ \widetilde{\Gamma }_{dec(dm)} \ &= \ 5.96 \, 10^{-24} s^{-1}\,
\g_*^{\frac{1}{2}}(T_{dec(dm)}) \ 
 \left(\Om\right)^{-1}\, \\
 &\hspace{1cm} \mbox{in the radiation era} \\
\Leftrightarrow  \ \widetilde{\Gamma }_{dec(dm)}  \ &= \  5.96 \, 10^{-24}s^{-1}\, 
\left(\Om\right)^{-1} \\ 
&\hspace{1cm} \mbox{in the matter era} \\
&\\ \hline 
a_{dec(dm)}  \ &= \ a_{nr} \\ 
\Leftrightarrow  \ \widetilde{\Gamma }_{dec(dm)} \ &= \,
1.48 \, 10^{-29}\, s^{-1}\, \g_*^{\frac{1}{2}}(T_{dec(dm)})\, 
\left(\frac{m_{dm} \, \kappa _{dm}}{1 MeV} \right)^{-1} \\ &\hspace{1cm} \mbox{in the radiation era} \\
\Leftrightarrow  \ \widetilde{\Gamma }_{dec(dm)} \ &= \ 2.33 \, 10^{-32}\, s^{-1}\, \left(\Om\right)^\frac{1}{2}\, 
\left(\frac{m_{dm} \, \kappa _{dm}}{1 MeV} \right)^{-\frac{3}{2}} \\ &\hspace{1cm} \mbox{in the matter era} \\ \ .\\ &\\ \hline \hline
\end{array}
\]
\end{table}
\end{center} 

\subsection{Borderlines of the  Regions in the neutrino  parameter space. \label{app:bordnu}}

The damping length due to neutrino induced-damping is to be evaluated at   Dark Matter-neutrino decoupling. All constraints on the interaction rates thus are obtained at this time.
{\it All relations given in this section hence are understood as being written at the epoch $a = a_{dec(dm-\nu)}$, the ``reference time'', corresponding to the decoupling of the   Dark Matter   with neutrinos.}

 Most important in this case are the interaction rates.
They determine whether $\nu-e$ or $\nu-dm$ scattering or simply neutrino free-streaming is responsible for the damping,  as discussed in 
Section \ref{sec:regnu}.
Depending on these rates, there are three regions in the 
[$ m_{dm}, \widetilde \Gamma_{dec(dm-\nu)}$] plane, namely  A, B and C in which the 
induced-damping scale results from different physical processes~: \begin{itemize} \item Region A~: $\Gamma_{\nu-e} \ge \Gamma_{\nu-dm}$, 
                 $\Gamma_{dm-\nu} \le \Gamma_{\nu} \sim \Gamma_{\nu-e}$, \\ \item Region B~: $\Gamma_{\nu-e} < \Gamma_{\nu-dm}$, 
                 $\Gamma_{dm-\nu} \le \Gamma_{\nu} \sim \Gamma_{\nu-dm}$, \\ \item Region C~: $\Gamma_{dm-\nu} > \Gamma_{\nu}$ . \end{itemize}

\begin{center}
\begin{table}[h]
\caption{Borderlines for  $dm-\nu$ induced-damping. \label{tab:nubor}}
\[ \begin{array}{ll} \hline 
\hline\\
 & \ \ \ \ \ \ \ \ \ \     \mbox{A-B borderline} \\
&
\widetilde{\Gamma }_{dm -\nu}   \le  \widetilde{\Gamma }_{\nu-dm} 
\ \ \  \ \ \  \ \ \  \ \ \  \ \ \  \ \ \  
\widetilde{\Gamma }_{\nu-dm}   =  \widetilde{\Gamma }_{\nu-e} 
\\
\\
&  \ \ \ \ \ \ \ \ \ \  \mbox{A-C borderline} \\
& 
\widetilde{\Gamma }_{dm -\nu}   \le  \widetilde{\Gamma }_{\nu-dm} 
  \ \ \  \ \ \  \ \ \  \ \ \  \ \ \  \ \ \  
 \widetilde{\Gamma }_{dm-\nu}   =  \widetilde{\Gamma }_{\nu-e}
 \\
\\
&  \ \ \ \ \ \ \ \ \ \  \mbox{B-C borderline} \\
& 
\widetilde{\Gamma }_{\nu-e }   <  \widetilde{\Gamma }_{\nu-dm } 
 \ \ \  \ \ \  \ \ \  \ \ \  \ \ \  \ \ \  \ \ 
\widetilde{\Gamma }_{dm-\nu}   =  \widetilde{\Gamma }_{\nu-dm}
\\
\mbox{and (NRFO):}
& 
\widetilde{\Gamma }_{\nu-e } < \widetilde{\Gamma }_{dm-\nu} < \widetilde H_{eq(dm-\nu)} \ \ \ 
 \widetilde{\Gamma }_{dm-\nu}   =  \widetilde{H}_{nr}
\\
\hline
\hline
\end{array}
\]
\end{table}
\end{center}

The borderlines A-B, A-C, and B-C which separate the three regions (Table \ref{tab:nubor}) may be directly inferred from these definitions.

It is however of interest to  express these relations 
in terms of $m_{dm}$ and 
$\widetilde \Gamma_{dec(dm-\nu)}$.
They may also be translated also in terms of 
 $m_{dm}$ and the cross-sections $\svb_{\nu-dm }$.
We will do this in the following two subsections (separating the URFO and the NRFO scenario) by using the relation between $\widetilde{\Gamma }_{\nu-dm }$ and $\widetilde{\Gamma }_{dm-\nu }$ established in Appendix \ref{app:Gamsym}.

\subsubsection{ {URFO} $dm-\nu$ scenario (Table \ref{tab:nuure}). }

\begin{center}
\begin{table}[h]
\caption{Explicit borderlines in the {URFO} $dm-\nu$ scenario. \label{tab:nuure}} 
\[ \begin{array}{ll} \hline 
\hline\\
\mbox{A-B} &
 \widetilde{\Gamma }_{dm -\nu}   \le  \widetilde{\Gamma }_{\nu-dm} 
 \ \ \   \ \ \  
\widetilde{\Gamma }_{\nu-dm}   =  \widetilde{\Gamma }_{\nu-e}
  \\ \\
  & a_{eq(dm-\nu)} \le a_{nr} \\
 &\widetilde{\Gamma }_{dm-\nu } = 
\left(\frac{a_{eq(dm-\nu)} }{a_{nr}} \right)^\frac{1}{3} 
\widetilde{\Gamma }_{dec(\nu-e) }  
\\
 & \widetilde{\Gamma }_{\nu-dm } =
\left(\frac{a_{eq(dm-\nu)} }{a_{nr}} \right)^{-\frac{2}{3}} \widetilde{\Gamma }_{dec(\nu-e) } 
\\
&\kappa_{dm} \svb_{\nu-dm } = 
\left(\frac{a_{eq(dm-\nu)} }{a_{nr}} \right)^\frac{1}{3} \snu  \\ &\\\hline &\\
\mbox{A-C}   &  \widetilde{\Gamma }_{dm-\nu}   >  \widetilde{\Gamma }_{\nu-dm}
 \ \ \   \ \ \  
\widetilde{\Gamma }_{dm-\nu}   =  \widetilde{\Gamma }_{\nu-e}
\\ \\
& a_{eq(dm-\nu)} > a_{nr} \\
&\widetilde{\Gamma }_{dm-\nu} = \widetilde{\Gamma }_{dec(\nu-e) } \\ &\\
 {a_{dec(dm-\nu)}} &> a_{nr}    \\ 
& \widetilde{\Gamma }_{\nu-dm} = 
\frac{a_{dec(\nu-e)}}{a_{eq(dm-\nu)}} \widetilde{\Gamma }_{dec(\nu-e) } \\ &\kappa_{dm} \svb_{\nu-dm} = \frac{a_{dec(\nu-e)}}{a_{nr}} \snu \\ \\
{a_{dec(dm-\nu)}} &< a_{nr}    \\ 
& \widetilde{\Gamma }_{\nu-dm} =
\frac{a_{nr}}{a_{eq(dm-\nu)} } \widetilde{\Gamma }_{dec(\nu-e) } \  \\ & \kappa_{dm} \svb_{\nu-dm} =  \snu 
\\
&\\ \hline &\\
\mbox{B-C} &
 \widetilde{\Gamma }_{\nu-e }   <  \widetilde{\Gamma }_{\nu-dm} 
 \ \ \   \ \ \  
\widetilde{\Gamma }_{dm-\nu}   =  \widetilde{\Gamma }_{\nu-dm}
\\ \\
& a_{dec(\nu-e)} < a_{dec(dm-\nu)} \\ 
\\
a_{dec(dm-\nu)} &> a_{nr} \\
&\widetilde{\Gamma }_{dm-\nu} = \widetilde H_{eq(dm-\nu)} \\ 
&\widetilde{\Gamma }_{\nu-dm} = \widetilde H_{eq(dm-\nu)} \\ &\kappa_{dm} \svb_{\nu-dm} = \frac{a_{eq(dm-\nu)}}{a_{nr}} \seq \\ \\ a_{dec(dm-\nu)} &< a_{nr} \\ & a_{eq(dm-\nu)} = a_{nr} 
\\
&\\
\hline
\hline
\end{array}
\]
\end{table}
\end{center}

\subsubsection{ {NRFO} $dm-\nu$ scenario (Table \ref{tab:nunre}). }

\begin{center}
\begin{table}[h]
\caption{Explicit borderlines for the NRFO $dm-\nu$ scenario (Excluding the case $a_{nr} > a_{eq(dm-\nu)}$, as discussed in the text). \label{tab:nunre}}

\[
\begin{array}{ll} \hline 
\hline\\
\mbox{A-B} & \hspace{0.5 cm}
 \widetilde{\Gamma }_{dm -\nu}   \le  \widetilde{\Gamma }_{\nu-dm} 
 \ \ \  \ \ \, 
 \widetilde{\Gamma }_{\nu-dm}   =  \widetilde{\Gamma }_{\nu-e} 
 \\
\\
&\hspace{0.5 cm}  a_{dec(dm-\nu)} \le a_{nr} \\
&\hspace{0.5 cm}  \widetilde{\Gamma }_{dm-\nu } = 
\left(\frac{\g_{*dm}}{\g_{*\nu}}\right)^\frac{1}{3} \widetilde{\Gamma }_{dec(\nu-e) } 
 \\
&\hspace{0.5 cm}\widetilde{\Gamma }_{\nu-dm } = 
 \frac{\g_{*dm}}{\g_{*\nu}} \widetilde{\Gamma }_{dm-\nu }  
\\
&\hspace{0.5 cm}\kappa_{dm} \svb_{\nu-dm}  =   \left(\frac{\g_{*dm}}{\g_{*\nu}}\right)^\frac{1}{3}  \snu \\
& \\ \hline\\
\mbox{A-C} & \hspace{0.5 cm} 
\widetilde{\Gamma }_{dm-\nu}  >  \widetilde{\Gamma }_{\nu-dm} \ \ \ \ \ \,
\widetilde{\Gamma }_{dm-\nu}   =  \widetilde{\Gamma }_{\nu-e}
 \\
\\
&\hspace{0.5 cm}
 a_{dec(dm-\nu)}  > {a_{nr}} \\
&\hspace{0.5 cm}
\widetilde{\Gamma }_{dm-\nu}  =  \widetilde{\Gamma }_{dec(\nu-e) } 
\\
& \hspace{0.5 cm}\widetilde{\Gamma }_{\nu-dm}  =  \frac{a_{dec(\nu-e)}}{a_{eq(dm-\nu)} } \widetilde{\Gamma }_{dec(\nu-e)} \\ & \hspace{0.5 cm}\kappa_{dm} \svb_{\nu-dm}  = \frac{a_{dec(\nu-e)}}{a_{nr}} \snu \\ \\ \hline \\ \mbox{B-C}  
&\hspace{0.5 cm} 
\widetilde{\Gamma }_{\nu-e }   <  \widetilde{\Gamma }_{\nu-dm} 
\ \ \ \ \ \ 
\widetilde{\Gamma }_{dm-\nu}   =  \widetilde{\Gamma }_{\nu-dm} 
 \\
\\
&\hspace{0.5 cm}
a_{dec(\nu-e)} < a_{dec(dm-\nu)} \\
&\hspace{0.5 cm}\widetilde{\Gamma }_{dm-\nu} = \widetilde H_{eq(dm-\nu)} 
\\
&\hspace{0.5 cm}\widetilde{\Gamma }_{\nu-dm} = \widetilde{\Gamma }_{dm-\nu} 
\\
&\hspace{0.5 cm}\kappa_{dm} \svb_{\nu-dm} = \frac{a_{eq(dm-\nu)}}{a_{nr}} \seq  \\ \\ \hline \\
\mbox{B-C}   &\hspace{0.5 cm} 
\widetilde{\Gamma }_{\nu-e } < \widetilde{\Gamma }_{dm-\nu} < \widetilde H_{eq(dm-\nu)} \ \ \ \ \ \,
 \widetilde{\Gamma }_{dm-\nu}   =  \widetilde{H}_{nr}
 \\
\\
&\hspace{0.5 cm} a_{dec(\nu-e)} < a_{dec(dm-\nu)} = a_{nr} < a_{eq(dm-\nu)} \\ \\ \mbox{B side:} \ &\hspace{0.5 cm} 
(\mbox{where   } \ \  \widetilde{\Gamma }_{\nu-e} < \widetilde{\Gamma }_{\nu} \sim \widetilde{\Gamma }_{\nu-dm}
\sim \widetilde{\Gamma }_{dm-\nu}) \\
&\hspace{0.5 cm} \widetilde{\Gamma }_{\nu-dm} =  \frac{\g_{*dm}}{\g_{*\nu}} \widetilde H_{nr}  \\ \\ \mbox{C side:} \ &\hspace{0.5 cm}
(\mbox{where   } \ \ \widetilde{\Gamma }_{\nu-e} < \widetilde{\Gamma }_{\nu} \sim \widetilde{\Gamma }_{\nu-dm}
< \widetilde{\Gamma }_{dm-\nu}) \\
&\hspace{0.5 cm}\widetilde{\Gamma }_{\nu-dm} = \frac{a_{nr}}{a_{eq(dm-\nu)}} \widetilde H_{nr} \\ \\ \mbox{B\&C}&\hspace{0.5 cm} \kappa_{dm} \svb_{\nu-dm} =  \frac{a_{nr}}{a_{eq(dm-\nu)}} \seq \\ \\ \hline \hline \end{array} \] \end{table} \end{center}

Here, we only consider the cases for which ${a_{eq(dm-\nu)} } > {a_{nr}}$. 
The case corresponding to ${a_{eq(dm-\nu)} } < {a_{nr}}$, 
is very unnatural as discussed in Section \ref{sec:VI}.

\paragraph{A-B borderline.} For $a_{dec(dm-\nu)} > a_{nr}$, as in the {URFO} scenario, we have 
${a_{eq(dm-\nu)} } > {a_{nr}}$, which is the case that we have decided to not consider explicitely. 

\paragraph{A-C borderline.} For ${a_{dec(dm-\nu)} } <  {a_{nr}}$, \textit{i..e.} relativistic 
  Dark Matter  , one would have $\widetilde{\Gamma }_{dm-\nu} =  \widetilde{\Gamma }_{\nu-dm}$, which is not possible in Region C.

\paragraph{B-C borderline.}

For $a_{dec(dm-\nu)} = a_{nr}$, that is for
$\widetilde{\Gamma }_{dm-\nu} = \widetilde H_{nr} $,
there is a borderline where
$\widetilde{\Gamma }_{\nu-dm}$ is discontinuous.
On the relativistic side, we are in Region B, with $\widetilde{\Gamma }_{dm-\nu} = \widetilde H_{nr}  \sim \widetilde{\Gamma }_{\nu-dm}$. On the 
non-relativistic side, we are in Region C with 
$\widetilde{\Gamma }_{dm-\nu} = \widetilde H_{nr} 
> \widetilde{\Gamma }_{\nu-dm} =
\frac{a_{nr}}{a_{eq(dm-\nu)}} \widetilde H_{nr}  $.
This however can only be the case for 
$a_{dec(\nu-e)} < a_{nr} $.

\subsection{Borderlines of the  Regions in the photon  parameter space.
 \label{app:bordphot}   }

There is now one single borderline since there are only two regions A and B  of interest to compute the photon induced-damping scale.

{\it All relations given in this section are understood as being written at the epoch $a = a_{dec(dm-\gamma)}$, the ``reference time'', corresponding to the decoupling of the   Dark Matter   with photons.}

\begin{center}
\begin{table}[h]
\caption{Borderlines for the $dm-\gamma$ scenario. 
\label{tab:gambor}}
\[
\begin{array}{ll} 
\hline 
\hline
\\
\mbox{A-B} &\hspace{.5cm} \widetilde{\Gamma}_{dm-\gamma} < \widetilde{H}_{rec} 
 \ \ \  \ \ \ \  \ 
\widetilde{\Gamma}_{\gamma-dm } \ = \   \widetilde{\Gamma }_{\gamma-e }
\\
\\
 \mbox{A-B} &\hspace{3cm} \ \ \ \ \widetilde{\Gamma}_{dm-\gamma} = \widetilde{H}_{rec} \\ \hline \hline \end{array} \] \end{table} \end{center}

Provided
$\widetilde{\Gamma}_{dm-\gamma } < \widetilde{H }_{rec}$,
the first borderline (Table \ref{tab:gambor}) corresponds to
\[
\widetilde{\Gamma}_{\gamma-dm } \ = \   \widetilde{\Gamma}_{\gamma-e } \equiv \widetilde{\Gamma}_{\Th }
 \ .
\]
This may be rewritten in terms of $\widetilde{\Gamma}_{dm-\gamma }$ by means of the relations of Section \ref{app:Gamsym}.
Relativistic   Dark Matter  , in the {URFO} scenario leads to a borderline, given in Table \ref{tab:gamure}, where we have $\widetilde{\Gamma }_{dm-\gamma } = 
\frac{a_{eq(dm-\gamma)} }{a_{nr}}
\widetilde{\Gamma}_{\Th }$. For all sensible values of $a_{nr}$ namely $ a_{nr} <     a_{eq(dm-\gamma}  \widetilde{\Gamma}_{\Th}/H_{eq(dm-\gamma)}$, it corresponds also to   Dark Matter   which decouples in the matter dominated era. It is also readily seen that $\widetilde{\Gamma}_{dm-\gamma } < \widetilde{H }_{rec}$ implies 
$a_{nr} > a_{eq(dm-\gamma)} \widetilde{\Gamma}_{\Th }/ H_{rec}$. Note that the limit is reached at a value $a_{nr} = a_{eq(dm-\gamma)} \widetilde{\Gamma}_{\Th }/ H_{rec} > a_{eq(dm-\gamma)} $. 

The second borderline (see Tables \ref{tab:gamure} and \ref{tab:gamnre}), in the URFO as well as the NRFO case, corresponds to \[ \widetilde{\Gamma}_{dm-\gamma } = \widetilde{H }_{rec} 
\]
and is relevant for the smaller values of $a_{nr}$, namely $a_{nr}   <  a_{eq(dm-\gamma)} \widetilde{\Gamma}_{\Th }/ H_{rec}$, that is for the larger masses $m_{dm}$.

\begin{center}
\begin{table}[h]
\caption{Explicit borderlines for the URFO $dm-\gamma$ scenario. \label{tab:gamure}} 
\[ 
\begin{array}{ll} \hline 
\hline\\
\mbox{A-B}&\hspace{.5cm} \widetilde{\Gamma}_{dm-\gamma} < \widetilde{H}_{rec} 
 \ \ \  \ \ \ \  \ 
\widetilde{\Gamma}_{\gamma-dm } \ = \   \widetilde{\Gamma }_{\gamma-e }
\\
\\
&\hspace{.5cm} a_{dec(dm-\gamma)} < a_{rec} \\
&\hspace{.5cm} \widetilde{\Gamma }_{dm-\gamma } = 
\frac{a_{eq(dm-\gamma)} }{a_{nr}}
\widetilde{\Gamma}_{\Th } \\
&\hspace{.5cm} \widetilde{\Gamma }_{\gamma-dm } = \widetilde{\Gamma}_{\Th } \\
&\hspace{.5cm}\kappa_{dm} \svb_{\gamma-dm } = 
\frac{a_{eq(dm-\gamma)} }{a_{nr}} \frac{\widetilde n_e}{ \nb_\gamma} \sth \\ \\ \hline \\ \mbox{A-B}&\hspace{.5cm} \widetilde{\Gamma}_{dm-\gamma} = \widetilde{H}_{rec} \\ \\ &\hspace{.5cm} a_{dec(dm-\gamma)} =  a_{rec}  \\  a_{dec(dm-\gamma)} > a_{nr} 
\\
&\hspace{.5cm} \widetilde{\Gamma }_{\gamma-dm } =
\frac{a_{dec(dm-\gamma)}}{a_{eq(dm-\gamma)} }   \widetilde{H }_{rec} 
\\
&\hspace{.5cm} \kappa_{dm} \svb_{\gamma-dm } = 
\frac{a_{dec(dm-\gamma)} }{a_{nr}} \srec
 \\
a_{dec(dm-\gamma)} < a_{nr} 
\\
&\hspace{.5cm} \widetilde{\Gamma }_{\gamma-dm } =
\frac{a_{nr}}{a_{eq(dm-\gamma)} }   \widetilde{H }_{rec} 
\\
&\hspace{.5cm} \kappa_{dm} \svb_{\gamma-dm } = \srec
\\
&\\
\hline
\hline
\end{array}
\]
\end{table}
\end{center}

\begin{center}
\begin{table}[h]
\caption{Explicit borderlines for the NRFO $dm-\gamma$ scenario (Evaluated only for $a_{nr} < a_{eq(dm-\gamma)}$ as discussed in the text).  \label{tab:gamnre}} 
\[ 
\begin{array}{ll} \hline 
\hline\\
\mbox{A-B}&\hspace{2.5cm} \widetilde{\Gamma}_{dm-\gamma} = \widetilde{H}_{rec} \\ \\ &\hspace{2.5cm} a_{dec(dm-\gamma)} =  a_{rec}  \\ \\ &\hspace{2.5cm} \widetilde{\Gamma }_{\gamma-dm } =
\frac{a_{dec(dm-\gamma)}}{a_{eq(dm-\gamma)} }   \widetilde{H }_{rec} 
\\
&\hspace{2.5cm} \kappa_{dm} \svb_{\gamma-dm } = 
\frac{a_{dec(dm-\gamma)} }{a_{nr}} \srec 
\\
&\\
\hline
\hline
\end{array}
\]
\end{table}
\end{center}

\section{Damping limits. \label{app:anndamp}}

We give in this section the analytical form  of the constraint one obtains by writing that the various damping lengths have to be smaller than a given a priori chosen limit $l_{struct}$. Note that in each case, the parameter space is constrained by the relevant damping limit {\it and} the condition to be in a given "Region" in which the expression used for the damping scale is valid.

\subsection{Self-damping. \label{app:anndampsdfs}}

\begin{center}
\begin{table}[h]
\caption{ Self-damping limits .} 
\[
\begin{array}{ll}
\hline
\hline 
\\
\mbox{Region I}
\\
l_{fs} & = l_{eq}  \frac{a_{nr} }{a_{eq}} \cr
[\ \  l_{fs} & = l_{eq}  \frac{a_{nr} }{a_{eq}} \frac{5+2{\rm ln} ({a_{eq}}/{a_{nr} })}{\pi}\ \  ]  \cr l_{sd} & = l_{eq}  r_{dm} \frac{a_{dec(dm)} }{a_{eq}} 
 \\
a_{nr}  &< a_{eq} \, \frac{l_{struct}} {l_{eq}}
\\
\\
\mbox{Region II}
\\
l_{fs}= \frac{l_{sd}}{r_{rm}} & = l_{eq}  (\frac{a_{nr}a_{dec(dm)}}{a^2_{eq}})^\frac{1}{2} \cr
[\ \   l_{fs} & = l_{eq} (\frac{a_{nr}a_{dec(dm)}}{a^2_{eq}})^\frac{1}{2}
 \frac{3+2{\rm ln} ({a_{eq}}/{a_{dec} })}{\pi}\ \   ]  
\\
a_{nr}  &<\frac{a^2_{eq}}{a_{dec(dm)}}\left( \frac{l_{struct}} {l_{eq}}\right)^2 \\ \\ \mbox{Region III} \\ l_{fs}= \frac{l_{sd}}{r_{rm}} & = l_{eq}  (\frac{a_{nr}}{a_{eq}})^\frac{1}{2} \cr
[\ \   l_{fs} & = l_{eq} (\frac{a_{nr}}{a_{eq}})^\frac{1}{2}  \frac{3}{\pi}\ \   ] 
 \\
a_{nr}  &< {a_{eq}}\left( \frac{l_{struct}} {l_{eq}}\right)^2 \\ \\ \mbox{Region III'} \\ \frac{l_{sd}}{r_{rm} } &= l_{eq} \left(\frac{\widetilde H_{nl}}{\widetilde \Gamma_{dm}}\right)^{\frac{1}{2}} (\frac{a_{nr} }{a_{eq}})^{1/2}\ , 
 \\
a_{nr}  &<\frac{\widetilde \Gamma_{dm}} {\widetilde H_{nl}} {a_{eq}}  \left( \frac{l_{struct}} {l_{eq}}\right)^2 \\ \\ \mbox{Region IV} \\ l_{fs} & = l_{eq} \frac{a_{nr}}{a_{eq}} \cr l_{sd} & = l_{eq}  r_{dm} \frac{a_{dec(dm)}}{a_{eq}} \\ a_{nr}  &< a_{eq} \, \frac{l_{struct}} {l_{eq}} \\ \\ \mbox{Region V} \\ l_{fs} & = l_{eq}  \frac{a_{nr}}{a_{eq}} \cr l_{sd} & = l_{eq}  r_{dm} \frac{a_{dec(dm)}}{a_{eq}}  \\ a_{nr}  &<  a_{eq} \,  \frac{l_{struct}} {l_{eq}} \\ \\ \mbox{Region VI} \\ l_{fs}= \frac{l_{sd}}{r_{rm}} & = l_{eq}  (\frac{a_{nr}}{a_{eq}})^\frac{1}{2} 
 \\
a_{nr}  &<{a_{eq}} \, \left( \frac{l_{struct}} {l_{eq}}\right)^2 \\ \\ \mbox{Region VI'} \\ \frac{l_{sd}}{r_{rm} } &= l_{eq} \left(\frac{\widetilde H_{nl}}{\widetilde \Gamma_{dm}}\right)^{\frac{1}{2}} (\frac{a_{nr}}{a_{eq}})^{1/2} \ ,  
\\
a_{nr}  &<\frac{\widetilde \Gamma_{dm}} {\widetilde H_{nl}} {a_{eq}}  \left( \frac{l_{struct}} {l_{eq}}\right)^2 \\ \\ \hline \hline \end{array} \] \label{tab:sda} \end{table} \end{center}


The damping scales in this case are given as a fraction of  the horizon at decoupling $\pi c t_r a_{dec(dm)}$, or $\pi c t_m a_{dec(dm)}^\frac{1}{2}$. Noting that $t_m$ differs from  $t_r$ basically by a factor $ a_{eq}^\frac{1}{2}$ (Sect. \ref{app:times}), one sees all damping lengths can be expressed as a fraction of 
\[ l_{eq} = \pi c t_r  a_{eq} = 207 \ Mpc  \,  \, \g_* ^{\frac{1}{2}}  \, \left(\Om \right)^{-1} \, . \]
The analytic expression of the damping scales and the associated limits on the interaction rates and the $\svb$ cross-sections are given in table \ref{tab:sda} for the six Regions which cover the parameter space.  The free-streaming damping scale has been calculated via the approximate expression (\ref{appintfs}), to be directly comparable to the expression of the self-damping length, estimated using a similar approximation (\ref{appintcd}) and to be related to the mass scale by (\ref{mlrel}). In Regions I,II and III where our limits are set, we give for completeness -within brackets- also the expression of the free-streaming scale when it is by convention taken to be given by (\ref{intfs}). {\it All relations given in this section are understood as being written at the epoch $a = a_{dec(dm)}$, the ``reference time'', corresponding to the decoupling of   Dark Matter   with every species including itself.}
\subsection{Neutrino induced-damping. \label{app:anndampnu}}

The typical scale in this case is the damping scale of the neutrinos at the time they decouple from the electrons \[ l_{dec(\nu-e)}  =  
\ \pi ct_r r_\nu \frac{\widetilde{\Gamma }_{dec(\nu-e) }}{H_r} =  97.1 \ pc \ r_{\nu} \ { \kappa^{-\frac{2}{3}}}{\g_*^{-\frac{2}{3}}}
 \  .
\]

For Region A, the damping is too small to be relevant to structure formation. For Region B, only the case $a_{dec(dm-\nu)} < a_{eq(dm-\nu)}$ has been considered explicitly: for $a_{dec(dm-\nu)} > a_{eq(dm-\nu)}$, the damping is too large. Needless to say, in the latter case, too, the damping may be readily evaluated by means of the same methods.

Analytical expression of the damping scales and limits on interaction rates and $\svb$ are given in tables \ref{tab:Anua}, \ref{tab:BnuURFOa}, \ref{tab:BnuNRFOa}, \ref{tab:CnuURFOa} and, \ref{tab:CnuNRFOa}.
{\it All relations given in this section are understood as being written at the epoch $a = a_{dec(dm-\nu)}$, the ``reference time'', corresponding to the decoupling of the   Dark Matter   with photons.}

\begin{center}
\begin{table}[h]
\caption{ Region A ($dm-\nu$ scenario).} 
\[
\begin{array}{ll}
\hline
\hline
 \\
 l_{\nu d} \,\,\,\,\,\,\,\,\,\,\,\,\,\,\,\,\,\,\,\,\,\,\,\,\,\,\,\,\,\,\, = \, l_{dec(\nu-e)} (\widetilde{\Gamma }_{dm-\nu}/\widetilde{\Gamma }_{dec(\nu-e)})^\frac{5}{2} \\ \\ \hline \hline \end{array} \] \label{tab:Anua} \end{table} \end{center}

\begin{center}
\begin{table}[h]
\caption{ Region B (URFO $dm-\nu$ scenario).} 
\[
\begin{array}{ll}
\hline
\hline
 \\
l_{\nu d} & = l_{dec(\nu-e)} ( \frac{a_{eq(dm-\nu)}}{a_{nr}})^{\frac{1}{2}} 
 \frac{\widetilde{\Gamma }_{dm-\nu}}{\widetilde{\Gamma }_{dec(\nu-e)}} 
 \\
\\
\widetilde{\Gamma }_{dm-\nu}  &<
\widetilde{\Gamma }_{dec(\nu-e)} 
 \left( \frac{a_{nr}} {a_{eq(dm-\nu)}}\right)^{\frac{1}{2}} 
\frac{l_{struct}}{l_{dec(\nu-e)}} 
\\
\\
\kappa_{dm} \svb_{\nu-dm}  &=
\frac{\widetilde{\Gamma }_{dm-\nu}}{\dslash n_\nu}
 \\
&<\snu  
\left( \frac{a_{nr}} {a_{eq(dm-\nu)}}\right)^{\frac{1}{2}} 
\frac{l_{struct}}{l_{dec(\nu-e)}} 
\\
\\
\hline
\hline
\end{array}
\]
\label{tab:BnuURFOa}
\end{table}
\end{center}

\begin{center}
\begin{table}[h]
\caption{Region B (NRFO  $dm-\nu$ scenario).} 
\[
\begin{array}{ll}
\hline
\hline \\
l_{\nu d} & = l_{dec(\nu-e)} 
\left( \frac{\g_{*\nu}}{ \g_{*dm}}\right)^{\frac{1}{2}}  
\frac{\widetilde{\Gamma }_{dm-\nu}}{\widetilde{\Gamma }_{dec(\nu-e)}} 
 \\
\\
\widetilde{\Gamma }_{dm-\nu}  &<
\widetilde{\Gamma }_{dec(\nu-e)}  
\left( \frac{ \g_{*dm}}{\g_{*\nu}}\right)^{\frac{1}{2}}  
\frac{l_{struct}}{l_{dec(\nu-e)}} 
\\
\\
\kappa_{dm} \svb_{\nu-dm}  &=
\frac{\widetilde{\Gamma }_{dm-\nu}}{\dslash n_\nu}
\nonumber 
<\snu 
\left( \frac{ \g_{*dm}}{\g_{*\nu}}\right)^{\frac{1}{2}} 
 \frac{l_{struct}}{l_{dec(\nu-e)}} 
\\
\\
\hline
\hline
\end{array}
\]

\label{tab:BnuNRFOa}
\end{table}
\end{center}

\begin{center}
\begin{table}[h]
\caption{ Region C (URFO  $dm-\nu$ scenario).} 
\[
\begin{array}{ll}
\hline
\hline 
\\
l_{\nu d} & = l_{dec(\nu-e)}  \frac{\widetilde{\Gamma }_{dm-\nu}}{\widetilde{\Gamma }_{dec(\nu-e)}} 
\\
\\
\widetilde{\Gamma }_{dm-\nu}  &<
\widetilde{\Gamma }_{dec(\nu-e)} \frac{l_{struct}}{l_{dec(\nu-e)}} 
\\
\\
a_{dec(dm-\nu)} > a_{nr}  \ \ \ &
\\
\kappa_{dm} \svb_{\nu-dm}  &= \frac{a_{dec(dm-\nu)} }{ a_{nr} } \frac{\widetilde{\Gamma }_{dm-\nu}}{\dslash{n}_\nu} \nonumber \\ &<\snu  
 \frac{a_{dec(\nu-e)}}{a_{nr}} \left(\frac{l_{struct}}{l_{dec(\nu-e)}}  \right)^2 \\ \\ a_{dec(dm-\nu)} < a_{nr} \ \ \  & \\ \kappa_{dm} \svb_{\nu-dm}  &= \frac{\widetilde{\Gamma }_{dm-\nu}}{\dslash{n}_\nu} \nonumber \\ &<\snu  \frac{l_{struct}}{l_{dec(\nu-e)}} 
\\
\\
\hline
\hline
\end{array}
\]
\label{tab:CnuURFOa}
\end{table}
\end{center}

\begin{center}
\begin{table}[h]
\caption{Region C (NRFO  $dm-\nu$ scenario).} 
\[
\begin{array}{ll}
\hline
\hline \\
l_{\nu d} & = l_{dec(\nu-e)} \frac{\widetilde{\Gamma }_{dm-\nu}}{\widetilde{\Gamma }_{dec(\nu-e)}} 
 \\
\\
\widetilde{\Gamma }_{dm-\nu}  &<
\widetilde{\Gamma }_{dec(\nu-e)}  \frac{l_{struct}}{l_{dec(\nu-e)}} 
\\
\\
a_{dec(dm-\nu)} > a_{nr}  \ \ \ &
\\
\kappa_{dm} \svb_{\nu-dm}  &= \frac{a_{dec(dm-\nu)} }{ a_{nr} } \frac{\widetilde{\Gamma }_{dm-\nu}}{\dslash{n}_\nu} \nonumber \\ &<\snu  
 \frac{a_{dec(\nu-e)}}{a_{nr}} \left(\frac{l_{struct}}{l_{dec(\nu-e)}}  \right)^2 \\ \\ a_{dec(dm-\nu)} < a_{nr} \ \ \  & \\ \hbox{does not exist in this case} \\ \\ \hline \hline \end{array} \] \label{tab:CnuNRFOa} \end{table} \end{center}

\subsection{Photon induced-damping.\label{app:anndampphot}}

A unit to characterize the damping scale in this case may be chosen as the horizon scale at equality, wheighted by a slightly time-dependent factor $r_\gamma$, to be taken at the proper time: \begin{eqnarray} l_{eq(dm-\gamma)} &=& \pi ct_r r_\gamma a_{eq(dm-\gamma)} \cr 
&=& 249 \, Mpc  \,  r_\gamma  \, \g_* ^{-\frac{1}{2}}   \frac{4\g_{*i}(T)}{3}  \, \left(\Odm \right)^{-1}
\nonumber
\, .
\end{eqnarray}

In Region A, rather than the URFO and NRFO scenarios which turn out here to yield the same damping limits, it turns out to be useful to distinguish two cases, depending on whether   Dark Matter   decoupling occurs in the radiation dominated era ($a_{dec(dm-\gamma)} < a_{eq}$), or in the matter dominated era ($a_{dec(dm-\gamma)} > a_{eq}$). We do not give the damping lengths when   Dark Matter   decouples from the photons in the matter dominated era since they are always prohibitive as compared to our requirements.

Analytical expression of the damping scales and limits on interaction rates and $\svb$ are given in Table \ref{tab:Agarda} for decoupling in the radiation dominated era. They are obtained from the expression of the damping length given in Section ref{sec:photlim} by means of the relations given in Table \ref{tab:irxs} and the definitions of Appendix \ref{app:def}. {\it All relations given in this section are understood as being written at the epoch $a = a_{dec(dm-\gamma)}$, the ``reference time'', corresponding to the decoupling of the   Dark Matter   with photons.}

\begin{center}
\begin{table}[h]
\caption{ Region A, radiation dominated 
(URFO and NRFO $dm-\gamma$ scenario).}
\[
\begin{array}{ll}
\hline
\hline 
\\
l_{\gamma d} & = l_{eq(dm-\gamma)} 
\left( \frac{\widetilde{H }_{eq(dm-\gamma)}} {\widetilde{\Gamma }_{\Th}}\right)^\frac{1}{2} \left( \frac{\widetilde{\Gamma }_{dm-\gamma}}{\widetilde{H }_{eq(dm-\gamma)}} \right)^\frac{3}{2} \\ \\ \widetilde{\Gamma }_{dm-\gamma}  &< \widetilde{H }_{eq(dm-\gamma)} \left( \frac{\widetilde{H }_{eq(dm-\gamma)}} {\widetilde{\Gamma }_{\Th}}\right)^{-\frac{1}{3}}  \left( \frac{l_{struct}}{l_{eq(dm-\gamma)}} \right)^\frac{2}{3} \\ \\ \\ a_{dec(dm-\gamma)} &>  a_{nr} 
\\
\kappa_{dm} \svb_{\gamma-dm}  &= \frac{a_{dec(dm-\gamma)} }{ a_{nr} } \frac{\widetilde{\Gamma }_{dm-\gamma}}{\dslash{n}_\gamma} \nonumber \\ &<\seq  
\left( \frac{\widetilde{H }_{eq(dm-\gamma)}} {\widetilde{\Gamma }_{\Th}}\right)^{-\frac{2}{3}}  \frac{a_{eq(dm-\gamma)}}{a_{nr}} \left(\frac{l_{struct}}{l_{eq(dm-\gamma)}}  \right)^\frac{4}{3} \\ \\ \\
a_{dec(dm-\gamma)} &<   a_{nr}
\\
\kappa_{dm} \svb_{\gamma-dm}  &=
\frac{\widetilde{\Gamma }_{dm-\gamma}}{\dslash{n}_\gamma}
 \\
&<\seq  
\left( \frac{\widetilde{H }_{eq(dm-\gamma)}} {\widetilde{\Gamma }_{\Th}}\right)^{-\frac{1}{3}} \left(\frac{l_{struct}}{l_{eq(dm-\gamma)}}  \right)^\frac{2}{3} \\ \\ \hline \hline \end{array} \] \label{tab:Agarda} \end{table} \end{center}


\end{document}